\documentclass[12pt]{amsart}
\pdfoutput=1

\usepackage{hyperref}

\usepackage{amsmath}
\usepackage[norelsize]{algorithm2e}
\usepackage{scalerel}
\usepackage{tikz-cd}
\usepackage{ulem}
\usepackage{float}
\restylefloat{figure}
\usepackage{stackengine}
\usepackage[english]{babel}
\usepackage[light,condensed,math]{anttor}
\usepackage[T1]{fontenc}
\usepackage{kpfonts}
\usepackage{comment}
\usepackage{array}
\usepackage{lmodern}
\usepackage[mathscr]{euscript}
\usepackage[utf8x]{inputenc}
\usepackage{breqn}
\usepackage{todonotes}
\usepackage{slashed}
\usepackage{youngtab}
\usepackage{amsmath}
\usepackage{amssymb}
\usepackage{amsxtra}
\usepackage{color}
\usepackage[margin=1.2in]{geometry}

\newcommand{\boxit}[1]{\vbox{\hrule\hbox{\vrule\kern8pt
\vbox{\hbox{\kern8pt}\hbox{\vbox{#1}}\hbox{\kern8pt}}
\kern8pt\vrule}\hrule}}
\newcommand{\mathboxit}[1]{\vbox{\hrule\hbox{\vrule\kern8pt\vbox{\kern8pt
\hbox{$\displaystyle #1$}\kern8pt}\kern8pt\vrule}\hrule}}

\newcommand{\picit}[2]{\includegraphics[width=#1cm]{#2.eps}}

\newcommand{\gcr}{{\mathfrak X}} 

 \newcommand{\beq}{\begin{equation}}
                \newcommand{\bea}{\begin{eqnarray}}
                \newcommand{\eea}{\end{eqnarray}}
                 \newcommand{\eeq}{\end{equation}}

\newcommand{\iM}{{\mathscr M}}
\newcommand{\Lf}{{\mathfrak{L}}}

\newcommand{\sV}{{\mathfrak{V}_{\gcr}}}
\newcommand{\sE}{{\mathfrak{E}_{\gcr}}}

\newcommand{\mM}{{\mathfrak M}}

\newcommand {\BC}   {\mathbb C}

\newcommand {\BN}   {\mathbb N}

\newcommand {\BR}   {\mathbb R}
\newcommand {\BP}   {\mathbb P}

\newcommand {\BT}   {\mathbb T}
\newcommand {\bB}   {\mathbf{B}}
\newcommand {\bI}   {\mathbf{I}}
\newcommand {\bN}   {\mathbf{N}}

\newcommand {\bJ}   {\mathbf{J}}

\newcommand {\qe} {\mathfrak q}

\newcommand {\Hf} {\mathsf{H}}

\newcommand {\ii} {\mathrm{i}}
\newcommand {\bT}   {\mathbf{T}}
\newcommand {\ba}  {\underline{\ac}}

\newcommand {\bn}{\underline{\mathbf{n}}}
\newcommand {\mm}{\underline{\mathbf{m}}}

\newcommand {\bk}{  \mathbf{k}}

\newcommand {\bw} {\underline{\mathbf{w}}}

\newcommand {\xt} {\tt x}

\newcommand {\ept} {\underline{\ec}}

\newcommand {\bqt} {\underline{\qe}}

\newcommand {\bkt} {\underline{\bk}}

\newcommand {\bX}{ \mathbf{X}}

\newcommand {\bS}{ \mathbf{S}}
\newcommand {\BS}   {\mathbb S}

\newcommand {\BZ}   {\mathbb Z}

\newcommand {\ac} {\mathfrak{a}}
\newcommand {\ec} {\mathfrak{e}}

\newcommand {\zb} {{\bar z}}

\newcommand {\CalA} {\mathcal A}

\newcommand {\CalE} {\mathcal E}

\newcommand {\CalG} {\mathcal G}
\newcommand {\CalH} {\mathcal H}
\newcommand {\CalI} {\mathcal I}

\newcommand {\CalN} {\mathcal N}

\newcommand {\CalR} {\mathcal R}
\newcommand {\CalS} {\mathcal S}
\newcommand {\CalT} {\mathcal T}

\newcommand {\CalX} {\mathcal X}

\newcommand {\CalZ} {\mathcal Z}

\newcommand {\es} {\mathscr E}

\newcommand {\hs} {\mathscr H}

\newcommand{\al}{\alpha}
\newcommand{\ve}{\varepsilon}
\newcommand{\ep}{\epsilon}

\renewcommand{\hat}{\widehat}

\newcommand{\Gg}{\mathsf{G}_{\mathbf{g}}}

\newcommand{\Gr}{\mathsf{G}_{\text{\tiny rot}}}

\newcommand{\Vg}{\mathsf{V}_{\gamma}}
\newcommand{\Vgp}{\mathsf{V}_{\gamma^{+}}}
\newcommand{\Vgm}{\mathsf{V}_{\gamma^{-}}}
\newcommand{\Vgmp}{\mathsf{V}_{\gamma^{\mp}}}

\newcommand{\Eg}{\mathsf{E}_{\gamma}}
\newcommand{\Egp}{\mathsf{E}_{\gamma^{+}}}
\newcommand{\Egpm}{\mathsf{E}_{\gamma^{\pm}}}
\newcommand{\Egm}{\mathsf{E}_{\gamma^{-}}}

\newcommand{\Tr}{\mathsf{Tr}\,}

\newcommand {\3}{\underline{\bf 3}}
\newcommand {\4}{\underline{\bf 4}}
\newcommand {\6}{\underline{\bf 6}}
\usepackage{float}
\restylefloat{figure}
\newcommand {\Hfr} {\mathfrak{H}}
\newcommand {\be}{\underline{\mathbf{e}}}
\newcommand {\fs} {\mathfrak{s}}

\newcommand{\fo}{\vert\kern -.03in\_}
\newcommand {\bht}{\underline{\mathbf{h}}}
\begin{document}
\title[BPS/CFT, Instantons at crossroads, Gauge origami]{BPS/CFT correspondence II :\\
 Instantons at crossroads, \\
 Moduli and Compactness Theorem}

\author{$\mathrm{Nikita\ Nekrasov}$}

\address{Simons Center for Geometry and Physics\\
Stony Brook University, Stony Brook NY 11794-3636, USA
\\
E-mail: nikitastring@gmail.com\footnote{on leave of absence from:
IHES, Bures-sur-Yvette, France\\
ITEP and IITP, Moscow, Russia}}

\begin{abstract}
Gieseker-Nakajima moduli  spaces ${\iM}_{k}(n)$ parametrize the charge $k$ noncommutative $U(n)$ instantons on ${\BR}^{4}$ 
and  framed rank $n$ torsion free sheaves $\mathcal{E}$
on ${\BC\BP}^{2}$ with ${\rm ch}_{2}({\mathcal{E}}) = k$. They also serve as local models of the moduli spaces of instantons on general four-manifolds. 
We study the generalization of gauge theory in which the four dimensional spacetime is a stratified space  $X$ immersed into a Calabi-Yau fourfold $Z$. The local model  ${\mM}_{k}({\vec n})$ of the corresponding instanton moduli space
 is the moduli space of charge $k$ (noncommutative) instantons on origami spacetimes. There, $X$ is modelled on a union  of (up to six)  
coordinate complex planes ${\BC}^{2}$ intersecting in $Z$ modelled on ${\BC}^{4}$.  The instantons are shared by the collection of four dimensional gauge theories sewn along two dimensional defect surfaces and defect points. We also define several
quiver versions ${\mM}_{\bk}^{\gamma}({\vec{\bn}})$ of ${\mM}_{k}({\vec n})$, motivated by the considerations of sewn gauge theories on orbifolds ${\BC}^{4}/{\Gamma}$. 

The geometry of the spaces ${\mM}_{\bk}^{\gamma}({\vec{\bn}})$, more specifically the compactness of the set of torus-fixed points, for various tori, underlies the non-perturbative Dyson-Schwinger identities recently found to be satisfied by the correlation functions of $qq$-characters viewed as local gauge invariant operators in the ${\CalN}=2$ quiver gauge theories.  

The cohomological and K-theoretic operations defined
using ${\mM}_{k}({\vec n})$ and their quiver versions as correspondences provide the geometric counterpart of the $qq$-characters, line and surface defects. 
\end{abstract}

\maketitle

\setcounter{tocdepth}{2} 
\tableofcontents

\section{Introduction}

Recently we introduced a set of observables in quiver ${\CalN}=2$ supersymmetric gauge theories which
are useful in organizing the non-perturbative Dyson-Schwinger equations, relating contributions of different instanton sectors to the expectation values of gauge invariant chiral ring observables. In this paper we shall provide the natural geometric
setting for these observables. We also explain the
gauge and string theory motivations for these considerations.

\uwave{$\mathbf{Notations.}$}
In our story we explore moduli spaces, which parametrize, roughly speaking, the sheaves $\CalE$ supported on a union of coordinate complex two-planes $\approx {\BC}^{2}$
inside ${\BC}^{4}$. The ${\BC}^4$ with a set of $\BC^2$'s inside is a local model $(Z^{loc}, X^{loc}$ of a Calabi-Yau fourfold $Z$
containing a possibly singular surface $X \subset Z$:
\beq
Z^{loc} = {\BC}^{4}, \qquad X^{loc} = \bigcup_{A \in {\6}}\, {\BC}^{2}_{A} , \qquad {\rm supp}({\mathcal E}) = \bigcup_{A \in {\6}} \ n_{A} \,{\BC}^{2}_{A}
\eeq
We denote by $\4$ the set of complex coordinates in ${\BC}^{4}$:
\beq
{\4} = \{  1, 2, 3, 4\} \, , \qquad a \in {\4} \ \leftrightarrow \  z_{a} \in {\BC}
\eeq and by 
\beq
{\6} = \left( \begin{matrix} {\4} \\ 2 \end{matrix} \right) = \Biggl\{ \ \{ 1, 2 \}, \{ 1, 3\} , \{ 1, 4 \} , \{ 2, 3 \}, \{ 2, 4\} , \{ 3, 4 \} \ \Biggr\} 
\eeq the set of two-element subsets of $\4$, i.e. the set of coordinate two-planes in ${\BC}^{4}$.

We shall sometimes denote the elements of ${\6}$ by the pairs $ab = ba\, \leftrightarrow \, \{ a, b \} \in {\6}$. We also define, for $A \in {\6}$, ${\bar A} = {\4} \backslash A$, and 
\beq
{\ep}(A) = {\ve}_{abcd}, \qquad A = \{ a , b \}, \ {\bar A} = \{ c , d\}, \qquad a< b, \, c< d
\eeq 
so that, e.g. $\overline{12} = 34$, ${\ep}(23) = {\ve}_{2314} = 1$, ${\ep}(24)= {\ve}_{2413} = -1$. 

The two-plane ${\BC}^{2}_{A} \subset {\BC}^{4}$ corresponding to $A \in {\6}$ is defined by the equations: $z_{\bar a} =0$, for all ${\bar a} \in {\bar A}$. 

{}We denote by $\3$ the quotient ${\6}/{\BZ}_{2}$ where ${\BZ}_{2}$
acts by the involution $A \mapsto {\bar A}$. The elements ${\sf a} \in {\3}$ are the unordered pairs $(A, {\bar A})$. 

{}We can visualize the sets $\3$, $\4$, $\6$, using 
the tetrahedron:
\begin{figure}[H]
\picit{3}{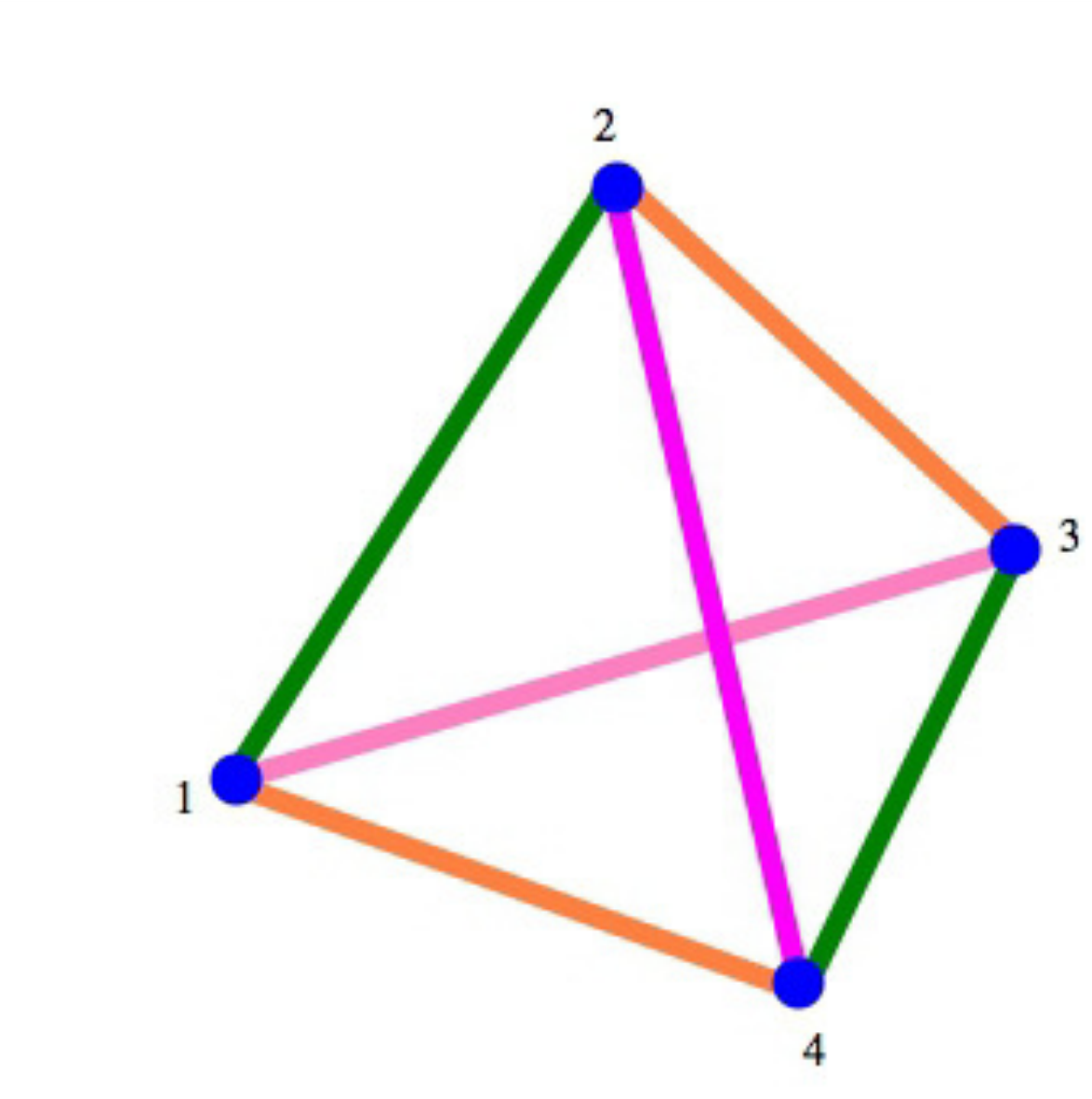}
\caption{\vbox{\hbox{Tetrahedron with the sets $\4$ and $\6$ of vertices and edges,}
\hbox{and the set ${\3} = \{ {\rm\color{magenta} red}, \, {\rm\color{green} green}, \, {\rm\color{orange} orange} \}$ of crossed edges}}}
\label{fig:tetra}
\end{figure}

{}Our story will involve four complex parameters $q_{a} \in {\BC}^{\times}$, $a \in {\4}$, obeying 
\beq
\prod_{a \in {\4}} \ q_{a} = 1
\eeq
We shall also use the additive variables ${\ec}_{a} \in {\BC}$, $a \in {\4}$, obeying
\beq
\sum_{a \in {\4}} {\ec}_{a} = 0
\eeq
Define the lattice
\beq
Z_{\ept} = {\BZ}{\ec}_{1} + {\BZ} {\ec}_{2} + {\BZ} {\ec}_{3} \subset {\BC}
\label{eq:zlattice}
\eeq
which is the image of the projection:
\beq
{\BZ}^{4} \to {\BC} \, , \qquad (i,j,k,l) \mapsto  {\ec}_{1} i + {\ec}_{2} j + {\ec}_{3} k + {\ec}_{4} l 
\eeq
We shall use the following functions on ${\6}$:
\beq
\begin{aligned}
& p_{A} = \prod_{a\in A} \, ( 1 -q_{a})\ , \\ 
& q_{A} = \prod_{a \in A} \, q_{a} \ , \\ 
& {\ec}_{A} = \sum_{a \in A} {\ec}_{a} = - {\ec}_{\bar A} \, . \\
 \label{eq:pqa}
\end{aligned}
\eeq
{}In what follows we denote by $[n]$, for $n \in {\BZ}_{>0}$, the set
$\{ 1, 2, 3, \ldots, n \} \subset {\BZ}_{>0}$. 

Let $S$ be a finite set, and $( V_{s} )_{s \in S}$ a collection of vector spaces. We use the notation
\beq
\sum_{s \in S} V_{s} 
\eeq
for the vector space which consists of all linear combinations
\beq
\sum_{s \in S} \, {\psi}_{s} \  , \qquad {\psi}_{s} \in V_{s}\ .
\label{eq:psis}
\eeq

\subsection{Organization of the paper}

We review
 the gauge and string theory motivation in the section $\bf 2$. The moduli space
 ${\mM}_{k}({\vec n})$ of spiked instantons is introduced in the section $\bf 3$.  The symmetries of spiked instantons are studied in the section $\bf 4$.
The moduli space of ordinary $U(n)$  instantons on (noncommutative) ${\BR}^{4}$ is reviewed in section $\bf 5$. The section $\bf 6$ discusses in more detail two particular cases of spiked instantons, the crossed instantons and the folded instantons. The crossed instantons live on two four-dimensional manifolds transversely intersecting in the eight-dimensional ambient manifold (a Calabi-Yau fourfold), the folded ones live on two four-dimensional manifolds intersecting transversely in the six dimensional ambient manifold.
The section $\bf 7$ constructs the spiked instantons out of the ordinary ones, and studies the toric spiked instantons in some detail. The section $\bf 8$ is the main result of this paper: the compactness theorem.  In section $\bf 9$ we enter the theory of integration over the spiked and crossed instantons, and relate the analyticity of the partition functions to the compactness theorem. 
The section $\bf 10$ discusses the $ADE$-quiver generalizations of crossed instantons. The section $\bf 11$ describes the spiked instantons on cyclic orbifolds, and the associated compactness
theorem.  The section $\bf 12$ is devoted to future directions and open questions. 

\subsection{Acknowledgements}

{}Research was supported in part by the NSF grant PHY 1404446. I am grateful to A.~Okounkov, V.~Pestun and S.~Shatashvili for discussions. I would also like to thank  Alex DiRe, Saebyeok Jeong, Xinyu Zhang, Naveen Prabhakar and Olexander Tsymbalyuk for their feedback and for painfully checking some of the predictions of the compactness theorem proven in this paper.

{}The constructions of this paper were reported in 2014-2016 in a series of lectures at the Simons Center for Geometry and Physics  \url{http://scgp.stonybrook.edu/video_portal/video.php?id=2202}, at the Institute for Advanced Studies at Hebrew University \url{https://www.youtube.com/watch?v=vGNfXQ3-Rjg}, 
at the Center for Mathematical Sciences and Applications at
Harvard University \url{http://cmsa.fas.harvard.edu/nikita-nekrasov-crossed-instantons-qq-character/} and at the String-Math-2015 conference in Sanya, China.

\section{Gauge and string theory motivations}

\subsection{Generalized gauge theory}

We study the moduli spaces ${\mM}_{X, G}$ of what might be called supersymmetric gauge fields in the generalized gauge theories, whose space-time $X$ contains several, possibly intersecting, components:  
\begin{figure}
\picit{5}{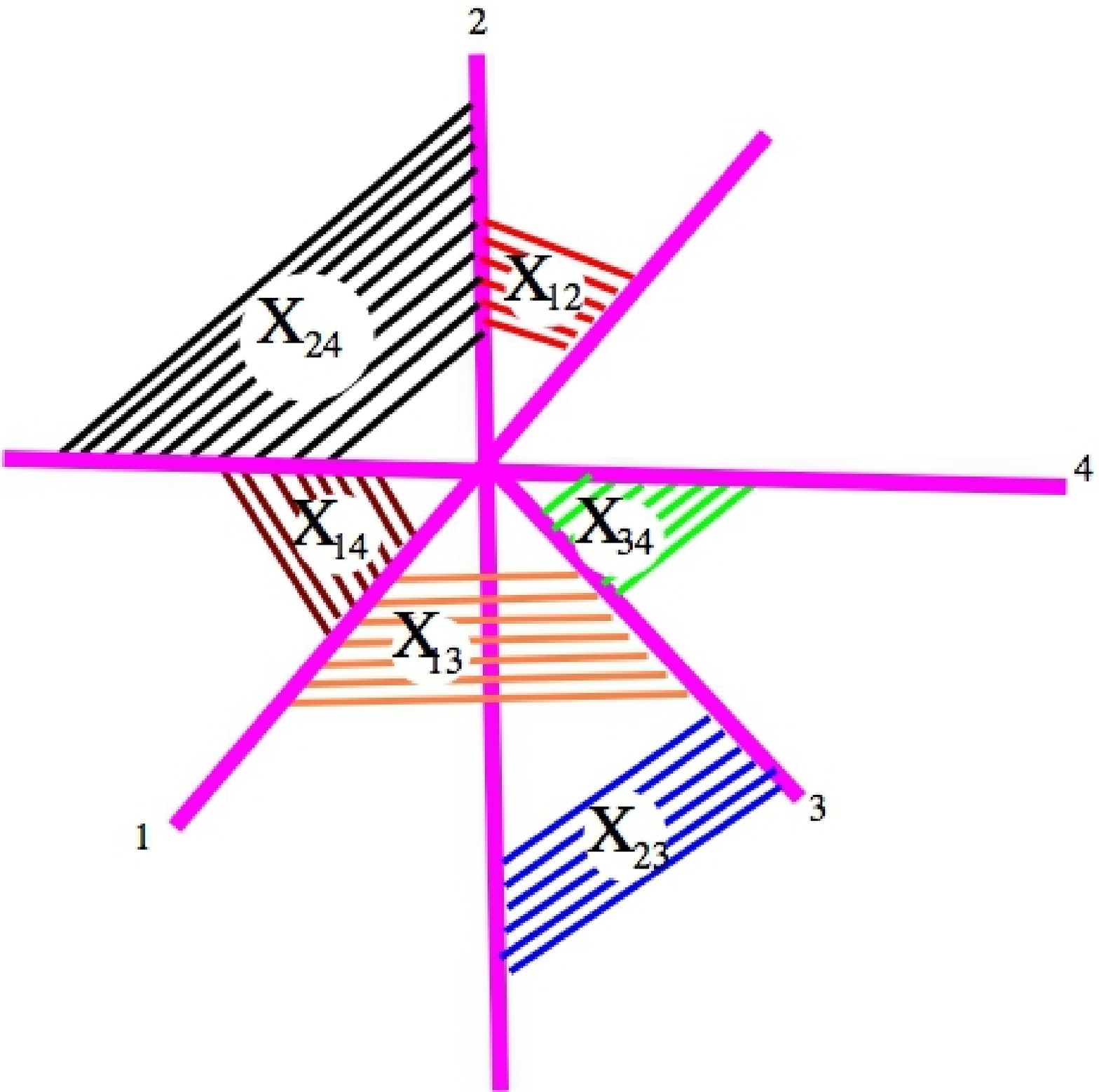}
\caption{The origami wolrdvolume $X = \bigcup\limits_{A}\ X_{A}$}
\label{fig:dbranesvac}
\end{figure}
see Fig. \ref{fig:dbranesvac}.
We call such $X$ the {\it origami worldvolume}. 
The gauge groups $G\vert_{X_{A}} = G_{A}$ on different components may be different. The intersections $X_{A} \cup X_{B}$ lead to the  bi-fundamental matter fields charged under $G_{A} \times G_{B}$.  The arrangement is motivated by the string theory considerations, where the open string Hilbert space, in the presence of several $D$-branes, splits into sectors labelled by the boundary conditions. It is well-known \cite{Seiberg:1999vs, Douglas:2001ba} that some features of the open string theory are captured by the noncommutative gauge theory. 
In fact, the theories we shall study descend from the maximally supersymmetric Yang-Mills theory, which is twisted and deformed. One can view the fields of this theory as describing the
deformations of the four dimensional stratified manifolds ${\bX} = (X_{A}, n_{A})$, i.e. singular, in general, spaces, which can be represented as unions $X = \cup_{A} X_{A}$ of manifolds with certain conditions on closures and intersections, endowed with multiplicities, i.e. the strata $X_{A}$ are allowed to have different multiplicity  $n_{A}$. The local gauge group $G_{A}$ is simply $U(n_{A})$. The particular twist of the super-Yang-Mills theory we study corresponds to $X \hookrightarrow Z \times E$, where $E$ is a two torus ${\BT}^2$, a cylinder ${\BR}^{1} \times {\BS}^{1}$, or a plane ${\BR}^{2}$, while $Z$ is a special holonomy eight dimensional manifold, e.g. the Calabi-Yau fourfold.

\subsection{Gauge origami}

Now suppose $Z$ has non-trivial isometries (it ought to be non-compact). It is natural, in this case, to deform the problem to take into account the symmetries of $Z$. The partition function of the theory of stratified multiple ${\bX}$'s localizes onto the set of fixed points, which are the configurations ${\bX} = (X_{A}, n_{A})$ where $X_{A}$'s are invariant under the isometries of $Z$. For example, when $Z$ is toric, with the three dimensional torus ${\bT}$ acting by isometries,
preserving the holomorphic top form, then at each vertex $z \in Z^{\bT}$ pass at most six strata $X_{A}$, $A \in {\6}$. 

We are interested in integrals over the moduli spaces ${\mM}_{{\bX}, G}$.  We shall view ${\mM}_{{\bX},G}$ as  the ``space, defined by some equations modulo symmetry''. More formally, ${\mM}_{{\bX},G}$ is the quotient of a set of zeroes of some $\Gg$-equivariant section $s: M \to V$ of $\Gg$-equivariant vector bundle $V \to M$ over some smooth space (vector space in our case) with $\Gg$-action, with some Lie group $\Gg$. 
If $M$ is compact the integral over ${\mM}_{{\bX},G}$ of a closed differential form
can be represented by the $\Gg$-equivariant integral over $M$ of the pull-back of the corresponding form times the Euler class of $V$. In the non-compact case one uses equivariant cohomology (mass deformation, in the physics language) with respect to both $\Gg$ and some global symmetry group $\Hf$, and Mathai-Quillen representatives of the Euler class. 

The resulting partition functions
\beq
{\CalZ}_{{\bX},G}({\xi}) \sim \int_{{\mM}_{{\bX},G}}^{{\Hf}_{\rm equiv}} 1 \sim \frac{1}{{\rm Vol}({\Gg})}\int_{{\rm Lie}({\Gg})} \int_{M}^{\left( {\Gg}\times {\Hf} \right)_{\rm equiv}} {\rm Euler}(V)
\label{eq:pf1}
\eeq
are functions on the Lie algebra of ${\Hf}_{\BC}$,  ${\xi} \in {\rm Lie}({\Hf}_{\BC})$. The analytic properties of ${\CalZ}_{{\bX},G}({\xi})$ reflect some of the geometric and topological features of ${\mM}_{{\bX},G}$. They are the main focus of this paper.

The equivariant localization expresses ${\CalZ}_{{\bX},G}$ as the sum over the fixed points
 of $\Hf$-action, which are typically labelled by multiple partitions, i.e. collections of Young diagrams. The resulting statistical mechanical model is called the \emph{gauge origami} and is studied in detail in the companion paper  \cite{Nekrasov:2015ii}.

\subsection{Symmetries, twisting, equivariance}

The partition functions ${\CalZ}_{X, G} ({\xi}) $ are analytic functions of  ${\xi} \in {\rm Lie}({\Hf}_{\BC})$, with possible singularities. Given ${\xi} \in {\rm Lie}({\Hf}_{\BC})$, the closure of the subgroup ${\exp}\, t{\xi}$, $t \in {\BC}$ defines a torus ${\sf T}_{\xi}$. The partition function ${\CalZ}_{X, G} ({\xi})$ can be computed, by Atiyah-Bott fixed point formula, as a sum over the ${\sf T}_{\xi}$-fixed points. Even though the moduli space
${\mM}_{X,G}$ may be noncompact (it is noncompact for noncompact $X$), the fixed point set, for suitable $\xi$, may still be compact, so that the integrals over ${\mM}_{X,G}$ of the equivariant differential forms converge. 
The set ${\mM}_{X,G}^{{\sf T}_{\xi}}$ of  ${\sf T}_{\xi}$-fixed points may have several connected components:
\beq
{\mM}_{X,G}^{{\sf T}_{\xi}} \, = \, \bigcup_{f} \  \left( {\mM}_{X,G}^{{\sf T}_{\xi}} \right)_{f}
\eeq
The contributions ${\CalZ}_{f}$ of $\left( {\mM}_{X,G}^{{\sf T}_{\xi}} \right)_{f}$
 are rational functions on ${\rm Lie}({\Hf}_{\BC})$, 
they have poles. In the nice situations the component $\left( {\mM}_{X,G}^{{\sf T}_{\xi}} \right)_{f}$ has a normal bundle in ${\mM}_{X,G}$ (or in the ambient smooth variety, as in the case of the obstructed theory), $N_{f}$, which inherits an action of ${\sf T}_{\xi}$, and  decomposes into the sum of complex line bundles (real rank two bundles) $L_{f, w}$, with $w$ going through the set of ${\sf T}_{\xi}$-weights. The fixed point formula states
\beq
{\CalZ}_{f} = \int_{\left( {\mM}_{X,G}^{{\sf T}_{\xi}} \right)_{f}} \frac{{\rm Euler}_{\xi}({\rm Obs}_{f})}{\prod\limits_{w} \left( w({\xi}) + c_{1}(L_{f, w}) \right)}
\label{eq:abf}
\eeq
The poles in ${\CalZ}_{f}$ occur when the Lie algebra element $\xi$ crosses the hyperplane $w({\xi}) = 0$ for some $w$ occuring in the decomposition of $N_{f}$. Geometrically this means that the ${\xi}$
belongs to a subalgebra of ${\mathrm{Lie}}{\sf T} \subset {\rm Lie}({\Hf}_{\BC})$ which fixes not only $\left( {\mM}_{X,G}^{{\sf T}_{\xi}} \right)_{f}$, but also (at least infinitesimally, at the linearized order) 
a two-dimensional surface passing through $f$, in the direction of $L_{f,w}$.

We shall be interested in the analytic properties of ${\CalZ}_{X,G}$ and one of the questions we shall be concerned with is whether the poles in ${\CalZ}_{f}$ are cancelled by the poles in the contribution of some other component $\left( {\mM}_{X,G}^{{\sf T}_{\xi}} \right)_{f'}$ of the fixed point set. More precisely, once ${\xi} \to {\xi}_{c}$ where ${\xi}_{c}$ belongs to the hyperplane
$w({\xi}) = 0$ defined relative to the weight decomposition of $N_f$, the component of the fixed point set may enhance, 
\beq
\left( {\mM}_{X,G}^{{\sf T}_{\xi}} \right)_{f} \subset \left( {\mM}_{X,G}^{{\sf T}_{{\xi}_{c}}} \right)_{f''}, \
\eeq
reaching out to the other component $\left( {\mM}_{X,G}^{{\sf T}_{\xi}} \right)_{f'}$
\beq
\left( {\mM}_{X,G}^{{\sf T}_{\xi}} \right)_{f'} \cap \left( {\mM}_{X,G}^{{\sf T}_{{\xi}_{c}}} \right)_{f''} \neq {\emptyset}
\eeq
If the enhanced component $\left( {\mM}_{X,G}^{{\sf T}_{{\xi}_{c}}} \right)_{f''}$ is compact, then the pole at ${\xi} = {\xi}_{c}$ in ${\CalZ}_{f}$ will be cancelled by the pole in ${\CalZ}_{f'}$. 

So the issue in question is the compactness of the fixed point set for the torus generated by the non-generic infinitesimal symmetries ${\xi}_{c}$.

In our case we shall choose a class of subgroups ${\Hfr} \subset {\Hf}$. We shall show that
the set of ${\Hfr}$-fixed points is compact. It means that
for generic choice of ${\hat\xi} \in {\rm Lie}({\Hfr}_{\BC})$ the partition function 
${\CalZ}_{X,G} ({\hat\xi} + {\xt})$
as a function of ${\xt} \in {\rm Lie}({\Hfr}_{\BC})^{\perp} \subset {\rm Lie}({\Hf})$ has no singularities.

The procedure of restricting the symmetry group of the physical system to a subgroup is well-known to physicists under the name of twisting \cite{Witten:1988xj}. It is used in the context of topological field theories, which are obtained from the
supersymmetric field theories having an $R$-symmetry group ${\Hf}_{R}$ such that the group of rotations 
${\Gr}$ of flat spacetime can be embedded nontrivially into the direct product 
\beq
{\Gr} \longrightarrow {\Gr} \times {\Hf}_{R}\ .
\label{eq:twist}
\eeq 
We shall encounter a lot of instances of the procedure analogous to \eqref{eq:twist} in what follows.

\subsection{Gauge theories on stacks of D-branes}

The maximally supersymmetric Yang-Mills theory in $p+1$-dimensions models \cite{Witten:1995im} the low energy behavior of a stack of parallel $Dp$-branes. This description can be made $p$-blind by turning on a background constant $B$-field. In the strong $B$-field the ``non-abelian Born-Infeld/Yang-Mills'' theory description of the low energy physics of the open strings connecting the $Dp$-branes crosses over to the noncommutative Yang-Mills description \cite{Seiberg:1999vs}. In this paper we shall use the noncommutative Yang-Mills to study the dynamics of intersecting stacks of $Dp$-branes. 

\subsubsection{The Matrix models} Recall the dimensional reductions of the maximally supersymmetric Yang-Mills theory down to $0+0$, $0+1$ and $1+1$ dimensions \cite{bfss}, \cite{ikkt}, \cite{Verlinde:1995}. We take the gauge group to be ${\Gg} = U({\bN})$ for some large $\bN$. Following \cite{Moore:1998et} we shall view the model of \cite{ikkt}  
\beq
\frac{1}{{\rm Vol}{\Gg}} \, \int\limits_{{\BR}^{10|16} \otimes {\rm Lie}{\Gg}} \ DX^{m} D{\theta}^{\alpha} \ {\exp} \, \left( \,  - \frac{1}{4} \sum_{m< n} {\Tr}  [ X^{m}, X^{n} ]^2 - \frac{1}{2} \sum_{m, {\alpha}, {\beta}} {\Gamma}_{\alpha\beta}^{m} {\Tr} {\theta}^{\alpha} [X^{m}, {\theta}^{\beta} ]  \, \right)
\label{eq:ikktlag}
\eeq
with the adjoint bosons $X^m$ and the adjoint fermions ${\theta}^{\alpha}$ transforming in the representations $\bf 10$ and $\bf 16$ of ${\mathrm{Spin}}(10)$, respectively, as the cohomological field theory in $0$ dimensions, while \cite{bfss} and \cite{Verlinde:1995} are obtained by the lift procedure of \cite{Baulieu:1997nj}.

The approach to the noncommutative gauge theory in which the gauge field $A_{m}$ is traded for the (infinite) matrix $X^m$ is used in the background-independent ($p$-uniform) formalism of \cite{Seiberg:2000zk}. 

\subsubsection{$(0,1)$-formalism, $\mathrm{{\mathrm{Spin}}}(7)$-instantons}

Let us start in the $1+1$-dimensional case. Let $\Sigma$ be the worldsheet of our theory, with the local complex coordinates $z, {\bar z}$. The theory has a gauge field $A = A_{z} dz + A_{\zb} d{\zb}$, $8$ Hermitian adjoint scalars $X^m$, and $16$ adjoint fermions, which split into $8$ right ${\psi}^{a}_{+}$, and $8$ left ${\chi}^{\dot a}_{-}$ ones.
Here $m$, $a$ and $\dot a$ are the indices of the ${\bf 8}_{v}$, ${\bf 8}_{s}$ and ${\bf 8}_{c}$ representations of the global symmetry group ${\mathrm{Spin}}(8)$, respectively. The formalism that we shall adopt singles out a particular spinor among ${\bf 8}_{c}$. The isotropy subgroup ${\mathrm{Spin}}(7)$ of that spinor has the following significance. The representation ${\bf 8}_{c}$, under ${\mathrm{Spin}}(7)$, decomposes as ${\bf 7} \oplus {\bf 1}$, the ${\bf 1}$ being the invariant subspace. Accordingly, we split $\left( {\chi}^{\dot a}_{-} \right)_{{\dot a} \in  {\bf 8}_{c}} \longrightarrow \left( {\chi}_{-}^{i} \right)_{i \in {\bf 7}} \oplus {\eta}_{-}$. The representations
${\bf 8}_{s}$ and ${\bf 8}_{v}$ become the spinor ${\bf 8}$ of ${\mathrm{Spin}}(7)$. We shall also need the auxiliary fields $h^i$, which are the worldsheet scalars, transform in the adjoint of the gauge group, and in the representation $\bf 7$ of ${\mathrm{Spin}}(7)$. 
The theory, in this formalism, has one supercharge ${\delta}_{+}$, which squares to the chiral translation on the worldsheet
\beq
{\delta}_{+}^{2} = D_{++} = {\bar\partial}_{\zb} + A_{\zb} 
\eeq
(the conjugate derivative $D_{--} = {\partial}_{z} + A_{z}$)
which acts on the fields of the model as follows:
\beq
\begin{aligned}
{\delta}_{+} X^{m} = {\psi}^{m}_{+}, \quad & \quad {\delta}_{+} {\psi}_{+}^{m} = D_{++} X^{m} = {\bar\partial}_{\zb} X^{m} + [ A_{\zb} , X^{m} ] \\
{\delta}_{+} {\chi}_{-}^{i}  = h^{i}, \quad & \quad {\delta}_{+} h^{i} = D_{++} {\chi}^{i}_{-} = {\bar\partial}_{\zb} {\chi}^{i}_{-} + [ A_{\zb} , {\chi}^{i}_{-} ] \\
{\delta}_{+} A_{z} = {\eta}_{-}, \quad & \quad {\delta}_{+} {\eta}_{-} = F_{z\zb} \end{aligned}
\label{eq:dplus}
\eeq
The Lagrangian 
\beq
L = {\delta}_{+} \, \int_{\Sigma} \, {\Tr} \left( {\psi}^{m}_{+} D_{--} X^{m} +
{\chi}^{i}_{-} \left( {\ii} {\wp}^{i}_{mn} [ X^{m}, X^{n} ] - h^{i} \right) + {\eta}_{-} F_{z\zb} \right)
\label{eq:lagsusy}
\eeq
becomes that of the standard ${\CalN}=8$ supersymmetric Yang-Mills once the auxiliary fields $h^i$ are eliminated by their equations of motion. Here ${\wp}^{i}_{mn}$ is the matrix of the projection  ${\wp}_{7} : {\Lambda}^{2} {\bf 8}_{v} = {\rm Lie}{\mathrm{Spin}}(8) \longrightarrow {\bf 7}$
onto the orthogonal complement to ${\rm Lie}{\mathrm{Spin}}(7) \subset {\rm Lie}{\mathrm{Spin}}(8) = {\rm Lie}{\mathrm{Spin}}(7) \oplus {\bf 7}$.  

 Upon the dimensional reduction to $0$ dimensions, the gauge field $A$ becomes a complex scalar ${\sigma} = A_{\zb}$ and its conjugate ${\bar\sigma}  = A_{z}$.

\subsubsection{$(0,2)$-formalism, ${\mathrm{SU}}(4)$-instantons}

In this formalism we have two supercharges $Q_{+}, {\bar Q}_{+}$, obeying
\beq
Q_{+}^{2} = {\bar Q}_{+}^{2} = 0\, , \qquad Q_{+}{\bar Q}_{+} + {\bar Q}_{+} Q_{+} = D_{++} 
\label{eq:02susy}
\eeq
so that ${\delta}_{+} = Q_{+} + {\bar Q}_{+}$. 
We split $8$ Hermitian adjoint scalars $X^m$ into $4$ complex adjoint scalars $Z^a$, $a \in {\4}$, and their conjugates ${\bar Z}^{\bar a}$, and the same for the fermions ${\psi}_{+}^{m} \longrightarrow {\psi}_{+}^{a}, {\bar\psi}_{+}^{\bar a}$. The $\bf 7$ ${\chi}_{-}^{i}$'s split as ${\6} \oplus {\bf 1}$: 
$({\chi}^{i}_{-})_{i \in {\bf 7}} \to \left( {\chi}_{A, -} = {\ve}_{A{\bar A}} {\bar\chi}_{\bar A, -} \right)_{A \in {\6}} \oplus {\chi}_{-}$. 
This splitting breaks the symmetry group ${\mathrm{Spin}}(8) \times {\mathrm{Spin}}(2) \subset {\mathrm{Spin}}(10)$ of \eqref{eq:ikktlag} down to ${\mathrm{SU}}(4)$. 
 
The Lagrangian \eqref{eq:lagsusy}, in this formalism, reads as follows:
\begin{multline}
L = {\delta}_{+} \, {\bf\Psi} \, , \qquad
\ {\bf\Psi} = \int_{\Sigma} \, {\Tr} \left( {\psi}^{a}_{+} D_{--} {\bar Z}^{\bar a} + {\bar\psi}^{\bar a}_{+} D_{--} Z^{a} + {\eta}_{-} F_{z\zb} \right) + \\ 
+ {\ii} \int_{\Sigma} \, {\Tr} \left( {\chi}_{ab, -} \left(  [ Z^{a}, Z^{b} ] + 
{\frac{1}{2}} {\ve}^{abcd} [ {\bar Z}^{\bar c}, {\bar Z}^{\bar d} ]  \right)  + {\chi}_{-} {\mu} \right)
- \\
- \int_{\Sigma} \, {\Tr} \left(  {\chi}_{-} h + 
{\chi}_{ab, -}   h^{ab}   \right)
\label{eq:lagsu4}
\end{multline}
where 
\beq
{\mu} = \sum_{a \in {\4}} [ Z^{a}, {\bar Z}^{\bar a} ] 
\eeq
The supersymmetric (for flat $\Sigma$) solutions of \eqref{eq:lagsu4} are the covariantly holomorphic matrices, solving the equations 
\beq
D_{\bar z} Z^{a} = 0, \qquad D_{\bar z} {\bar Z}^{\bar a} = 0, \quad a \in {\4}, \qquad {\mu} = 0
\label{eq:sdb}
\eeq
 and 
\beq
 [ Z^{a}, Z^{b} ] + \frac 12 {\ve}^{abcd} [ {\bar Z}^{\bar c}, {\bar Z}^{\bar d} ]  = 0, \qquad \{ a, b \} \in {\6}
\label{eq:sda}
\eeq
For finite dimensional ${\BC}^{\bN}$ these equations imply that all matrices commute and can be simultaneously diagonalized. 
 
\subsubsection{Noncommutative gauge theory}

 We now wish to consider a generalization of the model \cite{ikkt} in which the finite dimensional vector space ${\BC}^{\bN}$ is replaced by a  Hilbert space $\CalH$. In order to keep the action \eqref{eq:lagsusy} finite
 the combination
 \beq
 {\wp}^{i}_{mn} [ X^{m}, X^{n}]
 \eeq
 could be deformed to
 \beq
 {\wp}^{i}_{mn}  [ X^{m}, X^{n}] - {\ii} {\vartheta}^{i} \cdot {\bf 1}_{\CalH} 
 \label{eq:instmod}
 \eeq
 for some constants $\vartheta^{i}$. One possibility to have a finite action configuration (after $h^i$'s are integrated out) is to have the operators $X^m$ obey the Heisenberg algebra:
 \beq
 [X^m, X^n] = {\ii} {\vartheta}^{mn} \cdot {\bf 1}_{\CalH}
 \label{eq:Heis}
 \eeq
with the $c$-number valued matrix  ${\vartheta}^{mn}  = - {\vartheta}^{mn}$ obeying 
\beq
{\vartheta}^{i} = \sum_{m < n} {\wp}^{i}_{mn} {\vartheta}^{mn} \ .
\label{eq:thethe}
\eeq  
Since there are too many choices of $\vartheta^{mn}$  given $\vartheta^i$, the modification \eqref{eq:instmod} is not what we need. A more sensible modification is
to define the action (with the auxiliary fields eliminated) to have the bosonic potential:
\beq
\sum_{m < n} {\Tr}_{\CalH} \, \left( \, [ X^{m}, X^{n} ] - {\ii} {\vartheta}^{mn} \cdot 
{\bf 1}_{\CalH}\, \right)^{2}
\label{eq:bospot}
\eeq 
whose absolute minimuma are given by the representation of the Heisenberg algebra \eqref{eq:Heis} in $\CalH$. These are classified, for the non-degenerate ${\vartheta}^{mn}$, modulo the gauge group ${\Gg} = U({\CalH})$, by the Stone-von Neumann theorem. Fix a non-negative integer $N$, and a standard oscillator representation $H$ of the Heisenberg algebra $[ {\hat x}^{m}, {\hat x}^{n} ] = {\ii} {\vartheta}^{mn} \cdot {\bf 1}_{H}$.  Then $H = L^{2} ( {\BR}^{4} )$, ${\CalH} = {\BC}^{N} \otimes H$.  For example, let us choose a block diagonal basis for $\vartheta^{mn}$, in which
\beq
{\hat z}^{a} = {\hat x}^{2a-1} + {\ii} {\hat x}^{2a},  \quad 
 {\hat z}^{a\dagger} = {\hat x}^{2a-1} - {\ii} {\hat x}^{2a}, \qquad a \in {\4}
 \label{eq:zfromx}
 \eeq
 obey
\beq
\begin{aligned}
 & [ {\hat z}^{a} , {\hat z}^{b} ] = 0, \quad [ {\hat z}^{a\dagger} , {\hat z}^{b\dagger} ] = 0, \qquad \{ a , b \} \in {\6} \\
 & [ {\hat z}^{a}, {\hat z}^{b\dagger} ] = - {\theta}^{a} \, {\delta}^{ab} 
 \end{aligned}
 \label{eq:oscillator}
 \eeq
The supersymmetric solution of the matrix model, the $0$-dimensional reduction of \eqref{eq:lagsusy} is
 given by the operators 
 \beq
 \begin{aligned}
& X^{m} = {\bf 1}_{N} \otimes {\hat x}^{m}, \qquad m = 1, \ldots 8  \\
& {\sigma} = {\rm diag}( {\sigma}_{1} , \ldots , {\sigma}_{N} ) \otimes {\bf 1}_{H} \\
\end{aligned}
\label{eq:xsiops}
\eeq  
This solution, for ${\theta}_{a} \neq 0$  for all $a \in {\4}$ describes a stack of $N$ $D7$ branes whose worldvolume extends in the $1, \ldots, 8$ directions. They are localized in the remaining two dimensions, parametrized by the eigenvalues of the complex matrix ${\sigma}$. 

Now let us assume all $\theta_{a}$ equal to ${\zeta} > 0$. Take $H = L^{2}({\BR}^{2})$ to be the Fock space representation of the algebra 
\beq
\begin{aligned}
& [ c_{1}, c_{2} ] = [ c_{1}^{\dagger}, c_{2}^{\dagger} ] = 0 \\
& [ c_{i} , c_{j}^{\dagger} ] = {\zeta} {\delta}_{ij}, \qquad i, j = 1,2\\
\end{aligned}
\label{eq:twoosc}
\eeq
Define:
\beq
\begin{aligned}
& {\hat z}^{1} = \frac{1}{\sqrt{2}} \left( {\bf 1}_{N_{12}} \otimes c_{1}^{\dagger} + {\bf 1}_{N_{13}} \otimes c_{1}^{\dagger} + {\bf 1}_{N_{14}} \otimes c_{1}^{\dagger} \right) \\
& {\hat z}^{2} = \frac{1}{\sqrt{2}} \left( {\bf 1}_{N_{12}} \otimes c_{2}^{\dagger} + {\bf 1}_{N_{23}} \otimes c_{1}^{\dagger} + {\bf 1}_{N_{24}} \otimes c_{1}^{\dagger} \right) \\
& {\hat z}^{3} = \frac{1}{\sqrt{2}} \left( {\bf 1}_{N_{13}} \otimes c_{2}^{\dagger} + {\bf 1}_{N_{23}} \otimes c_{2}^{\dagger} + {\bf 1}_{N_{34}} \otimes c_{1}^{\dagger} \right) \\
& {\hat z}^{4} = \frac{1}{\sqrt{2}} \left( {\bf 1}_{N_{14}} \otimes c_{2}^{\dagger} + {\bf 1}_{N_{24}} \otimes c_{2}^{\dagger} + {\bf 1}_{N_{34}} \otimes c_{2}^{\dagger} \right) \\
\end{aligned}
\label{eq:zc}
\eeq
These operators obey:
\beq
\begin{aligned}
& [ {\hat z}^{a}, {\hat z}^{b} ] = [ {\hat z}^{a\dagger}, {\hat z}^{b\dagger} ] = 0  \\
& \sum_{a \in {\4}} [ {\hat z}^{a}, {\hat z}^{a\dagger} ] = - {\zeta} {\bf 1}_{N \otimes H} \\
\end{aligned}
\label{eq:zzbps}
\eeq
where
\beq
N = \bigoplus_{A \in {\6}} N_{A}
\eeq
The solution \eqref{eq:zzbps} describes six stacks of $D3$-branes spanning the coordinate two-planes ${\BC}^{2} \subset {\BC}^{4}$, with $n_{A} = {\rm dim}N_{A}$ branes spanning the two-plane ${\BC}^{2}_{A}$. This is a generalization of the ``piercing string'' and ``fluxon'' solutions of \cite{Gross:2000ss, Gross:2000ph}.
  
We can easily produce more general solutions of the BPS equations. Take six solutions ${\hat C}^{1}_{A}, {\hat C}^{2}_{A}$ of non-commutative instanton equations in ${\BR}^{4}$, viewed as operators in $N_{A} \otimes H$, obeying:
\beq
[  {\hat C}^{1}_{A}, {\hat C}^{2}_{A} ] = 0 \, , \qquad
[ {\hat C}^{1}_{A}, {\hat C}^{1\dagger}_{A} ] + [ {\hat C}^{2}_{A}, {\hat C}^{2\dagger}_{A} ] = {\zeta} 
\label{eq:2inst}
\eeq
Define operators in ${\CalH} = \bigoplus\limits_{A \in {\6}} N_{A} \otimes H$:
\beq
{\hat Z}^{a} = \frac{1}{\sqrt{2}} \bigoplus_{A \ni a} \, {\hat C}^{h_{A}(a)\dagger}_{A}
\label{eq:hzc}
\eeq
where $h_{\{ a, b \}}(a) = 1$ for $a< b$, and $h_{\{ a, b \}}(a) = 2$ for $a > b$. These operators satisfy 
 the higher dimensional analogues of the noncommutative instanton equations \eqref{eq:sda}, \eqref{eq:sdb}, \cite{Nekrasov:2000ih}:
\beq
\begin{aligned}
& [ {\hat Z}^{a}, {\hat Z}^{b} ] + \frac 12 \, \sum_{c, d} \, {\ve}_{abcd} [ {\hat Z}^{d\dagger}, {\hat Z}^{c\dagger} ] = 0 , \qquad a, b, c, d \in {\4} \\
& \qquad\qquad \sum_{a \in {\4}} [ {\hat Z}^{a}, {\hat Z}^{a\dagger} ] = - {\zeta} {\bf 1}_{N \otimes H} \\
\end{aligned}
\label{eq:zzbpsii}
\eeq
The aim of the next section is to produce the (almost) finite-dimensional model of the moduli space of finite action solutions to \eqref{eq:zzbpsii}. Some of these solutions are of the form \eqref{eq:hzc}. 

\bigskip
\centerline{$\dots$}
\bigskip

Recently the field theory description of two stacks of intersecting $D3$ branes in IIB string theory sharing a common $1+1$-dimensional worldvolume was explored in \cite{Constable:2002xt, Mintun:2014aka}. The theories exibit unusual holographic and renormalization properties. 

\bigskip
\centerline{$\dots$}
\bigskip

The string theory of two stacks of transversely intersecting $D3$ branes in IIB theory has been recently studied in \cite{Tong:2014cha, Tong:2014yna}, albeit in the ${\zeta}=0$ case. None of the beauty (to the trained eye) of the picture presented below seem to survive in this limit.

\section{Spiked  instantons} 

\bigskip
We are going to work with the collections of vector spaces and linear maps between them. 
The vector spaces will be labelled by the coordinate complex two-planes in the four dimensional complex vector space ${\BC}^{4}$.

\subsection{Generalized ADHM equations}

We start by fixing seven Hermitian vector spaces: $K$ and $N_A$, $A \in {\6}$. Let $k = {\rm dim}_{\BC}(K)$, $n_{A} = {\rm dim}_{\BC}(N_{A})$.  Consider
the vector space ${\CalA}_{k}({\vec n})$
of linear maps $({\bB}, {\bI}, {\bJ})$

\beq
\begin{aligned}
& {\bB} = (B_{a})_{a\in {\4}} \, , \quad
B_{a}: K \to K \, , \\ 
& {\bI} = (I_{A})_{A \in {\6}} \ , \qquad I_{A} \, :\,  N_{A} \to K \, , \\
& {\bJ} = (J_{A})_{A\in {\6}}\ , \qquad J_{A} \, :\,  K \to N_{A} \, . 
\label{eq:bijmaps}
\end{aligned}
\eeq
\begin{figure}[H]
\picit{6}{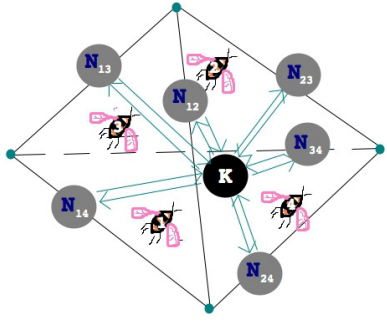}
\caption{Seven vector spaces and maps between them}
\label{fig:tetrabees}
\end{figure}
The vector spaces and the maps are conveniently summarized by the
tetrahedron diagram on Fig. \ref{fig:tetrabees}. The choice of the matrices can be motivated by the string theory considerations.  
Namely, consider $k$ $D(-1)$-branes in the vicinity of the six stacks of $D3$-branes (some of these stacks could be $\overline{D3}$-branes) spanning the coordinate two-planes ${\BC}^{2}_{A} \subset {\BC}^{4}$. The number of branes spanning ${\BC}^{2}_{A}$ is $n_{A}$. 
\medskip
\begin{figure}[H]
\picit{8}{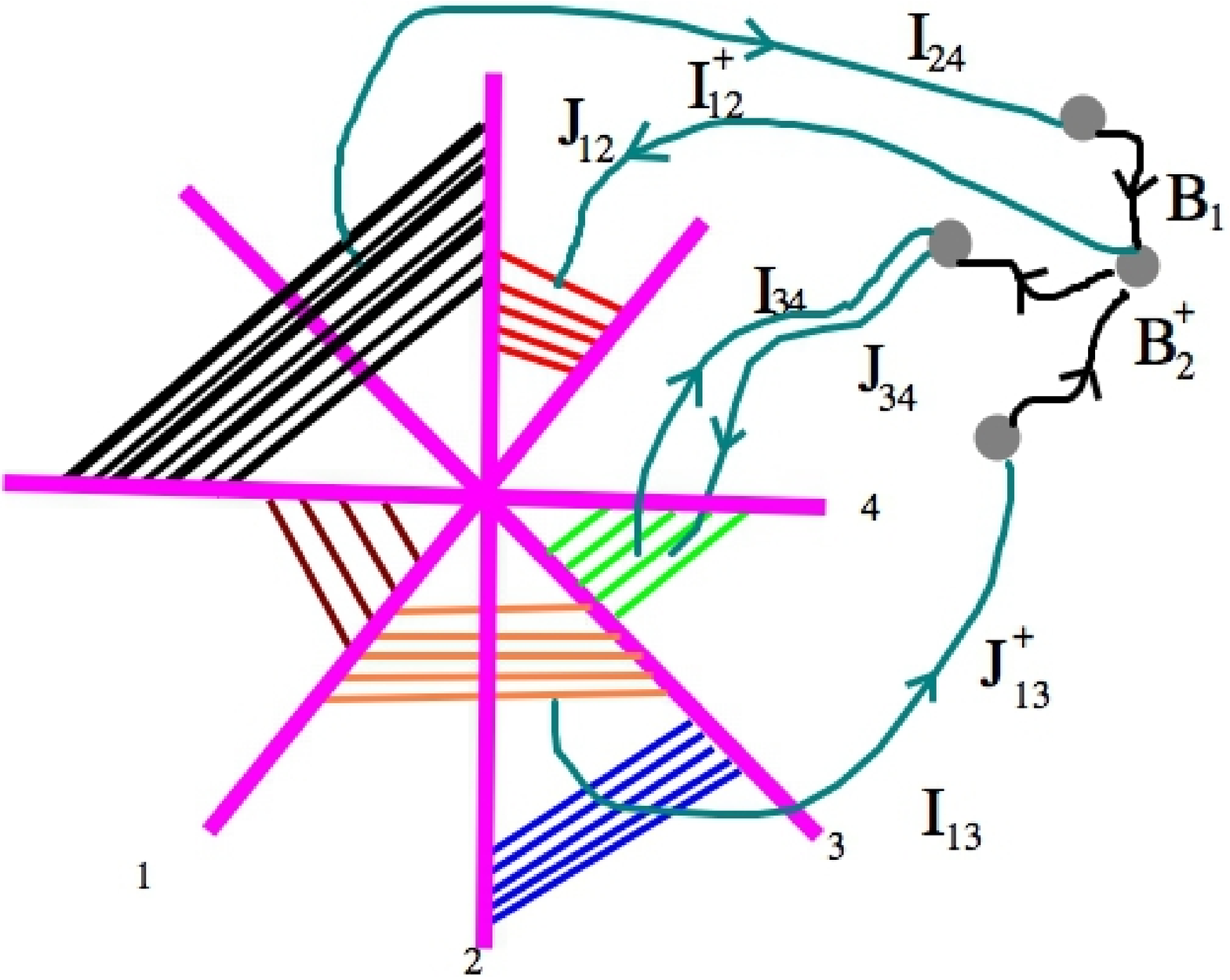}
\caption{Open string sectors: fields $\bB$, $\bI$, $\bJ$}
\label{fig:dbranesij}
\end{figure}

\bigskip

Then the open strings stretched between the $D(-1)$ and $D(-1)$'s produce, upon quantization, the matrices $B_{a}, B_{a}^{\dagger}$, together with their superpartners, and some auxiliary fields, which enter the effective Lagrangian in such a way so as to impose the following 
\subsubsection{KK equations}

{}Define, for $A = \{ a, b \}$, $a < b$, 
\beq
{\mu}_{A} = [ B_{a}, B_{b} ] + I_{A}J_{A}, \qquad 
\label{eq:commm}
\eeq
and
\beq
s_{A} = {\mu}_{A} + {\ep}(A){\mu}_{\bar A}^{\dagger} \ : \ K \to K, \qquad A \in {\6}
\label{eq:smm}
\eeq
obeying
\beq
s_{A}^{\dagger} = {\ep}(A) s_{\bar A}
\label{eq:ssymm}
\eeq
Define the {\it real moment map}
\beq
{\mu} = \sum_{a \in {\4}}\  [ B_{a}, B_{a}^{\dagger} ] + 
\sum_{A \in {\6}} \ \left( I_{A}I_{A}^{\dagger} - J_{A}^{\dagger}J_{A} \right)
\label{eq:mur}
\eeq 
The symmetry \eqref{eq:ssymm} allows to view the collection 
${\vec s} = (s_{A} )_{A \in {\6}} \oplus {\mu}$ as the $U(K)$-equivariant map
\beq
{\vec s} : {\CalA}_{k} ({\vec n}) \longrightarrow {\mathrm{Lie}}U(K)^{*} \otimes {\BR}^{7}\, , 
\eeq
as a sort of an octonionic version of the hyperk\"ahler moment map \cite{Hitchin:1986ea}.

\bigskip
\centerline{$\dots$}
\bigskip

{}Likewise, the open strings stretched between the $D(-1)$ and $D3$'s produce, upon quantization, the matrices $I_{A}, J_{A}, I_{A}^{\dagger}, J_{A}^{\dagger}$, together with their superpartners, and some auxiliary fields,

\begin{figure}[H]
\picit{8}{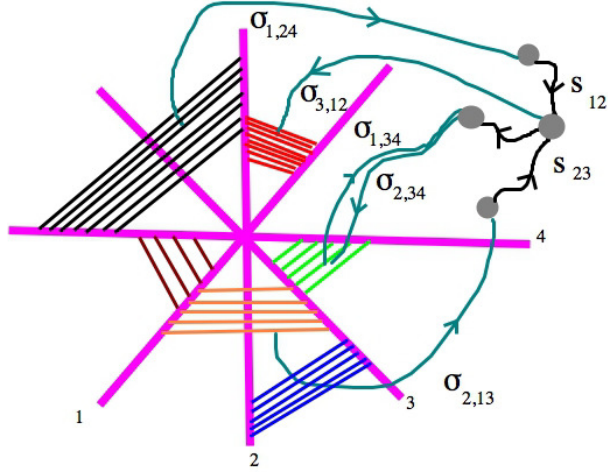}
\caption{Open string sectors: mutiplets of the equations ${\sigma}_{{\bar a}A}, s_{A}$}
\label{fig:dbraneseqs}
\end{figure}

\noindent
{}which enter the effective Lagrangian in such a way so as to impose the following 

\subsubsection{KN equations}

For each pair $({\bar a}, A)$, where $A \in {\6}$, and ${\bar a} \in {\bar A}$, define
\beq
{\sigma}_{{\bar a}A} = B_{\bar a} I_{A} + {\ve}_{{\bar a}{\bar b} A} B_{\bar b}^{\dagger} J_{A}^{\dagger} \ : \  
N_{A} \to  K \, 
\label{eq:sigmaa}
\eeq
where ${\bar b} \in {\bar A}$, and ${\bar b}  \neq {\bar a}$. 

\subsubsection{NN equations}

Now, for each $A \in {\6}$ define
\beq
{\Upsilon}_{A} = J_{\bar A}I_{A} - {\ep}(A) I_{\bar A}^{\dagger} J_{A}^{\dagger} \ : \ N_{A} \to  N_{\bar A}
\eeq
which obey 
\beq
{\Upsilon}_{A}^{\dagger} = - {\Upsilon}_{\bar A} \ .
\label{eq:nnsym}
\eeq
Because of the symmetry \eqref{eq:nnsym}  the collection of the maps $({\Upsilon}_{A})_{A \in {\6}}$ takes values in 
the real vector space of dimension 
\beq
\sum_{A \in {\6}} n_{A}n_{\bar A}
\eeq
The equations \eqref{eq:nnsym} result from integrating out the open strings connecting the two stacks of $D3$-branes which intersect only at a point, the origin in ${\BC}^{4}$. 

\bigskip
\centerline{$\dots$}
\bigskip

For
each pair $A', A'' \in {\6}$, such that $A' \cap A'' = \{ a \}$, and 
$i \geq 1$,
define
\beq
{\Upsilon}_{A',A''; i} = J_{A'} B_{a}^{i-1} I_{A''} \ .
\eeq
These equations result (conjecturally) from integrating out the $3-3$ strings connecting the neighbouring stacks ${\BC}^{2}_{A'}$ and ${\BC}^{2}_{A''}$, intersecting along a real two-dimensional plane ${\BC}^{1}_{a}$. 

\bigskip
\centerline{$\dots$}
\bigskip

{}Finally, for each $A = \{ a', a'' \} \in {\6}$, e.g. $a' < a''$, and $i, j \geq 1$, define
\beq
{\Upsilon}_{A; i,j} = J_{A} B_{a'}^{i-1}B_{a''}^{j-1} I_{A}
\eeq
\subsubsection{A very useful identity}

Let us compute
\begin{multline}
\sum_{A \in {\6}} {\Tr} s_{A} s_{A}^{\dagger} + \sum_{A \in {\6}, {\bar a} \in {\bar A}} {\Tr} {\sigma}_{{\bar a}A} {\sigma}_{{\bar a}A}^{\dagger} + \sum_{A \in {\6}} {\Tr} {\Upsilon}_{A} {\Upsilon}_{A}^{\dagger} = \\
2 \sum_{A \in {\6}} {\Tr} {\mu}_{A} {\mu}_{A}^{\dagger} + \sum_{A \in {\6}}
{\ve} (A) \left( {\Tr} {\mu}_{A} {\mu}_{\bar A} + {\Tr} {\mu}_{A}^{\dagger} {\mu}_{\bar A}^{\dagger}  \right) + \\
 \sum_{A \in {\6}, a \in {\bar A}} {\Tr} \left( B_{a}^{\dagger}  B_{a} {\Pi}^{I}_{A} + B_{a} B_{a}^{\dagger} {\Pi}^{J}_{A} \right)  + 2 \sum_{A \in {\6}} {\Tr} {\Pi}^{J}_{A} {\Pi}^{I}_{\bar A} + \\
 - 2 \sum_{A \in {\6},  {\bar A} = \{ {\bar a}, {\bar b} \} } {\ep}(A) {\Tr}  \left(  [B_{\bar a}, B_{\bar b}] I_{A}J_{A} + c.c. \right) 
 - \sum_{A \in {\6}} {\ep}(A) {\Tr} \left( I_{A} J_{A}I_{\bar A} J_{\bar A}  + c.c. \right)   = \\
 = 2 \sum_{A\in {\6}}  \left(  \Vert {\mu}_{A} \Vert^2 + \Vert J_{\bar A} I_{A} \Vert^2 \right)+ \sum_{A \in {\6}, a \in {\bar A}} 
 \Vert B_{a} I_{A} \Vert^2 + \Vert J_{A} B_a \Vert^2 
 \label{eq:bps}
\end{multline}
where
\beq
{\Pi}^{J}_{A} = J_{A}^{\dagger}J_{A}, \qquad {\Pi}^{I}_{A} = I_{A} I_{A}^{\dagger} 
\eeq

\subsection{Holomorphic equations}

Using the identity \eqref{eq:bps} it is easy to show that 
the equations 
\beq
\begin{aligned}
& s_{A} = 0, \quad {\Upsilon}_{A} = 0, \quad A \in {\6} \\
& {\sigma}_{{\bar a}A} = 0, \quad {\bar a} \in {\bar A} \, , \\
\label{eq:spik}
\end{aligned}
\eeq
which are not holomorphic in the variables ${\bB}, {\bI}, {\bJ}$, imply  stronger holomorphic
equations: for each $A \in {\6}$, 
\beq
\begin{aligned}
& {\mu}_{A} = 0,  \quad  J_{\bar A} I_{A} = 0 \\
& B_{\bar a}I_{A} = 0\, , \quad J_{A} B_{\bar a} = 0 \, , \qquad {\bar a} \in {\bar A}, \\
\end{aligned}
\label{eq:bbij}
\eeq

\subsection{The moduli spaces ${\mM}_{k}^{*}({\vec n})$}

Define ${\mM}_{k}^{i}({\vec n})$ to be the $U(k)$-quotient of the space of solutions
to \eqref{eq:spik} (which imply, by the above argument,  \eqref{eq:bbij}), the additional equations
\beq
{\Upsilon}_{A', A'' ; j} = 0, \qquad 1 \leq j \leq i, 
\label{eq:upab}
\eeq
for all $A', A'' \in {\6}$ with $\# A' \cap A'' = 1$, and the ``moment map''
equation 
\beq
{\mu} = {\zeta} \cdot {\bf 1}_{K}
\label{eq:rmom}
\eeq 
The group $U(k)$ acts by:
\beq
({\bB}, \, {\bI}, \, {\bJ}) \mapsto ( g^{-1} {\bB} g, \, g^{-1} {\bI}, \, {\bJ} g )\, , \qquad
g \in U(k)
\label{eq:bij}
\eeq
It is clear that, as a set
\beq
{\mM}_{k}^{\infty}({\vec n})  \subset \ldots \subset {\mM}_{k}^{i}({\vec n}) \subset {\mM}_{k}^{i-1}({\vec n}) \subset \ldots \subset {\mM}_{k}^{1}({\vec n}) 
\subset {\mM}_{k}^{0}({\vec n})
\label{eq:spikgrad}
\eeq
and that the sequence stabilizes at $i \geq k$ (use the fact that a $k\times k$ matrix obeys the degree $k$ polynomial equation). 

\subsection{Stability}

Imposing \eqref{eq:rmom} with $\zeta > 0$ and dividing by $U(k)$ is equivalent to imposing the stability condition and dividing by the action \eqref{eq:bij} with $g \in  GL(k) \equiv GL(k, {\BC})$.
Note that we deal with the equations  \eqref{eq:bbij} when talking about the $GL(k)$ symmetry. 
The stability condition reads:
\beq\mathboxit{\begin{aligned}
& {\rm Any\ subspace}\ K' \subset K \, , \ {\rm such\ that} \\
&  \qquad\qquad \color{red}{I_{A}(N_{A}) \subset K'\, , \quad {\rm for\ all}\ A \in {\6}} \\
& \qquad\qquad\qquad {\color{black}\rm and}\\
& \qquad\qquad \color{blue}{B_{a}(K') \subset K' \, , \quad {\rm for\ all}\ a \in {\4}}  \\
& \qquad\qquad\qquad \color{black}{\rm coincides\ with\ all\ of\ } K \, , \ K' = K \\
& \\
& {\rm in\ other\ words\, , }\quad\color{green} \sum_{A \in {\6}}
{\BC} [ B_{1}, B_{2}, B_{3}, B_{4} ]  \, I_{A} (N_{A}) = K \\
\end{aligned}}   
\label{eq:stab}
\eeq
The proof is standard. In one direction, let us prove  \eqref{eq:stab} holds given that the $GL(k)$-orbit of the tuple $({\bB}, {\bI}, {\bJ})$ of matrices crosses the locus ${\mu} = {\zeta} {\bf 1}_{K}$. Indeed, assume there is $K'$ which is ${\bB}$-invariant, and contains the image of $I_{A}$'s. Let $K''$ be the orthogonal complement $K'' = ( K')^{\perp}$. Let $P'$, $P''$ be the orthogonal projections onto $K'$, $K''$, respectively:
\beq
{\bf 1}_{K} = P' + P'', \qquad P' P'' = P'' P' = 0, \qquad (P')^2 = (P')^{\dagger} = P', \qquad (P'')^2 = (P'')^{\dagger} = P''
\eeq
Since the images of ${\bI}$'s are in $K'$, we have:
\beq
P'' I_{A} = 0, \qquad A \in {\6}
\eeq
Since ${\bB}$ preserve $K'$, we have:
\beq
P'' B_{a} P' = 0, \qquad a \in {\4}
\eeq
Define 
\beq
\begin{aligned}
& b_{a} = P'' B_{a} P'', \quad b_{a}^{\dagger} = P'' B_{a}^{\dagger} P'', \qquad a \in {\4} \, , \\ 
& j_{A} = J_{A} P'', \quad P'' J_{A}^{\dagger} = j_{A}^{\dagger}, \qquad A \in {\6} \\
\end{aligned}
\eeq
Thus:
\beq
{\zeta} P'' = P'' {\mu} P'' = \sum_{a\in {\4}} [ b_{a}, b_{a}^{\dagger} ] - \sum_{a \in {\4}} P'' B_{a}^{\dagger} P' B_{a} P'' - \sum_{A \in {\6}} j_{A}^{\dagger} j_{A} 
\label{eq:stab''}
\eeq
Now, taking the trace of both sides of \eqref{eq:stab''} we arrive at the conclusion $K'' = 0$:
\beq
0 \leq {\zeta} {\rm dim}K'' = - \sum_{a \in {\4}} \Vert P ' B_{a} P'' \Vert^2   - \sum_{A \in {\6}} \Vert j_{A} \Vert^{2} \leq 0 \ \Longrightarrow \ {\rm dim}K'' = 0
\eeq
Conversely, assume \eqref{eq:stab} holds. 
Let 
\beq
f = \frac 12 {\Tr} \left( {\mu} - {\zeta} {\bf 1}_{K} \right)^{2}
\eeq
Consider the gradient flow, generated by $f$ with respect to the flat K\"ahler metric 
\beq
ds^2 = \Vert d{\bB} \Vert^2 + \Vert d{\bI} \Vert^2 + \Vert d{\bJ} \Vert^2
\eeq
The function $f$ decreases along the gradient trajectory. Moreover, the trajectory belongs to the $GL(k)$-orbit. Eventually, the trajectory stops at a critical point of $f$. Either it is the absolute minimum, i.e. the solution to \eqref{eq:rmom}, or the higher critical point, where
\beq
\langle {\xi} , {\nabla}{\mu} \rangle = 0, \qquad {\xi} = {\mu}  - {\zeta} {\bf 1}_{K} \neq 0
\eeq
The one-parametric subgroup $\left( {\exp}\, t {\xi} \right)_{t \in {\BC}} \subset GL(K)$, preserves $({\bB}, {\bI}, {\bJ})$, 
\beq
\begin{aligned}
& [ B_{a}, {\xi} ] = 0, \qquad a \in {\4} \ , \\
&  {\xi} I_{A} = 0, \quad J_{A} {\xi} = 0, \qquad A \in {\6}  \\
\end{aligned}
\label{eq:iaja}
\eeq
Define $K' = {\rm ker}{\xi}$. The Eq. \eqref{eq:iaja} implies  $K'$ is ${\bB}$-invariant, and contains the image of ${\bI}$. Therefore, by \eqref{eq:stab}, $K' = K$, ${\xi} \equiv 0$, i.e. \eqref{eq:rmom} is satisfied.

\uwave{$\mathbf{Notation.}$}

{}We denote by $[{\bB}, {\bI}, {\bJ}]$ the $GL(k)$-orbit
$\left( g^{-1}B_{a} g, g^{-1}I_{A}, J_{A} g \right)_{a \in {\4}, A \in {\6}, g \in GL(k)}$.

\section{The symmetries of spiked instantons}

The moduli spaces ${\mM}^{*}_{k}({\vec n})$ are acted on by a group ${\Hf} = {\Hf}_{\vec n}$ of symmetries, defined below. 
The symmetry of ${\mM}^{*}_{k}({\vec n})$ will be used in several ways. First, we shall be  studying $\Hf$-equivariant integration theory of the spiked instanton moduli, in cohomology and equivariant $K$-theory. Second, the shall define new moduli spaces by studying the $\Gamma$-fixed loci $\left( {\mM}^{*}_{k}({\vec n}) \right)^{\Gamma}$ in ${\mM}^{*}_{k}({\vec n})$, for subgroups $\Gamma \subset \Hf$. These moduli spaces have the commutant $C_{\Gamma}({\Hf})$ as the symmetry group. Finally, the connected components ${\mM}_{\bk}^{*,\gamma}({\vec{\bn}}) \subset \left( {\mM}^{*}_{k}({\vec n}) \right)^{\Gamma}$ can be defined using only the quiver of $\Gamma$, not the group $\Gamma$. The definition can be then generalized to define more general quiver spiked instantons. Their symmetry ${\Hf}_{\gamma}$ generalizes the commutant $C_{\Gamma}({\Hf})$. 

\subsection{Framing and spatial rotations}

First of all, we can act by a collection
${\bht} = (h_{A})_{A \in {\6}}$  of unitary matrices $h_{A} \in U(n_{A})$, defined up to an overall $U(1)$ multiple:
\beq
{\bht} \cdot
\left[ B_{a}\, , \ I_{A}\, , \ J_{A} \right]= \left[ B_{a}\, , \ I_{A}h_{A}\, , \
 h_{A}^{-1} J_{A} \right] 
\label{eq:hbij}
\eeq
We call the symmetry \eqref{eq:hbij} the framing rotation. 

Secondly, we can multiply the matrices $B_{a}$ by the phases $B_{a} \mapsto q_{a} B_{a}$, as long as their product is equal to $1$:
\beq
\prod_{a\in {\4}} \, q_{a} = 1 
\label{eq:prodq}
\eeq
and we supplement this transformation with the transformation $J_{A} \mapsto q_{A} J_{A}$:
\beq
{\bqt} \cdot \left[ B_{a}\, , \ I_{A}\, , \ J_{A} \right]= \left[ q_{a} B_{a}\, , \ I_{A}\, , \
 q_{A} J_{A} \right] 
 \label{eq:qbij}
 \eeq
We can view $\bqt$ as the diagonal matrix
\beq
{\bqt} = {\rm diag} \left( q_{1}, q_{2}, q_{3}, q_{4} \right)  \in 
U(1)^{3}_{\ept} \subset SU(4)
\label{eq:qdm}
\eeq
which belongs to the maximal torus $U(1)^{3}_{\ept}$ of the group $SU(4)$ of rotations of ${\BC}^{\4}$ preserving some supersymmetry. We call \eqref{eq:qbij} the spatial rotations.

The group
\beq
{\Hf} = \ {\mathrm P} \left( \, \varprod\limits_{A \in {\6}} U(n_{A}) \, \right) \, {\times} \, U(1)^{3}_{\ept}
\eeq
is the symmetry of the moduli space of spiked instantons for generic
$\zeta$ and $\vec n$.  The complexification ${\Hf}_{\BC}$ preserves the holomorphic equations \eqref{eq:bbij} and the stability condition 
\eqref{eq:stab}.

The center $Z_{\Hf}$ of ${\Hf}$ is the eight dimensional torus
\beq
Z_{\Hf} = U(1)^{5}_{\xt} \times U(1)^{3}_{\ept}
\label{eq:zhf}
\eeq 
The maximal torus $T_{\Hf}$ of $\Hf$ is the torus
\beq
T_{\Hf} = \left( \left( \times_{A \in {\6}} \, T_{A} \right)
\, {\bigg\slash}\, U(1)  \right) \, \times U(1)^{3}_{\ept}
\label{eq:thf}
\eeq
where
\beq
T_{A} \subset U(n_{A})
\label{eq:tatorus}
\eeq
is the group of diagonal $n_{A} \times n_{A}$ unitary matrices, 
the maximal torus of $U(n_{A})$, $T_{A} \approx U(1)^{n_{A}}$. In the Eq. \eqref{eq:thf}
we divide by the $U(1)$ embedded diagonally into the product of all 
$\sum_{A} n_{A}$ $U(1)$'s.

\subsubsection{Coulomb parameters}

Let $( {\ba}, {\ept}) \in {\mathrm{Lie}} (T_{\Hf}) \otimes {\BC}$, 
\beq
\begin{aligned}
& {\ept} = ({\ec}_{1}, {\ec}_{2}, {\ec}_{3}, {\ec}_{4}) \, , \qquad {\ec}_{a} \in {\BC}, \quad \sum_{a\in {\4}} {\ec}_{a} = 0 \ , \\
& {\ba} = ({\ac}_{A})_{A \in {\6}}\, , \qquad {\ac}_{A} = {\rm diag} \left({\ac}_{A, 1} , \ldots , {\ac}_{A, n_{A}} \right)  \in
{\rm Lie}(T_{A})\otimes {\BC}
\end{aligned}
\eeq
The eigenvalues ${\ac}_{A, {\alpha}} \in {\BC}$ are defined modulo the overall shift ${\ac}_{A, {\alpha}} \mapsto {\ac}_{A,{\alpha}} + {\xt}$, ${\xt} \in {\BC}$. 
 
The integrals \eqref{eq:pf1} which we define below are meromorphic functions of $( {\ept}, {\ba})$.

\subsubsection{Symmetry enhancements}

Sometimes the symmetry of the spiked ADHM equations enhances. First of all, if all ${\bI} = {\bJ} = 0$ (for $N_{A} = 0$, for all $A$), then the $\bqt$-transformations can be generalized to the action of the full ${\mathrm{SU}}(4) = {\mathrm{Spin}}(6)$: 
\beq
B_{a} \mapsto \sum_{c\in {\4}} \, g_{a{\bar c}} B_{c}\, ,  \qquad\,  gg^{\dagger} = 1, \ {\rm det}(g) = 1 
\label{eq:su4}
\eeq
In the case of less punitive restrictions on $N_{A}$'s, e.g. in the crossed instanton case, 
the symmetry enhances to ${\mathrm{SU}}(2) \times U(1) \times {\mathrm{SU}}(2)$, and, if $\zeta = 0$, to
${\mathrm{SU}}(2)^3$. 
Let us assume, for definiteness, that only $N_{12}$ and $N_{34}$ are non-zero. 
Then the transformations:
\beq
\begin{aligned}
& (B_{1}, B_{2},  B_{3}, B_{4},  I_{12}, J_{12}, I_{34}, J_{34}) \mapsto \\
& \quad \mapsto \left(u aB_{1} + u b B_{2}, -u {\bar b} B_{1}  + u {\bar a} B_{2}, {\bar u} cB_{3} + {\bar u} dB_{4}, -{\bar u}{\bar d} B_{3} + {\bar u}{\bar c} B_{4}, u I_{12}, u J_{12}, {\bar u} I_{34}, {\bar u} J_{34} \right) \, , \\
& \  \left( \begin{matrix} a & b \\ - {\bar b} & {\bar a} \\ \end{matrix} \right) \in {\mathrm{SU}}(2)_{12} \, , \quad \left( \begin{matrix} c & d \\ - {\bar d} & {\bar c} \\ \end{matrix} \right) \in {\mathrm{SU}}(2)_{34} \, , \qquad \left( \begin{matrix} u & 0 \\ 0 & {\bar u} \\ \end{matrix} \right) \in U(1)_{\Delta} \subset {\mathrm{SU}}(2)_{\Delta}\\
& \qquad\qquad\qquad a {\bar a} + b {\bar b} = c {\bar c} + d {\bar d} = u {\bar u} = 1\\
\end{aligned}
\label{eq:abcdu}
\eeq
preserve the crossed instanton equations \eqref{eq:spik}. When ${\zeta} = 0$
the $U(1)_{\Delta}$ symmetry enhances to the full ${\mathrm{SU}}(2)_{\Delta}$, acting by:
\beq
\begin{aligned}
& (B_{1}, B_{2},  B_{3}, B_{4}), \ (I_{12}, J_{12}, I_{34}, J_{34}) \mapsto \\
& \qquad\qquad \mapsto \left(u  B_{1} - v B_{2}^{\dagger}, \,  v B_{1}^{\dagger}  + u B_{2}, \, {\bar u} B_{3} - {\bar v} B_{4}^{\dagger}, \, {\bar v}B_{3}^{\dagger} + {\bar u} B_{4} \right), \\
& \qquad\qquad\qquad\qquad \left(u I_{12} - v J_{12}^{\dagger},\, u J_{12} + v I_{12}^{\dagger}, \, {\bar u} I_{34} - {\bar v} J_{34}^{\dagger}, \, {\bar u} J_{34} + {\bar v} I_{34}^{\dagger} \right) \, , \\
& \qquad\qquad  \left( \begin{matrix} u & {\bar v} \\ -v & {\bar u} \\ \end{matrix} \right) \in  {\mathrm{SU}}(2)_{\Delta}\, , 
 \qquad    u {\bar u} + v {\bar v} = 1\\
\end{aligned}
\label{eq:uv}
\eeq
The equation 
${\Upsilon}_{12} \equiv J_{34} I_{12} - I_{34}^{\dagger} J_{12}^{\dagger}$, the equations
$s_{13} = - s_{24}^{\dagger},  s_{14} = s_{23}^{\dagger}$, 
 the equations ${\sigma}_{3, 12}, {\sigma}_{4, 12}$ as well as the equations
 ${\sigma}_{1, 34}, {\sigma}_{2, 34}$
are ${\mathrm{SU}}(2)_{\Delta}$-invariant, while the equations $s_{12}, {\mu}, s_{34} = s_{12}^{\dagger}$  form a doublet.

\subsection{Subtori}

In what follows we shall encounter the arrangement of hyperplanes ${\mathfrak H}_{l}$ in ${\mathrm{Lie}} (T_{\Hf}) \otimes {\BC}$ defined by the system of linear equations:
\beq
L_{i}({\ba}, {\ept}) = \sum_{A \in {\6}} \sum_{{\alpha} \in [n_{A}]} \ {\varpi}_{i; A, {\alpha}} {\ac}_{A, {\alpha}} + \sum_{a \in {\4}} n_{i; a} {\ec}_{a} = 0
\label{eq:hyper}
\eeq
with ${\varpi}_{i; A, \alpha} \in \{ -1, 0, +1 \}$, $n_{i; a} \in {\BZ}$ and the matrix ${\varpi}_{i; A, {\alpha}}$ of maximal rank. Such equations  \eqref{eq:hyper} can be interpreted as defining a subtorus
${\Hfr} = T_{L} \subset T_{\Hf}$: simply solve \eqref{eq:hyper} for the subset of  ${\ac}_{A, \alpha}$'s for which the matrix ${\varpi}_{i; A, \alpha}$ is invertible. We shall not worry about the integrality of the inverse matrix in this paper, by using the covering tori, if necessary. 

One of the reasons we need to look at the subtori $T_{L}$ is the following construction. 

\subsection{Orbifolds, quivers,  defects}

In this section the global symmetry group $\Hf$ is equal to 
\[ {\Hf} = P\left( \varprod_{A \in {\6}} U(n_{A}) \right) \times {\Gr} \]
where  
\begin{enumerate}
\item{}
\[ {\Gr} = U(1)^{3}_{\ept} \]
if there are at least two $A' \neq A'' \in {\6}$ with non-empty intersection with $n_{A'} n_{A''} \neq 0$, and to
\item{}
\[ {\Gr} =  {\mathrm{SU}}(2)_{A} \times U(1)_{\Delta} \times {\mathrm{SU}}(2)_{\bar A} \]
otherwise, i.e. there is at most one pair $A, {\bar A}$ with $n_{A}n_{\bar A} \neq 0$. 
\end{enumerate}
In all cases
\[ {\Gr} \subset {\mathrm{SU}}(4) \,  , \]
so that to every ${\gamma} \in {\Gamma}$ one associates a unitary
$4 \times 4$ matrix ${\bqt} = \Vert q_{a}^{b}({\gamma}) \Vert_{a, b \in {\4}}$ with unit determinant. In the first case this matrix is diagonal, in the second case it is a $2 \times 2$ block-diagonal matrix with unitary $2\times 2$ blocks of inverse determinants.

The symmetry of ${\mM}^{*}_{k}({\vec n})$ can be used to define new moduli spaces. Suppose $\Gamma \subset {\Hf}$ is a discrete subgroup. Let ${\Hf}^{\Gamma} \subset {\Hf}$ be the maximal subgroup commuting with $\Gamma$, the centralizer of $\Gamma$. Let
${\Gamma}^{\vee}$ be the set of irreducible unitary representations $\left( {\CalR}_{\omega} \right)_{{\omega} \in {\Gamma}^{\vee}}$ of $\Gamma$,  and ${\vec k} \in {\BZ}_{\geq 0}^{{\Gamma}^{\vee}}$. 
The representations $N_{A}$, $A \in {\6}$ of $\Hf$ decompose as representations of $\Gamma$
\beq
N_{A} = \bigoplus_{{\omega} \in {\Gamma}^{\vee}} \, N_{A, \omega} \otimes {\CalR}_{\omega}
\label{eq:naom}
\eeq
Let $\vec{\bn}$ now denote the collection $(n_{A, \omega})_{A \in {\6}, {\omega} \in {\Gamma}^{\vee}}$ of dimensions
\beq
n_{A , \omega} = {\rm dim}N_{A, \omega} 
\label{eq:naomdim}
\eeq
of multiplicity spaces. 
The vector ${\bk} = (k_{\omega} )_{{\omega} \in {\Gamma}^{\vee}}$ defines a representation of $\Gamma$:
\beq
{\gamma} \in {\Gamma} \mapsto g_{\gamma} \in U(K) \, , \qquad
K = \bigoplus\limits_{{\omega} \in {\Gamma}^{\vee}} \, K_{\omega} \otimes {\CalR}_{\omega} \, , \qquad k_{\omega} = {\rm dim}K_{\omega}
\label{eq:kom}
\eeq
We call the components $k_{\omega}$ of the vector ${\bk}$
fractional instanton charges. 
The moduli space of $\Gamma$-folded spiked instantons of charge $\vec k$ is the component ${\mM}_{\bk}^{*, \Gamma} ({\vec{\bn}})$ set of $\Gamma$-fixed points $ \left( {\mM}^{*}_{k}({\vec n}) \right)^{\Gamma}$. The representation \eqref{eq:kom}
enters the realization of the $\Gamma$-fixed locus in the space of matrices $({\bB}, {\bI}, {\bJ})$:
\beq
{\gamma} \cdot \left( {\bB}, {\bI}, {\bJ} \right) = \left( g_{\gamma} {\bB} g_{\gamma}^{-1}, g_{\gamma} {\bI}, {\bJ} g_{\gamma}^{-1} \right)\, , \qquad g_{\gamma} \in U(K)
\label{eq:gamfix}
\eeq
where
\begin{multline}
{\gamma} \cdot \left( {\bB}, {\bI}, {\bJ} \right) \equiv \left( q_{a}^{b}({\gamma}) B_{b}, I_{A} h_{A} ({\gamma})^{-1} , q_{A} ({\gamma})h_{A}({\gamma}) J_{A} \right) \, , \\
{\gamma} \in {\Gamma} \mapsto \left( h_{A} ({\gamma}) \right)_{A \in {\6}}  \times  \Vert q_{a}^{b} ({\gamma}) \Vert_{a,b \in {\4}}  \ \in {\Hf}
\end{multline}
is the defining representation of $\Gamma$, with
$q_{A}({\gamma})$ given by \eqref{eq:pqa} in the case (1), and by the
projection to $U(1)_{\Delta}$ in the case (2). The equations
\eqref{eq:gamfix} are invariant under the subgroup
\beq
\varprod_{{\omega} \in {\Gamma}^{\vee}} U(K_{\omega}) \ \subset \ U(K)
\label{eq:ukga}
\eeq
of unitary transformations of $K$ commuting with $\Gamma$. The holomorphic equations \eqref{eq:bbij} restricted onto the locus
of $\Gamma$-equivariant i.e. obeying \eqref{eq:gamfix} matrices ${\bB}, {\bI}, {\bJ}$ become the holomorphic equations defining ${\mM}^{\Gamma}_{\bk}({\vec\bn})$. The stability condition \eqref{eq:stab}
can be further refined, analogously to the refinement of the real moment map equation ${\mu} = \sum_{{\omega} \in {\Gamma}^{\vee}} \, {\zeta}_{\omega} {\bf 1}_{K_{\omega}} \otimes {\bf 1}_{{\CalR}_{\omega}}$. We shall work in the chamber where all ${\zeta}_{\omega} > 0$. 

The moduli spaces ${\mM}^{\Gamma}_{\bk}({\vec\bn})$ in the case  (1)
parametrize the spiked instantons in the presence of $U(1)^{3}_{\ept}$-invariant surface operators, while in the case (2) they parametrize 
the instantons in supersymmetric quiver gauge theories on the ALE spaces, with additional defect. 

The commutant ${\Hf}^{\Gamma}$ acts on ${\mM}^{\Gamma}_{\bk}({\vec\bn})$, so that the partition functions we study are meromorphic functions on ${\mathrm{Lie}}\left( {\Hf}^{\Gamma} \right) \otimes {\BC}$. 

Note that if $\Gamma$ has trivial projection to $\Gr$ then the moduli space of $\Gamma$-folded instantons is simply the product of the moduli spaces of spiked instantons for $N_{\omega}$'s. In what follows we assume the projection to $\Gr$ to be non-trivial. 

\subsubsection{Subtori for $\Gamma$-folds}

Let us now describe the maximal torus $T_{{\Hf}^{\Gamma}}$ of the $\Gamma$-commutant as $T_L$. In other words, the choice
of a discrete subgroup $\Gamma \subset {\Hf}$ defines the hyperplanes 
$L_{i}({\ac}, {\ept}) =0 $ in ${\mathrm{Lie}}(T_{\Hf})$. 

In the case (1) the ${\Gr}$-part of ${\Gamma}$ is abelian, i.e. it is a product of cyclic groups (if $\Gamma$ is finite) or it is a torus itself. In either case there is no restriction on the $\ept$-parameters. The framing
part of $\Gamma$ reduces $P\left( \times_{A} U(n_{A}) \right)$
to
$P \left( \times_{A, \omega} U(n_{A, \omega}) \right)$ which means that
some of the eigenvalues ${\ac}_{A, {\alpha}}$, viewed as the generators of ${\mathrm{Lie}}U(n_{A})$, must coincide, more precisely to be of multiplicity ${\rm dim}{\CalR}_{\omega}$. 
The minimal case, when $\Gamma$ is abelian, imposes no restrictions
on $({\ac}, {\ept})$, so that $T_{{\Hf}^{\Gamma}} = T_{\Hf}$.

In the case (2) the ${\Gr}$-part of $\Gamma$ need not be abelian. 
Let us assume, for definiteness, that $A = 12$, ${\bar A} = 34$. If the image of $\Gamma$ in $SU(2)_{12}$ is non-abelian, then ${\ec}_{1} = {\ec}_{2}$. Likewise if the image of $\Gamma$ in $SU(2)_{34}$ is non-abelian then ${\ec}_{3} = {\ec}_{4}$. The non-abelian discrete subgroups
of $SU(2)$ have irreducible representations of dimensions $2$ and higher, up to $6$. Thus the corresponding ${\ac}_{A, \alpha}$ eigenvalues will have the multiplicity up to $6$.

\subsubsection{Subtori for sewing} Let us specify the integral data for the subtori, i.e. the explicit solutions to the constraints \eqref{eq:hyper}.  
Let ${\be} = (e_{a})_{a\in {\4}}$, $e_{a} \in {\BZ}_{\neq 0}$,  be a $4$-tuple of non-zero integers, with no common divisors except for $\pm 1$, which sum up to zero:
\beq
\sum_{a \in {\4}} e_{a} = 0 \ . 
\label{eq:eaz}
\eeq
Such a collection $\be$ defines a split ${\6} = {\6}^{+} \amalg {\6}^{-}$, where ${\6}^{\pm}$ being the set of  $A = \{ a , b\}$ such that $\pm e_{a}e_{b} > 0$. 

For $A = \{ a, b \} \in {\6}^{-}$, i.e. $e_{a}e_{b}<0$, let $p_{A} = {\rm gcd}(|e_{a}|, |e_{b}|) > 0$. 
Let us also fix for such $A \in {\6}^{-}$ a partition ${\nu}_{A} = \left( {\nu}_{A, \iota} \right)$ of size $n_{A}$, whose parts do not exceed $e_{A}$: $1\leq {\nu}_{A, \iota} \leq p_{A}$. Let ${\ell}_{A} = {\ell}({\nu}_{A})$ be its length. 

Given ${\nu}_{A}$ we partition the set $[n_{A}]$  
as the union of nonintersecting subsets
\beq
[n_{A}] \, = \, \bigcup\limits_{\iota} \ [n_A]_{\iota} \ , \qquad  {\#}[n_A]_{\iota} = {\nu}_{A,\iota} \ ,
\label{eq:partna}
\eeq
$[n_A]_{\iota'}  \cap [n_A]_{\iota''} = {\emptyset}$ for $\iota' \neq \iota''$.  
Fix a map $c_{A, \iota}: [n_A]_{\iota} \to {\BZ}$
obeying, for any $a', a'' \in [n_A]_{\iota}$, $a' \neq a''$:
\beq
c_{A, \iota} (a' ) - c_{A, \iota} ( a'' ) \, \neq 0 \, ({\rm mod} \, p_{A} ) \, 
\label{eq:cagab}
\eeq
When $p_{A} =1$ the condition \eqref{eq:cagab} is empty.

For $A = \{ a , b \} \in {\6}^{+}$,  let
us fix a map $c_{A}: [n_{A}] \to {\BZ}$, obeying, for any $a', a'' \in [n_{A}]$,
\beq
c_{A}(a') - c_{A}(a'') \, \notin {\BZ}_{>0} e_{a} + {\BZ}_{>0} e_{b}
\label{eq:cagagb}
\eeq
Note that \eqref{eq:cagagb} does not forbid the situation where $c_{A}(a') = c_{A}(a'')$ for some $a', a'' \in [n_{A}]$. 
 To make the notation uniform we assign to such $A$, ${\nu}_{A} = (n_{A})$, ${\ell}_{A} = 1$. 
 
 The final piece of data is the choice ${\iota}_{A} \in [{\ell}_{A}]$ for each $A \in {\6}^{-}$. Define the set ${\lambda}_{A} = [{\ell}_{A}]\backslash \{ {\iota}_{A} \}$, of cardinality ${\ell}_{A}-1$.

Now, we associate to the data 
\beq
{\Lf} = ({\nu}, {\lambda}, c,{\be}) \ ,
\eeq
the torus
 \beq
 {\Hfr} = T_{\Lf} = \left( \varprod_{A \in {\6}} U(1)^{{\ell}_{A}-1} \right) \ \times \ U(1)
 \label{eq:mintor}
 \eeq
 Note that only $A \in {\6}^{-}$ contribute to the product in \eqref{eq:mintor}. 
 This torus is embedded into $T_{\Hf}$ as follows:
 the element  
 \beq
( e^{{\ii}{\xi}}, e^{{\ii}t} ) \equiv \varprod_{A \in {\6}^{-}} \left( e^{{\ii}{\xi}_{A, i}} \right)_{i \in {\lambda}_{A}} \times e^{{\ii}t}  \in T_{\Lf}
 \eeq
is mapped to
 \beq
 \Biggl[ \ \varprod_{A \in {\6}^{-}} \, {\rm diag}_{A}^{-}  \ \times   \, \varprod_{A \in {\6}^{+}} {\rm diag}^{+}_{A} \Biggr] \ \times \ {\rm diag}\left( e^{{\ii}e_{a}t} \right)_{a \in {\4}}  \in {\mathrm P}\left( \varprod_{A \in {\6}} U(n_{A}) \right) \times {\mathrm{SU}}(4)
 \eeq
 where ${\rm diag}_{A}^{\pm} \in U(n_{A})$ are the diagonal matrices with the eigenvalues
 \beq
 {\rm Eigen} \left( {\rm diag}_{A}^{-} \right) = 
 \left\{  \  e^{{\ii}c_{A, {\iota}_{A}}({\alpha})t} \ \vert\ {\alpha} \in [n_{A}]_{{\iota}_{A}} \right\} \, \cup \,  \left\{ \ e^{{\ii} \left( {\xi}_{A, \iota} + c_{A,i} ({\alpha}) t \right) } \ \vert\ {\iota} \in {\lambda}_{A}, {\alpha} \in [n_{A}]_{\iota}  \ \right\} 
 \eeq
 \beq
 {\rm Eigen} \left( {\rm diag}_{A}^{+} \right) = \left\{ \  e^{{\ii}   c_{A} ({\alpha}) t   } \vert\ {\alpha} \in [n_{A}] \ \right\}
 \eeq
Thus, the torus $T_{\Lf}$ corresponds to the solution of the Eqs. \eqref{eq:hyper} with
\begin{multline}
{\ec}_{a} = e_{a}u, \ a \in {\4}, \\
 {\ac}_{A,  {\alpha}} = c_{A}({\alpha})u,  \, A \in {\6}^{+}, {\alpha} \in [n_{A}]  \\
 {\ac}_{A, {\alpha}} = c_{A, {\iota}_{A}}({\alpha})u, \, A \in {\6}^{-}, {\alpha} \in [n_{A}]_{{\iota}_{A}} \\
 {\ac}_{A, {\alpha}} = {\xi}_{A, {\iota}} + c_{A, {\iota}}({\alpha}) u , \, A \in {\6}^{-} , \ {\alpha} \in [n_{A}]_{\iota} \, , \ {\iota} \in {\lambda}_{A}  
 \end{multline}
In other words, the $\Omega$-background parameters
are maximally rationally dependent (the worst way to insult the rotational parameters), the framing of the spaces $N_{A}$, $A \in {\6}^{+}$ is completely locked with space rotations (spin-color locking), while the framing of the spaces $N_{A}$, $A \in {\6}^{-}$ is locked partially.

\subsection{Our goal: compactness theorem}
  
Our goal is to establish the compactness of the fixed point sets ${\mM}_{k}({\vec n})^{T_{\Lf}}$ and 
$\left( {\mM}^{\Gamma}_{\bk}({\vec\bn}) \right)^{T_{{\Hf}^{\Gamma}}}$. Before we attack this problem we shall discuss a little bit the ordinary instantons, then look at a
few examples of the particular types of spiked instantons: the crossed and the folded instantons, and then proceed with the analysis of the general case. The reader interested only in the compactness theorem can skip the next two sections at the first reading. 

\section{Ordinary instantons}

In this section we discuss the relations between the ordinary four dimensional $U(n)$ instantons and the spiked instantons. 

\subsection{ADHM construction and its ${\rm fine\ print}$}

In the simplest case  only one of six vector spaces is non-zero, e.g.
\beq
N_{A} = 0\, , \qquad A \neq  \{ 1 , 2 \} . 
\label{eq:ordi}
\eeq
Let $n = n_{12}$. We shall now show that, set theoretically, the moduli space of spiked instantons in this case is ${\iM}_{k}(n)$, the ADHM moduli space (more precisely, its Gieseker-Nakajima generalization).

Recall the ADHM construction of the $U(n)$ framed instantons of charge $k$  on (noncommutative) ${\BR}^{4}$ \cite{Atiyah:1978ri, Nakajima:1994r, Nekrasov:1998ss}.  
It starts by fixing Hermitian vector spaces $N$ and $K$ of dimensions $n$ and $k$, respectively. Consider the space of quadruples $(B_{1}, B_{2}, I, J)$, 
\beq
I: N \to K, \ J: K \to N, \ B_{\al} : K \to K, \ {\al} = 1, 2
\label{eq:ijbb}
\eeq
obeying 
\beq
{\vec\mu}_{12} \equiv
 \left( \ 2{\mu}_{12}^{\BR}, \ {\mu}_{12}^{\BC} + {\mu}_{12}^{\BC\dagger} , \ {\ii} \left( {\mu}_{12}^{\BC} - {\mu}_{12}^{\BC\dagger}\right)\  \right) = ( {\zeta}, 0, 0 ) \cdot {\bf 1}_{K} \ ,
 \label{eq:adhmeq}
 \eeq
 where 
\beq
{\mu}_{12}^{\BC} = [ B_{1}, B_{2} ] + IJ, \qquad 
2 {\mu}_{12}^{\BR} = [ B_{1}, B_{1}^{\dagger}] + [B_{2}, B_{2}^{\dagger} ] + I I^{\dagger} - J^{\dagger} J
\label{eq:adhmmom}
\eeq 
Note that the number of equations \eqref{eq:adhmeq} plus the number of symmetries is less then the number of variables. The moduli space ${\iM}_{k}(n)$ of solutions to \eqref{eq:adhmeq} modulo the $U(k)$ action
\beq
( B_{1}, B_{2}, I, J) \mapsto (g^{-1} B_{1} g, g^{-1} B_{2} g, g^{-1} I, J g)\, ,  \quad g \in U(k)
\label{eq:ukac}
\eeq
has the positive dimension 
\beq
{\rm dim}_{\BR} {\iM}_{k} (n) = 4k ( n + k) - 3 k^2 - k^2 = 4 n k 
\eeq
Again, the ${\mu}_{12}^{\BR}$-equation, with ${\zeta} > 0$, can be replaced by the stability condition, and the $GL(k)$-symmetry:

\beq\mathboxit{\begin{aligned}
& {\rm Any\ subspace}\ K' \subset K \, , \ {\rm such\ that} \  \color{red}{I(N) \subset K'\, \color{black} , } \\
& \qquad\qquad {\color{black}\rm and}\  \color{blue}{B_{\alpha}(K') \subset K' \, , \quad \color{black}{\rm for\ all}\ \color{blue}{\alpha} = 1,2}  \\
& \qquad\qquad \color{black}{\rm coincides\ with\ all\ of\ } K \, , \ K' = K \\
& \qquad{\rm in\ other\ words\, , }\quad\color{green} {\BC} [ B_{1}, B_{2}]  \, I (N ) = K \\
\end{aligned}}   
\label{eq:stab12}
\eeq

\uwave{\bf Notation.}

{}We denote by $[B_{1}, B_{2}, I, J]$ the $GL(k)$-orbit
$\left( g^{-1}B_{1} g, g^{-1}B_{2}g, g^{-1}I, J g \right)_{g \in GL(k)}$.

\subsection{Ordinary instantons from spiked instantons}

Now, to show that the spiked instantons reduce to the ordinary instantons when \eqref{eq:ordi} is obeyed, we need to show that $B_{3} = B_{4} = 0$ on the solutions of our equations \eqref{eq:bbij}. This is easy:
\beq
B_{3} f (B_{1}, B_{2} ) I = f (B_{1}, B_{2} ) B_{3}I = 0
\eeq
where we used $ [ B_{1}, B_{3} ] = {\mu}_{13} = 0$, $ [B_{2}, B_{3}] = {\mu}_{23} = 0$, and
$B_{3} I = 0$. Therefore $B_{3}$ acts by zero on all of $K$. The same argument proves the vanishing of $B_4$. 

\subsection{One-instanton example}

Let $k=1$. We can solve the equations \eqref{eq:adhmmom} explicitly. The matrices $B_{1}, B_{2}$ are just complex numbers, e.g. $b_{1}, b_{2} \in {\BC}$. The pair $I, J$ obeys $IJ = 0$, $ \Vert I \Vert^{2} - \Vert J \Vert^2 = {\zeta}$. Assuming $\zeta > 0$ define the vectors $w_{1} = J(K) \in N$, $w_{2} = \frac{1}{\sqrt{{\zeta} + \Vert J \Vert^{2}}} \, I^{\dagger}(K) \in N$. They obey $\langle w_{2}, w_{1} \rangle = 0$, $\langle w_{2}, w_{2} \rangle  = 1$. Dividing by the $U(1) = U(K)$ symmetry we arrive at the conclusion:
\beq
{\iM}_{1}(n) = {\BC}^{2} \times T^{*}{\BC\BP}^{n-1}
\label{eq:1inst}
\eeq   
The first factor parametrizes $(b_{1}, b_{2})$, the base ${\BC\BP}^{n-1}$ of the second factor is the space of $w_{2}$'s obeying $\Vert w_{2} \Vert^{2} =1$ modulo $U(1)$ symmetry.

 \subsection{$U$ versus $PU$}
 
 In describing the action of  ${\Hf}_{12}$ in \eqref{eq:hbij} specified to the case of ordinary instantons we use an element $h$ of the group $U(n)$ yet it is the group $PU(n) = U(n)/U(1) = SU(n)/{\BZ}_{n}$ which acts faithfully on ${\iM}_{k}(n)$. Indeed, multiplying $h$ by a scalar matrix
 \beq
 h \to h {\tilde u},  \qquad {\tilde u} \in U(1)
 \label{eq:htou}
 \eeq
does not change the effect of the transformation \eqref{eq:hbij} since it can be undone by the $U(k)$-transformation \eqref{eq:ukac} with $g = {\tilde u}^{-1} {\bf 1}_{K} \in U(k)$. 
 
 \subsection{Tangent space}
  
Let $m \in {\iM}_{k}(n)$. Let $(B_{1}, B_{2}, I, J)$ be the representative
of $m = [B_{1}, B_{2}, I, J]$.  
 Consider the nearby quadruple
\beq
 ( B_{1} + {\delta}B_{1}, B_{2} + {\delta}B_{2}, I + {\delta}I, J + {\delta}J )
 \eeq
Assuming it solves the ADHM equations to the linear order, 
 the variations ${\delta}B_{1}, {\delta}B_{2}, {\delta}I, {\delta}J$ are subject to the linearized ${\mu}_{12}^{\BC}$ equation:
 \beq
 d_{2} \left( {\delta} B_{1}, {\delta}B_{2} , {\delta}I, {\delta} J \right) : = [B_{1}, {\delta}B_{2}] + [{\delta}B_{1} , B_{1}] + ({\delta}I) J + I ({\delta}J) = 0
 \eeq
 and we identify the variations which differ by an infinitesimal $GL(K)$-transformation of $(B_{1}, B_{2}, I, J)$: 
 \begin{multline}
 \left( {\delta} B_{1}, \, {\delta}B_{2} , \, {\delta}I, \, {\delta} J \right) \sim \left( {\delta} B_{1}, \, {\delta}B_{2} , \, {\delta}I, \, {\delta} J \right) + d_{1} \left( {\delta}{\sigma} \right), \\
 d_{1} \left( {\delta}{\sigma} \right) : = \left( \, [ B_{1}, {\delta}{\sigma} ]\, , \,  
  [ B_{2}, {\delta}{\sigma} ], \, - {\delta}{\sigma} \cdot I \, ,  \, J \cdot {\delta}{\sigma} \, \right) 
  \end{multline}
  Since $d_{2} \circ d_{1} = 0$, the tangent space is the degree $1$ cohomology, $T_{m}  {\iM}_{k}(n) = {\ker} d_{2} / {\rm im} d_{1}\, = \, H^{1} {\CalT}_{m}{\iM}_{k}(n)$, of the complex
\begin{multline}
{\CalT}_{m}{\iM}_{k}(n)  = \\
 [ 0 \to {\rm End}(K)  \longrightarrow^{\kern -.2in d_{1} \kern .2in}\qquad\qquad\qquad\qquad\qquad\qquad \qquad\qquad\qquad\qquad    \\
 \longrightarrow^{\kern -.2in d_{1} \kern .2in} {\rm End}(K) \otimes {\BC}^{2}_{12} \oplus {\rm Hom}(N, K) \oplus {\rm Hom}(K, N)  \longrightarrow^{\kern -.2in d_{2} \kern .2in} \\ \qquad\qquad\qquad\qquad \qquad\qquad \qquad\qquad  
\longrightarrow^{\kern -.2in d_{2} \kern .2in} {\rm End}(K) \otimes \wedge^{2}{\BC}_{12}^{2}   \to 0 ]\\
\label{eq:tangcomplex}
\end{multline}

 \subsection{Fixed locus}
  
In applications we will be interested in the fixed point set ${\iM}_{k}(n)^{\rm T_{12}}$ with ${\rm T}_{12} \subset {\Hf}_{12}$
a commutative subgroup. The maximal torus ${\rm T}_{12} \subset {\Hf}_{12}$ is the product of the maximal torus ${\bf T}_{n} \subset PU(n)$ and the two dimensional torus $U(1)_{12} \times U(1)_{12}^{\prime} \subset SU(2)_{12} \times U(1)_{12}^{\prime}$. Let 
\beq
{\ac} = {\ii}\, {\rm diag} ({\ac}_{1}, \ldots , {\ac}_{n}) \, , \qquad {\ac}_{\alpha} \in {\BR}
\eeq
be the generic element of ${\mathrm{Lie}}\left( {\bf T}_{n}\right)$. It means that the numbers ${\ac}_{\alpha}$ are defined up to the simultaneous shift
\beq
{\ac}_{\alpha} \sim {\ac}_{\alpha} + a, \qquad a \in {\BR}
\label{eq:pushi}
\eeq
and we assume ${\ac}_{\alpha} \neq {\ac}_{\beta}$, for ${\alpha} \neq {\beta}$.  
Let
\beq
{\ept}_{12} = \frac 12 \left(  \, {\ec}_{1} -{\ec}_{2} ,\,
\, {\ec}_{1} + {\ec}_{2} \,  \right)
\eeq be the generic element of
${\mathrm{Lie}} \left( U(1)_{12} \times U(1)_{12}^{\prime} \right)$. 
The pair $({\ac}, {\ept}_{12})$ generates an infinitesimal transformation \eqref{eq:hbij}, \eqref{eq:qbij} of the quadruple $(B_{1}, B_{2}, I, J)$:
\beq
{\delta}_{{\ac}, {\ept}_{12}} (B_{1}, B_{2}, I, J) = \left( \, {\ec}_{1}B_{1}, {\ec}_{2} B_{2}, I{\ac}, \left( {\ec}_{1} + {\ec}_{2} - {\ac} \right) J \, \right)
\eeq 
For the $U(k)$-equivalence class $f = [ B_{1}, B_{2}, I, J]$ to be fixed under the  infinitesimal transformation generated by the generic pair $( {\ac}, {\ept})$ there must exist an infinitesimal $U(k)$-transformation \eqref{eq:ukac}
\beq
 {\delta}_{\sigma} (B_{1}, B_{2}, I, J) =  \left( \, [B_{1}, {\sigma} ],\, [B_{2}, {\sigma}] , \,  - {\sigma} I, \, J {\sigma}\, \right)
 \eeq
 undoing it: ${\delta}_{\sigma}(B_{1}, B_{2}, I, J) + {\delta}_{{\ac}, {\ept}_{12}} (B_{1}, B_{2}, I, J) = 0$. In other words, there must exist the operators
 ${\hat d}_{\alpha}, {\hat{\rm d}}_{1}, {\hat {\rm d}}_{2} \in {\rm End}(K)$, such that:
 \beq
 \begin{aligned}
 {\sigma} = \sum_{{\alpha} \in [n]} {\ac}_{\alpha} {\hat d}_{ \alpha} + {\ec}_{1} {\hat {\rm d}}_{1} + {\ec}_{2} {\hat {\rm d}}_{2}\, , \\
 \sum\limits_{{\alpha} \in [n]} {\hat d}_{\alpha} = {\bf 1}_{K} 
 \label{eq:sial}
 \end{aligned}
 \eeq
 obeys:
 \beq
 \begin{aligned}
 & {\ec}_{a} B_{a} =  [ {\sigma} , B_{a} ] \, , \ a \in \{ 1, 2 \} \\
 &  I {\ac}  = {\sigma} I \, , \quad
 - \left( {\ec}_{1} + {\ec}_{2} \right) J + {\ac} J =  J {\sigma}  \, , \label{eq:compen}
 \end{aligned}
 \eeq
 or, in the group form:
 \beq
 q_{a} B_{a} = g_{t} B_{a} g_{t}^{-1}\, , \quad q_{1}q_{2} h_{t}^{-1} J = J g_{t}^{-1} \, , \quad I h_{t}  = g_{t} I 
 \label{eq:grcompen}
 \eeq
 where $q_{a} = e^{t{\ec}_{a}}, g_{t} = e^{t {\sigma}}, h_{t} = e^{t {\ac}}$. Here $t$ is an arbitrary complex number, the map $t \mapsto g_{t}$ defines the representation ${\bT}_{12} \to GL(k)$.
 The Eqs. \eqref{eq:grcompen} imply:
 \beq
 g_{t} \, \left( B_{1}^{i-1}B_{2}^{j-1} I \right) = q_{1}^{i-1}q_{2}^{j-1}\, \left( B_{1}^{i-1}B_{2}^{j-1} I \right) h_{t} 
 \eeq
  
The Eqs. \eqref{eq:compen} for generic $( {\ac}, {\ept})$ imply:
 \beq
 K   = \bigoplus_{{\alpha} \in [n]} \, K_{\alpha}\, , \qquad
 {\hat d}_{\alpha} \vert_{K_{\beta}} = {\delta}_{\alpha, \beta} \label{eq:kab}
 \eeq
 The eigenspace $K_{\alpha}$ is generated by $I_{\alpha} = I(N_{\alpha})$, where $N_{\alpha} \subset N$ is the eigenline of ${\ac}$ with the eigenvalue ${\ac}_{\alpha}$:
 \beq 
 K_{\alpha} = \sum_{i,j \geq 1} \,  B_{1}^{i-1}B_{2}^{j-1} I_{\alpha} \ .
 \eeq
 The subspace $I_{\alpha} \subset K$ (it is one-dimensional for generic $\ac$) obeys:
 \beq
 {\hat d}_{\beta} I_{\alpha} = {\delta}_{\alpha, \beta}\, , \ {\hat{\rm d}}_{1} I_{\alpha} = {\hat{\rm d}_{2}} I_{\alpha} = 0
 \eeq
 On $K_{\alpha}$ the operators ${\hat{\rm d}}_{1}, {\hat{\rm d}}_{2}$ have non-negative spectrum:
 \beq
 {\sigma} \left( B_{1}^{i-1} B_{2}^{j-1} I_{\alpha} \right) = \left(  {\ac}_{\alpha} + {\ec}_{1}(i-1)+ {\ec}_{2} (j-1) \right) \left( B_{1}^{i-1} B_{2}^{j-1} I_{\alpha} \right)
 \eeq
 Therefore, as long as ${\ac}_{\beta} - {\ac}_{\alpha} \notin {\ec}_{1} {\BZ}_{> 0} + {\ec}_{2} {\BZ}_{>0}$, 
 \beq
 J \left( B_{1}^{i-1} B_{2}^{j-1} I_{\alpha} \right) = 0\, , 
 \eeq
 as follows from the last Eq. in \eqref{eq:compen}.
 
Thus, we have shown that
 \beq
 J  = 0, \qquad [ B_{1}, B_{2} ] = 0
 \eeq
Define an ideal ${\CalI}^{({\alpha})} \subset {\BC}[x,y]$ in the ring of polynomials in two variables by:
\beq
P (x,y) \in {\CalI}^{({\alpha})}\, \Leftrightarrow \, P(B_{1}, B_{2}) I_{\alpha} = 0
\label{eq:ideal}
\eeq
Define the partition ${\lambda}^{({\alpha})} = \left( {\lambda}_{1}^{({\alpha})} \geq {\lambda}_{2}^{({\alpha})} \geq \ldots \geq {\lambda}^{({\alpha})}_{{\ell}_{{\lambda}^{({\alpha})}}} \right)$ by
\beq
{\lambda}_{i}^{({\alpha})} = {\rm min} \{ \, j \, | \, B_{1}^{i-1} B_{2}^{j} I_{\alpha} = 0\, \} 
 \eeq
Thus, $K_{\alpha} = {\BC}[z_{1}, z_{2}] / {\CalI}_{{\lambda}^{({\alpha})}}$.
 Here we denote by ${\CalI}_{\lambda} \subset {\BC}[x, y] $ the ideal generated by the monomials $x^{i-1}y^{{\lambda}_{i}}$, $i = 1, 2, \ldots , {\ell}_{\lambda}$. 
 
{}Conversely, given the monomial ideal 
${\CalI}_{{\lambda}^{({\alpha})}}$
 define the vector $I_{\alpha} \subset K_{\alpha}$ to be the image of the polynomial $1$ in the quotient ${\BC}[x, y] / {\CalI}_{{\lambda}^{({\alpha})}}$. The operators $B_{1}, B_{2}$ act by multiplication by the coordinates $x$, $y$, respectively. 
Furthermore, 
\beq
K_{\alpha} = \bigoplus_{i=1}^{{\ell}_{\lambda}} \, \bigoplus\limits_{j=1}^{{\lambda}_{i}}\ K_{{\alpha}; i,j}
\eeq
where
\beq
{\hat{\rm d}}_{1} \vert_{K_{{\alpha}; i,j}} = i-1, \qquad
 {\hat{\rm d}}_{2} \vert_{K_{{\alpha}; i,j}} = j-1
 \eeq
The map  $({\ac}, {\ept}_{12}) \mapsto {\sigma} \in {\rm End}(K)$ makes the space $K$ a ${\bT}_{12}$-representation. 
 Its character is easy to compute:
 \beq
 K_{\chi} : = {\Tr}_{K} \, g_{t} \, = \,  \sum_{{\alpha} \in [n]} e^{t{\ac}_{\alpha}} \, \sum_{i=1}^{{\ell}_{{\lambda}^{({\alpha})}}} q_{1}^{i-1} \sum_{j=1}^{{\lambda}^{({\alpha})}_{i}} q_{2}^{j-1}
 \label{eq:kchar}
 \eeq
 The set of eigenvalues of $\sigma$ is a union of $n$ collections
 of centers of boxes of Young diagrams ${\lambda}^{(1)}, \ldots , {\lambda}^{(n)}$
 
 \begin{figure}[H]
 \centerline{\picit{8}{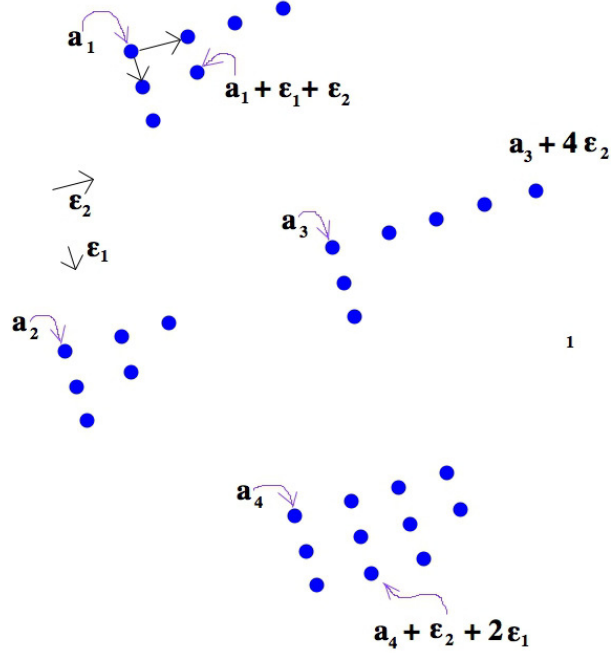}}
 \label{fig:sigma12}
  \caption{Eigenvalues of $\sigma$}
  \end{figure}
 {}The space $N$ is a ${\bT}_{12}$-representation by definition:
 \beq
 N_{\chi} : = {\Tr}_{N} \, h_{t}  \, = \, \sum_{{\alpha} \in [n]} \ e^{t{\ac}_{\alpha}}
 \label{eq:nchar}
 \eeq
We also define: 
\beq
K_{\chi}^{*} = {\Tr}_{K} \, g_{t}^{-1}\ , \ N_{\chi}^{*} = {\Tr}_{N} \, h_{t}^{-1}
\eeq
\subsection{Tangent space at the fixed point}

Finally, the tangent space $T_{f}{\iM}_{k}(n)$ to the moduli space at
the fixed point $f$ is also a ${\bT}_{12}$-representation. Let us compute its character. Let $f = [B_{1}, B_{2}, I, J]$. The quadruple $\left( B_{1}, B_{2}, I, J \right)$ is fixed by the composition of the ${\bT}_{12}$ transformation $(e^{t{\ac}}, e^{t{\ept}_{12}}) \in {\bT}_{12}^{\BC}$ and the $GL(k)$-transformation $e^{t{\sigma}} \in  GL(k)$, for any complex number $t$, cf. \eqref{eq:compen}. 
Now take the nearby quadruple 
$$
\left( {\tilde B}_{1} = B_{1} + {\delta}B_{1}, {\tilde B}_{2} = B_{2} + {\delta}B_{2}, {\tilde I} = I+{\delta}I, {\tilde J} = J + {\delta}J \right)
$$ and act on it by the combination of the ${\bT}_{12}^{\BC}$ transformation $(e^{t{\ac}}, e^{t{\ept}_{12}})$: 
$$\left({\tilde B}_{1}, {\tilde B}_{2}, {\tilde I}, {\tilde J}\right) \mapsto  \left( \, q_{1} {\tilde B}_{1}, \, q_{2} {\tilde B}_{2}, \, {\tilde I}, \, q_{1}q_{2} {\tilde J}\right)$$
{\sl and} the $GL(k)$-transformation $g_{t}$: 
$$\left( \, {\tilde B}_{1}, \, {\tilde B}_{2}, \, {\tilde I}, \, {\tilde J}\right) \mapsto  \left( \, g_{t}^{-1} {\tilde B}_{1} g_{t}, \, g_{t}^{-1} {\tilde B}_{2} g_{t},\, g_{t}^{-1}  {\tilde I},  \, {\tilde J} g_{t} \right)\, ,
$$ 
defining the ${\bT}_{12}$-action on 
$({\delta}B_{1}, {\delta}B_{2} ,{\delta}I, {\delta}J)$: 
 \beq
 e^{t} \cdot \left[   {\delta}B_{1},   {\delta}B_{2},   {\delta}I,   {\delta}J \right] = 
 \left[ q_{1} g_{t}^{-1}    {\delta}B_{1}  g_{t} \, , \, q_{2} g_{t}^{-1}   {\delta}B_{2}  g_{t}\, , \, g_{t}^{-1}  {\delta}I   h_{t} , q_{1}q_{2} h_{t}^{-1}   {\delta}J   g_{t} \right] \, .
 \eeq
 So the space ${\CalT}^{1}_{f} {\iM}_{k}(n)$ of variations $({\delta}B_{1}, {\delta}B_{2}, {\delta}I, {\delta}J)$ is a ${\bT}_{12}$ representation, with the character:
 \beq
 {\Tr}_{{\CalT}^{1}_{f} {\iM}_{k}(n)} \, \left( h, q \right) = \left( q_{1} + q_{2} \right) K_{\chi} K_{\chi}^{*} + N_{\chi} K_{\chi}^{*} + N_{\chi}^{*}K_{\chi} q_{1}q_{2} 
 \eeq 
 The first two terms on the right hand side account for ${\delta}B_{1}, {\delta}B_{2}$, the third term corresponds to the ${\delta}I$ variations, and the last term accounts for the ${\delta}J$ variations. 
 
 Now, the tangent space $T_{f}{\iM}_{k}(n)$ is the degree $1$ cohomology $H^{1}{\CalT}_{f}{\iM}_{k}(n)$ of the complex \eqref{eq:tangcomplex}, which has no $H^0$ or $H^2$ cohomology (for ${\zeta}>0$). The character of  $T_{f}{\iM}_{k}(n)$ can be therefore computed by taking the alternating sum of the characters of ${\CalT}^{0}_{f} {\iM}_{k}(n)$, ${\CalT}^{1}_{f} {\iM}_{k}(n)$ and ${\CalT}^{2}_{f} {\iM}_{k}(n)$, giving:
 \beq
 {\Tr}_{T_{f}{\iM}_{k}(n)} (h, q) = N_{\chi}  K_{\chi}^{*} + q_{12} N_{\chi}^{*} K_{\chi}  
- p_{12} K_{\chi}K_{\chi}^{*}
 \eeq
  Thus, 
 \beq
 {\Tr}_{T_{f}{\iM}_{k}(n)} (h, q) = \sum_{{\alpha} ,  {\beta} \in [n]} e^{t({\ac}_{\alpha} - {\ac}_{\beta})} \ T_{\chi} ({\lambda}^{(\alpha)}, {\lambda}^{(\beta)}) 
 \label{eq:nkabchar}
 \eeq
 where 
 \beq
 T_{\chi}({\mu}, {\lambda})  = \sum_{(i,j) \in {\lambda}} q_{1}^{i - {\lambda}_{j}^{t}} q_{2}^{{\mu}_{i}+1-j} + \sum_{(i,j)\in {\mu}} q_{1}^{{\mu}_{j}^{t}+1-i} q_{2}^{j-{\lambda}_{i}}
 \label{eq:tablm}
 \eeq
 We see that, as long as there is no rational relation between ${\ec}_{1}$ and ${\ec}_{2}$, and ${\ac}_{\alpha} - {\ac}_{\beta} \notin {\ec}_{1} {\BZ} + {\ec}_{2}{\BZ}$ the weights which appear in the character of the tangent space are non-zero. In other words, the tangent space $T_{f}{\iM}_{k}(n)$ does not contain trivial representations of ${\bT}_{12}$, i.e. $f$ is an isolated fixed point.

\subsection{Smaller tori}
\label{sec:smtori}

Let $({\Gamma}^{\vee},d)$ be a pair consisting of a finite or a countable set ${\Gamma}^{\vee}$ (the meaning of the notation will become clear later), and a function $d: {\Gamma}^{\vee} \to {\BN}$, which we shall call the dimension. We assign to each ${\omega} \in {\Gamma}^{\vee}$ a vector space
\beq
{\CalR}_{\omega} = {\BC}^{d({\omega})}
\eeq
of  the corresponding dimension. 

Let  $\bn$ be a $d$-partition of $n$, 
\beq
 n = \sum_{{\omega} \in {\Gamma}^{\vee}} \, n_{\omega} d({\omega})\, , \qquad n_{\omega} \geq 0
\eeq
with only a finite number of $n_{\omega} > 0$. Let
\beq
{\ell}_{\bn} = \# \{ \, {\omega} \, | \, n_{\omega} > 0 \, \}
\eeq
We associate to $\bn$ a
decomposition
of $N$ into the direct sum of tensor products:
\beq
N = \bigoplus_{{\omega} \in {\Gamma}^{\vee}} \, N_{\omega} \otimes {\CalR}_{\omega}
\label{eq:ndeco}
\eeq
with $n_{\omega}$-dimensional complex vector spaces $N_{\omega}$.

Define, for the $d$-partition $\bn$  
and a pair  $(e_{1}, e_{2})$ of non-zero integers,
 the sub-torus  
\beq
{\rm T}_{{\bn}; {\be}} \approx {\bT}_{\bn} \times U(1)_{\be} \subset PU(N) \times {\mathrm{Spin}}(4)_{12}\, .
\label{eq:smalltorus}
\eeq 
Here
 $U(1)_{\be}$ is embedded into $U(1)_{12} \times U(1)_{12}^{\prime} \subset U(2)_{12} \subset {\mathrm{Spin}}(4)_{12}$ by
 \beq
 U(1)_{\be}\, : \, \ e^{{\ii}{\vartheta}} \mapsto \left( e^{\frac{\ii}{2} (e_{1}+e_{2}){\vartheta}}, \ e^{\frac{\ii}{2}(e_{1}-e_{2}){\vartheta}} \right) \, , 
 \eeq
 in other words, it acts on ${\BC}^{2}_{12}$ by:
 \beq
 \left( z_{1}, z_{2} \right) \mapsto \left( e^{{\ii}e_{1}{\vartheta}} z_{1}, e^{{\ii}e_{2}{\vartheta}} z_{2} \right) \ .
 \eeq
 The torus ${\bT}_{\bn} \subset {\bT}_{n}$ is defined to be a quotient of the product of the maximal tori of $U(n_{\omega})$ by the overall center $U(1)$: 
\beq
h = \bigoplus_{{\omega} \in {\Gamma}^{\vee}} \ h_{\omega}  \otimes {\bf 1}_{{\CalR}_{\omega}} \in U(N)\ ,
 \quad \left( h_{1} : \ldots : h_{{\ell}_{\bn}} \right) = \left( h_{1} u : \ldots : h_{{\ell}_{\bn}} u \right) \in {\bT}_{\bn}\, , 
\label{eq:htut}
\eeq
where $h_{\omega} h_{\omega}^{\dagger} = 1$, ${\omega} \in {\Gamma}^{\vee}$, $h_{\omega} = {\rm diag} \left( u_{{\omega}, 1} , \ldots , u_{{\omega}, n_{\omega}} \right)$, and  $| u |^{2} = 1$.

\subsection{Fixed points of smaller tori}

Let us start with $n=1$, so that ${\bT}_{12} = {\rm T}_{\be}$. 
The ${\rm T}_{\be}$-fixed points on ${\iM}_{k}(1) = {\rm Hilb}^{[k]}({\BC}^{2})$ are isolated for $e_{1}e_{2} < 0$ and non-isolated for $e_{1}e_{2} > 0$, as
we see from the $T_{\chi}({\lambda}, {\lambda})$ character \eqref{eq:tablm}. Indeed, as soon as the partition $\lambda$ has a box
${\square} = (i,j)$ whose arm plus one-to-leg, or leg plus one-to-arm ratio is equal to $e_{1}:e_{2}$, 
\beq
e_{1} (i - {\lambda}_{j}^{t}) + e_{2} ( {\lambda}_{i} + 1 - j)  = 0\, ,\qquad {\rm or}
\qquad e_{1} ({\lambda}_{j}^{t}+1 -i ) + e_{2} ( j - {\lambda}_{i} ) = 0
\eeq
then $T_{\lambda}{\iM}_{k}(1)$ contains trivial ${\rm T}_{\be}$-representations, i.e. $\lambda$ is not an isolated fixed point. Geometrically, 
the fixed points of the ${\rm T}_{\be}$-action
for $e_{1}e_{2} >0$ are the $(e_{1},e_{2})$-graded ideals ${\CalI}$ in ${\BC}[x,y]$, i.e. the ideals which are invariant under the ${\BC}^{\times}$-action:
\beq
(x,y) \to (t^{e_{1}}x, t^{e_{2}}y)
\label{eq:pqgrad}
\eeq 
For such an ideal ${\CalI}$ the quotient $K = {\BC}[x,y]/{\CalI}$ is also a graded vector space:
\beq
K = \bigoplus_{{\fs}=0}^{d_{K}} K_{\fs}
\eeq
For general $\Gamma^{\vee}$ and the general partition $\bn$ the ${\bT}_{{\bn}; {\be}}$-fixed point set is a finite union of finite product 
\beq
{\iM}_{k}(n)^{{\bT}_{{\bn}; {\be}}} = \bigcup\limits_{ \sum\limits_{{\omega} \in {\Gamma}^{\vee}, \alpha \in [n_{\omega}]} k_{\alpha, \omega} = k} \quad  \varprod_{{\omega} \in {\Gamma}^{\vee}} \ \varprod_{{\alpha} \in [n_{\omega}]}\ {\iM}_{k_{\alpha, \omega}} (d({\omega}))^{{\bT}_{\be}}
\label{eq:imknfixp}
\eeq 
of the ${\bT}_{\be}$-fixed point sets on the moduli spaces ${\iM}_{k'}(n')$. 
This is easy to show using the same methods as we employed so far.
It suffices then to analyze the structure of ${\iM}_{k}(n)^{{\bT}_{\be}}$
where the torus ${\bT}_{\be} \approx U(1)$ acts on the matrices 
$(B_{1}, B_{2}, I, J)$ via:
\beq
e^{{\ii}t} \cdot \left[ B_{1}, B_{2}, I, J \right] =  \left[ e^{{\ii}e_{1}t}B_{1}, e^{{\ii}e_{2}t} B_{2}, I, e^{{\ii}(e_{1}+e_{2})t} J \right]
\label{eq:pqbijn}
\eeq
As usual, the $GL(K)$-equivalence class of the quadruple $[B_{1}, B_{2}, I, J]$ is ${\bT}_{\be}$-invariant if for every $e^{t} \in {\BC}^{\times}$ there is an operator $g_{t} \in GL(K)$ which undoes \eqref{eq:pqbijn}, i.e. 
\beq
\begin{aligned} 
& e^{{\ii}e_{1}t} g_{t}^{-1} B_{1} g_{t} = B_{1}\, , \quad e^{{\ii}e_{2}t} g_{t}^{-1} B_{2} g_{t} = B_{2} \\
& g_{t}^{-1} I = I \, , \quad e^{{\ii}(e_{1}+e_{2})t} J g_{t} = J \\
\label{eq:tpqgt}
\end{aligned}
\eeq 
The correspondence $e^{t} \mapsto g_{t}$ splits $K$ as the sum of irreducible representations of ${\bT}_{\be}$ 
\beq
K = \bigoplus_{{\fs} \geq 0} \ K_{\fs} \otimes {\CalR}_{\fs}
\label{eq:ksplitks}
\eeq
with $K_{\fs}$ being the multiplicity space of the charge $\fs$ representation ${\CalR}_{\fs}$: $e^{t} \mapsto e^{t\fs}$. 
Let 
\beq
k_{\fs} = {\dim}K_{\fs} \ . 
\label{eq:mults}
\eeq  We have:
\beq
\sum_{\fs} k_{\fs}  = k 
\label{eq:smsk}
\eeq
The grade $0$ component is $1$-dimensional:
\beq
K_{0} = I(N) \, .
\eeq
The operators $B_{1}, B_{2}$ raise the grading by $e_{1}$ and $e_{2}$, respectively:
\beq
B_{a}: K_{\fs} \to K_{{\fs}+e_{a}}\, , \ a = 1,2 
\eeq
The complex ${\CalS}_{0}$ and its cohomology $P_{12}^{\pm}$ are also graded:
\beq
P_{12}^{\pm} = \bigoplus_{\fs} \ P_{12, \fs}^{\pm}\, . \eeq
The dimensions $k_{\fs}$ are constant throughout the connected component of the set of $(e_{1}, e_{2})$-homogeneous ideals. In fact, for $e_{1}=e_{2}$ the component is a smooth projective variety, \cite{iarrobino:1977}. See also \cite{Iarrobino:1997}, \cite{Loginov:2014}.  
 
\subsection{Compactness of the fixed point set}

The topology of the fixed point set ${\iM}_{k}(n)^{{\bT}_{12}}$ depends on the choice of the torus ${\bT}_{12}$. In other words, it depends on how non-generic the choice of $({\ba}, {\ept})$ is. 

If there is no rational relation between ${\ba}$ and ${\ec}_{1}, {\ec}_{2}$, 
more precisely, if for any ${\alpha}, {\beta} \in [n]$ and $p, q \in {\BZ}$
\beq
{\ac}_{\alpha} - {\ac}_{\beta} + {\ec}_{1} p + {\ec}_{2} q =  0\, \qquad \Longrightarrow \qquad {\alpha} = {\beta}\,  , \quad p = q = 0
\label{eq:noratrel}
\eeq
then the fixed points $f$ are isolated, $f \leftrightarrow \left( {\lambda}^{({\alpha})} \right)_{{\alpha} \in [n]}$. Their total number, for fixed $k$ is finite, therefore the set of fixed points is compact.

What if there is a rational relation between 
${\ac}_{\alpha} - {\ac}_{\beta}$ and ${\ec}_{1}, {\ec}_{2}$? That is for some non-trivial ${\alpha}, {\beta} \in [n]$ and $p,q \in {\BZ}$, 
\beq
{\ac}_{\alpha} - {\ac}_{\beta} + {\ec}_{1} p + {\ec}_{2} q = 0\, .
\label{eq:ratrel}
\eeq
We shall assume all the rest of the parameters ${\ac}_{\gamma}, {\ec}_{1}, {\ec}_{2}$ generic. In particular we assume both ${\ec}_{1}, {\ec}_{2}$ non-zero. 
There are three cases to consider:
\begin{enumerate}
\item{}
${\alpha}\neq {\beta}$ and $pq > 0$ ; 

\item{}
${\alpha} = {\beta}$ and $pq > 0$; 

\item{} 
$pq < 0$ and no restriction on ${\alpha}, {\beta}$ ;

\end{enumerate}
In the case $(1)$ the fixed locus is non-compact. It is parametrized by the value of the invariant 
\beq
J_{\beta} B_{1}^{p-1} B_{2}^{q-1} I_{\alpha} 
\label{eq:invf}
\eeq
We therefore must make sure, in what follows, that the eigenvalues $({\ac}_{\alpha})_{{\alpha} \in [n]}$ of the infinitesimal framing rotations and the parameters $({\ec}_{1}, {\ec}_{2})$ of the spatial rotation do not land on the hyperplanes:
\beq
{\ac}_{\alpha} - {\ac}_{\beta} + {\ec}_{1} p + {\ec}_{2} q \neq 0, 
\label{eq:nowall}
\eeq
for all ${\alpha} \neq {\beta}$, and integer $p, q \geq 1$.

In the case $(2)$ the fixed points corresponding to the monomial ideals are isolated, since the weights in \eqref{eq:tablm}  have the form ${\ec}_{1} p' + {\ec}_{2} q'$ with $p'q' \leq 0$. We shall show below that 

\bigskip
\centerline{\boxit{\centerline{}
\centerline{
the $U(1)$-fixed points in the case $pq > 0$ correspond to the monomial ideals.} 
\centerline{In other words they are $U(1) \times U(1)$-invariant}}}

For fixed $k$ the sizes of the Young diagrams ${\lambda}^{({\alpha})}$
 are bounded above, since
 \beq
 \sum_{{\alpha}=1}^{n}\ \vert {\lambda}^{({\alpha})} \vert  \ = \ k
 \label{eq:instcha}
 \eeq
 Since the number of collections of Young diagrams which 
 obey \eqref{eq:instcha} is finite, the set of points fixed by the 
 action of the maximal torus ${\bf T}$ is compact. This set, as we just showed, is in one-to-one correspondence with the collections
 \beq
 f \leftrightarrow \left( {\lambda}^{({\alpha})} \right)_{{\alpha}\in [n]}
 \label{eq:fpts}
 \eeq
 obeying \eqref{eq:instcha}.

 In the case $(3)$ the fixed points are not isolated, but the fixed point set is compact. 
Let us show the ${\bT}_{{\bn}; {\be}}$-fixed point set is compact. There are two cases:

\begin{enumerate}

\item{} $e_{1}e_{2} > 0$. In this case the minimal torus corresponds to 
 ${\Gamma}^{\vee} = \{ 0 \}$, $d(0) = n$, $n_{0} = 1$, i.e. for ${\bT}_{\bn}  = 1$, ${\bT}_{{\bn}; {\be}} = {\bT}_{\be}$. 
The corresponding Coulomb parameter vanishes, ${\ac}=0$. 

We are going to demonstrate that for all ${\bT}_{{\bn}; {\be}}$-fixed points on ${\iM}_{k}(n)$ the $L^2$-norm of  $(B_{1}, B_{2}, I, J)$ is bounded above by a constant which depends only on $n$, $k$, and $\zeta$. 
We use the real moment map equation \eqref{eq:mur}:
\beq
\begin{aligned}
& k \zeta = {\Tr}_{K} \left( {\mu} \right) = \Vert I \Vert^{2} - \Vert J \Vert^{2} \\
& \zeta {\Tr}_{K} {\sigma} = {\Tr}_{K} \left( {\sigma}{\mu} \right) = 
e_{1} \Vert B_{1} \Vert^{2} + 
 e_{2}  \Vert B_{2}  \Vert^{2} + (e_{1}+e_{2}) \Vert J \Vert^{2} \\
 \end{aligned}
 \label{eq:momsig}
 \eeq
 where we used the Eqs.  \eqref{eq:compen} with the specialization 
 ${\ec}_{1} = e_{1}, \ {\ec}_{2} = e_{2}$:
 \beq
 \begin{aligned}
 \label{eq:infinitfix}
 & e_{a}   B_{a} = [ {\sigma} , B_{a} ]\, , \quad 
 e_{a}  B_{a}^{\dagger} =  [ B_{a}^{\dagger}, {\sigma} ] \, , \quad
  a = 1, 2 \,  , \\
&  0 = {\sigma} I\, , \quad  0 = I^{\dagger} {\sigma} \, , \quad
 ( e_{1}+e_{2}  ) J =   - J  {\sigma} \, , \quad J^{\dagger}  ( e_{1}+e_{2} ) =  - {\sigma} J^{\dagger}\, , \\
\end{aligned}
\eeq
The Eqs. \eqref{eq:infinitfix} imply, by the same arguments as before, that
the spectrum of $\sigma$ has the form:
\beq
{\fs} = e_{1} (i-1) +  e_{2} (j-1), \qquad (i,j) \in {\Sigma}
\label{eq:ijeig}
\eeq
for a finite set ${\Sigma}$ of pairs $(i,j)$ of positive integers, and that
$J$ maps the eigenvectors of ${\sigma}$ in $K$ to zero, unless
the eigenvalue is equal to $-e_{1}-e_{2}$. Now, the eigenvalues of $\sigma$ are of the form \eqref{eq:ijeig}, which are never equal to $-e_{1}-e_{2}$. Thus, $J \vert_{K} = 0$,  therefore $B_1$ and $B_2$ commute on $K$. 

Now,
${\sigma} \vert_{I(N)} = 0$, i.e. $m_{0} = {\rm dim}\left( {\rm im}I \right) \leq n$. Now, the vector spaces
\beq
K_{i,j} ={\BC}\cdot B_{1}^{i-1} B_{2}^{j-1}\, I(N) 
\eeq
if non-zero, contribute ${\rm dim}K_{i,j} \leq n$ to $k_{\fs}$ with ${\fs} = e_{1} (i-1) +  e_{2} (j-1)$. It is clear that
\begin{multline}
k_{\fs} = {\rm dim} \sum_{e_{1} (i-1) +  e_{2} (j-1) = {\fs}} \, {\BC}\left( B_{1}^{i-1} B_{2}^{j-1} I(N) \right) \qquad\leq \\  \leq \sum_{e_{1} (i-1) +  e_{2} (j-1) = {\fs}} {\rm dim}\, {\BC}\left( B_{1}^{i-1} B_{2}^{j-1} I(N) \right) \ \leq \ n \ {\rm Coeff}_{t^{\fs}} \frac{(1-t^{ke_{1}})(1-t^{ke_{2}})}{(1-t^{e_{1}})(1-t^{e_{2}})}\, ,
\end{multline}
since both $i$ and $j$ cannot be greater then $k$. 
The trace ${\Tr}_{K}{\sigma}$ can be estimated by
 \begin{multline}
 {\Tr}_{K}\, {\sigma} = 
\sum_{{\fs} \in {\Sigma}} \,  {\fs}\, k_{\fs}\  \leq  \\
\ n\, \sum_{{\fs}=0}^{\infty} \ {\fs}\ {\rm Coeff}_{t^{\fs}} \frac{(1-t^{ke_{1}})(1-t^{ke_{2}})}{(1-t^{e_{1}})(1-t^{e_{2}})} = n\, t \frac{d}{dt} \Biggr\vert_{t=1} \frac{(1-t^{ke_{1}})(1-t^{ke_{2}})}{(1-t^{e_{1}})(1-t^{e_{2}})} = \\
 = \frac{1}{2} k^{2}(k-1) n (e_{1}+e_{2})
\end{multline}
Thus, $J = 0$, the norms $\Vert B_{1,2} \Vert^{2}$ of the operators $B_{1,2}$ are bounded above, while the norm
of the operator $I$ is fixed: 
\beq
\Vert I \Vert^2 = {\zeta} k \ , \quad \frac{e_{1}}{e_{1}+e_{2}} \Vert B_{1} \Vert^{2} + \frac{e_{2}}{e_{1}+e_{2}} \Vert B_{2} \Vert^{2} \ \leq \ \frac{\zeta}{2} k^{2}(k-1) n 
\label{eq:b1b2inorm}
\eeq

\item{} $e_{1}e_{2} < 0$. In this case we take ${\Gamma}^{\vee} = [n]$, $d({\omega}) =1$ for all ${\omega} \in [n]$. The Coulomb parameters are the generic $n$ complex numbers ${\ac}_{\alpha} \in {\BC}$, ${\alpha} \in [n]$, defined up to an overall shift. Below we further restrict the parameters ${\ac}_{\alpha}$ to be real, so that they belong to the Lie algebra of the 
compact torus ${\bT}_{{\bn}; {\be}}$. The equations \eqref{eq:infinitfix} generalize to:
 \beq
 \begin{aligned}
 \label{eq:infinitfixii}
 & e_{a}   B_{a} = [ {\sigma} , B_{a} ]\, , \quad 
 e_{a}  B_{a}^{\dagger} =  [ B_{a}^{\dagger}, {\sigma} ] \, , \quad
  a = 1, 2 \,  , \\
&  I {\ac} = {\sigma} I\, , \quad  {\ac} I ^{\dagger} = I^{\dagger}{\sigma} \, , \quad
 ( e_{1}+e_{2}  - {\ac}) J =   - J  {\sigma} \, , \quad J^{\dagger}  ( e_{1}+e_{2}   - {\ac}) =  - {\sigma} J^{\dagger}\, , \\
\end{aligned}
\eeq
 The fixed point set ${\iM}_{k}(n)^{{\bT}_{{\bn}; {\be}}}$ splits:
\beq
{\iM}_{k}(n)^{{\bT}_{{\bn}; {\be}}} = \bigcup_{k_{1} + \ldots + k_{n} = k} \ \varprod_{{\alpha} \in [n]} \ {\iM}_{k_{\alpha}}(1)^{{\bT}_{\be}}
\label{eq:factpart}
\eeq
The fixed points are isolated, these are our friends $({\lambda}^{({\alpha})})_{{\alpha} \in [n]}$, the $n$-tuples of partitions with the total size equal to $k$. Since it is a finite set, it is compact. 

Note that we couldn't restrict the torus ${\bT}_{{\bn}; {\be}}$ any further in this case. Indeed, the crucial ingredient in arriving at \eqref{eq:factpart} is vanishing of the $J$ matrix for the ${\bT}_{{\bn}; {\be}}$-invariant solutions of the ADHM equations. The argument below the Eq. \eqref{eq:ijeig} we used before would not work for $e_{1}e_{2} < 0$, since ${\ac}_{\alpha}-(e_{1}+e_{2})$ may be equal to ${\ac}_{\beta} + e_{1}(i-1) + e_{2}(j-1)$ for some ${\alpha}, {\beta} \in [n]$, $i,j \geq 1$. In this case $J$ may have a non-trivial matrix element, giving rise to a non-compact fixed locus. Now, insisting on the ${\bT}_{{\bn}; {\be}}$-invariance with ${\bT}_{\bn} = U(1)^{n-1}$ means 
${\ac}_{\alpha}$'s in \eqref{eq:infinitfixii} are completely generic, in particular, for
${\alpha} \neq {\beta}$, ${\ac}_{\alpha} - {\ac}_{\beta} \notin {\BZ}$. This still leaves the case ${\alpha} = {\beta}$ as a potential source of noncompactness. But this is the case of the ${\bT}_{\be}$-action on ${\iM}_{k}(1) = {\rm Hilb}^{[k]}({\BC}^{2})$. In this case $J$ vanishes not because of the toric symmetry, but rather because of the stability condition \cite{Nakajima:1999}: 
\begin{multline}
JI = {\Tr} IJ = {\Tr} [B_{2}, B_{1} ] = 0, \\
 J (x B_{1} + y B_{2})^l I = {\Tr}  (x B_{1} + y B_{2})^l [ B_{2}, B_{1} ] = \qquad\qquad\qquad\qquad \\
 \qquad\qquad\qquad\qquad {\Tr}  (x B_{1} + y B_{2})^l [ x B_{1} + y B_{2}, x' B_{1} + y' B_{2} ] = 0 , \\
 \qquad\qquad\qquad\qquad\qquad\qquad {\rm for\ any}\ x, y, x', y', \ {\rm s.t.}\ x' y - x y' = 1\,  ,  \\
J f(B_{1}, B_{2})  B_{1} B_{2} g(B_{1}, B_{2}) I = J f(B_{1}, B_{2})  B_{2} B_{1} g(B_{1}, B_{2}) I + \\
+ \left( J f(B_{1}, B_{2}) I  \right) \left( J g(B_{1}, B_{2}) I  \right) = 0\, , 
\qquad {\rm by\ induction} \\
\Longrightarrow J {\BC}[B_{1}, B_{2}] I = 0 \Longrightarrow J = 0, \quad {\rm by\ stability}
\end{multline}

\end{enumerate}

{}The compactness of ${\iM}_{k}(n)^{{\bT}_{{\bn}; {\be}}}$ is thus established.

\subsection{Ordinary instantons as the fixed set}

Let us now consider the particular ${\bT}_{\xt} = U(1)^{5}$ symmetry of the spiked 
instanton equations, 
\beq
\left( \, I_{A} \, , \, J_{A} \, \right) \
\longrightarrow  \
\left( \, e^{{\ii}{\vartheta}_{A}} I_{A} \, , \, e^{-{\ii}{\vartheta}_{A}}J_{A} \, \right) 
\label{eq:pu5}
\eeq
where ${\vec\vartheta} = ({\vartheta}_{1}, {\vartheta}_{2}, {\vartheta}_{3},
{\vartheta}_{4}, {\vartheta}_{5}, {\vartheta}_{6}) \sim ({\vartheta}_{1}+ {\vartheta}, {\vartheta}_{2}+ {\vartheta}, {\vartheta}_{3}+ {\vartheta},
{\vartheta}_{4}+ {\vartheta}, {\vartheta}_{5}+ {\vartheta}, {\vartheta}_{6}+ {\vartheta})$
for any $\vartheta$.   The ${\bT}_{\xt}$-invariant configuration
$[{\bB}, {\bI}, {\bJ}]$ defines a homomorphism of the covering torus ${\tilde\bT}_{\xt} \approx U(1)^6\longrightarrow U(k)$, 
via the compensating $U(k)$-transformation $g({\vec\vartheta})$
obeying:
\beq
e^{{\ii}{\vartheta}_{A}} I_{A} = g({\vec\vartheta}) I_{A} \, , \ e^{-{\ii}{\vartheta}_{A}} J_{A} =  J_{A} g({\vec\vartheta})^{-1} \, , \
g({\vec\vartheta}) B_{a} g({\vec\vartheta})^{-1} = B_{a}
\label{eq:ijbinv}
\eeq
The space
$K$ splits as the orthogonal direct sum
\beq
K = \bigoplus_{A \in {\6}} \, K_{A} \, , \qquad g({\vec\vartheta}) \vert_{K_{A}} = e^{{\ii}{\vartheta}_{A}} \, , \qquad K_{A} = {\BC} [ B_{a}, B_{b} ] I_{A}(N_{A})\ , \qquad {\rm for}\ A = \{ a, b \}
\label{eq:kasp}
\eeq
This decomposition is preserved by the matrices ${\bB}, {\bI}, {\bJ}$. Thus the solution is the 
direct sum of the solutions of ADHM equations:
\beq
{\mM}_{k}^{*}({\vec n})^{{\bT}_{\xt}} = \bigcup\limits_{\sum_{A \in {\6}} k_{A} = k}\
\varprod\limits_{A \in {\6}} \ {\iM}_{k_{A}}(n_{A})
\label{eq:xprod}
\eeq
 
\section{Crossed and folded instantons}

\bigskip
\hfill\vbox{
\hbox{\tiny Distorted shadows fell}
\hbox{\tiny Upon the lighted ceiling:}
 \hbox{\tiny Shadows of crossed arms, of crossed legs-}
 \hbox{\tiny Of crossed destiny.$^{\dagger}$}}\let\thefootnote\relax\footnote{\hbox{$^{\dagger}$}
 \hbox{$\scriptstyle{Winter\ night,\ from\ ``Dr. Zhivago'',}$}
 \hbox{$\scriptstyle{B.Pasternak,\ English\ translation\ by\ Bernard\ G. Guerney}$}}
\bigskip

The next special case is where only two e.g. $N_{A'}$ and $N_{A''}$ out of six vector spaces $N_{A}$ are non-zero. There are two basic cases.

\subsection{Crossed instantons} 

Suppose $A' \cup A'' = {\emptyset}$,  e.g.  $A' = \{ 1, 2\}$ and $A'' = \{ 3, 4\}$. In this case we define
\beq
{\mM}^{+}_{k} (n, w) = {\mM}^{0}_{k}(n_{12} = n, 0, 0, 0, 0, n_{34} = w)
\label{eq:cross}
\eeq
We call the space ${\mM}^{+}_{k} (n, w)$ the space of $(n,w)$-{\emph{crossed instantons}}.

The virtual dimension of the space ${\mM}^{+}_{k} (n, w)$ is
independent of $k$, it is equal to  $- 2nw$.  As a set, ${\mM}^{+}_{k}(n,w)$ is stratified
\beq
{\mM}^{+}_{k}(n,w) = \bigcup_{k' + k'' \geq k} \,  
{\mM}^{+}_{k',  k''; k}(n,w) \ .
\label{eq:mmstrat}
\eeq
The stratum 
\beq
{\mM}^{+}_{k',  k''; k}(n,w) = \Biggl\{ \ [ {\bB}, {\bI}, {\bJ} ] \ | \ {\rm dim}K_{12} = k', \ {\rm dim}K_{34} = k'' \ \Biggr\}
\eeq
parametrizes the crossed instantons, whose ordinary instanton components have the charges $k'$ and $k''$, respectively: the crossed instanton $ [ {\bB} , {\bI}, {\bJ} ] $ defines two ordinary instantons, $[ B_{1}, B_{2}, I_{12}, J_{12} ]$ on ${\BC}^{2}_{12}$ and
 $[ B_{3}, B_{4}, I_{34}, J_{34} ]$ on ${\BC}^{2}_{34}$, of the charges 
 \beq
 k' = {\dim}K_{12}\, ,  \quad k'' = {\dim} K_{34}
 \eeq
 
 \subsection{One-instanton crossed example}

When $k=1$ the matrices $B_{a}$ are just complex numbers $b_{a} \in {\BC}$, $a \in {\4}$. The equations $b_{1} I_{34} = b_{2} I_{34} = 0$
and $b_{3} I_{12} = b_{4} I_{12} = 0$ imply that if $(b_{1}, b_{2} ) \neq 0$, then $I_{34} = 0$, $I_{12} \neq 0$, $(b_{3}, b_{4}) = (0,0)$, $K = K_{12}$ and the rest of the matrices define the ordinary charge $1$ $U(n)$ instanton, parametrized by the space 
\eqref{eq:1inst}. Likewise, if $(b_{3}, b_{4}) \neq 0$, then $I_{12} = 0$, 
$I_{34} \neq 0$, $(b_{1}, b_{2}) = (0,0)$, $K = K_{34}$ and the rest of the matrices define the ordinary charge $1$ $U(w)$ instanton. Finally, if $(b_{1}, b_{2}, b_{3}, b_{4}) = 0$, then both $I_{12}, I_{34}$ need not vanish. If, indeed, both $I_{12}, I_{34}$ do not vanish, then $J_{12}$ and $J_{34}$ vanish, by the ${\upsilon}$-equaitons, while $I_{12}, I_{34}$ obey
\beq
\Vert I_{12} \Vert^2 + \Vert I_{34} \Vert^2 = {\zeta} \, ,
\eeq
which, modulo $U(1) = U(K)$ symmetry, define a subset in ${\BC\BP}^{n+w-1}$, the complement to the pair of ``linked'' projective spaces, ${\BC\BP}^{n-1}$ and ${\BC\BP}^{w-1}$, corresponding to the vanishing of $I_{34}$ and $I_{12}$, respectively. The result is, then
\beq
{\mM}^{+}_{1}(n, w) = \ {\BC}^{2} \times T^{*}{\BC\BP}^{n-1} \ \cup \ {\BC\BP}^{n+w-1} \ \cup \ {\BC}^{2} \times T^{*}{\BC\BP}^{w-1} \, , 
\label{eq:1instcr}
\eeq
the first and second components intersect along $(0,0) \times {\BC\BP}^{w-1}$
the second and the third components intersect along $(0,0) \times {\BC\BP}^{n-1}$, where ${\BC\BP}^{n-1} \cup {\BC\BP}^{w-1} \subset {\BC\BP}^{n+w-1}$
are non-intersecting ${\BC\BP}^{n-1} \cap {\BC\BP}^{w-1} = {\emptyset}$ projective subspaces. 

\begin{figure}[H]
\picit{8}{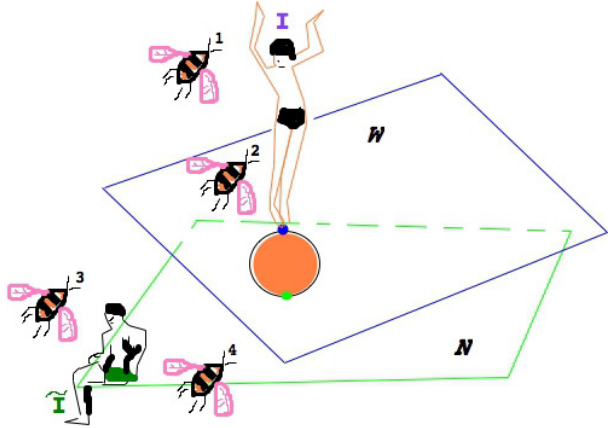}
\caption{Charge one crossed instanton moduli space: the planes represent the complex plane factors ${\BC}^{2}_{12}$ and ${\BC}^{2}_{34}$, the girl represents the $T^{*}{\BC\BP}^{w-1}$ factor, the man represents the $T^{*}{\BC\BP}^{n-1}$ factor, the orange ball is the ${\BC\BP}^{n+w-1}$ component, the blue and green dots are the ${\BC\BP}^{n-1}$ and ${\BC\BP}^{w-1}$ loci of the intersections of components}
\end{figure}

\subsection{Folded instantons}

In this case $A' \cap A'' = \{ a \}$, e.g. $A' = \{ 1, 2\}$, $A'' = \{ 1, 3 \}$, $a = 1$. 

We define:
\beq
{\mM}^{\fo}_{k} (n, w) = {\mM}^{k}_{k}(n_{12} = n, n_{13} = w, 0, 0, 0, 0)
\label{eq:fold}
\eeq
We call the space ${\mM}^{\fo}_{k} (n, w)$ the space of $(n,w)$-{\emph{folded instantons}}.

There exists an analogue of the stratification \eqref{eq:mmstrat} for the
folded instantons. Again, the folded instanton data
$ \left[ B_{1}, B_{2} , B_{3}, B_{4} , I_{12}, I_{13}, J_{12}, J_{13} \right]$
defines two ordinary noncommutative instantons on ${\BR}^{4}$, one on
${\BC}^{2}_{12}$, $\left[ B_{1}, B_{2}, I_{12}, J_{12} \right]$, another on ${\BC}^{2}_{13}$, $\left[ B_{1}, B_{3}, I_{13}, J_{13} \right]$. The stability implies that $B_{4}$ vanishes. The spaces
$K_{12} = {\BC}[B_{1}, B_{2}] I_{12}(N_{12})$ and $K_{13} = {\BC} [ B_{1}, B_{3} ] I_{13} (N_{13})$ generate all of $K$, 
\beq
K = K_{12} + K_{13}
\eeq

\subsection{One-instanton folded example}

When $k=1$, as before, the matrices $B_{a}$ are the complex numbers $b_{a}$, $a \in {\4}$, except that $b_{4}$ vanishes. Now, the equation
$b_{2} I_{13} = 0$ implies that if $b_{2} \neq 0$ then $I_{13} = 0$, and we have the charge one ordinary $U(n)$ instanton on ${\BC}^{2}_{12}$. Likewise the equation $b_{3} I_{12} = 0$ implies that if $b_{3} \neq 0$ then $I_{12} =0$ and we have the charge one ordinary $U(w)$ instaton
on ${\BC}^{2}_{13}$. Finally, when both $b_{2} = b_{3} = 0$, 
the remaining equations $J_{12} I_{13} = J_{13} I_{12} = J_{12} I_{12} = J_{13} I_{13} = 0$, and
$\Vert I_{12} \Vert^2 + \Vert I_{13} \Vert^2 - \Vert J_{12} \Vert^2 - \Vert J_{13} \Vert^2 = {\zeta}$, define the variety which is a product of a copy of ${\BC}^{1}$ (parametrized by $b_{1}$) and our friend the union of three pieces: 
${\BC\BP}^{n+w-1}$ (this is the locus where $J_{12} = J_{13} =0$), $T^{*}{\BC\BP}^{n-1}$ (the locus where $I_{13} = J_{13} = 0$) and $T^{*} {\BC\BP}^{w-1}$ (the locus where $I_{12} = J_{12} = 0$):
\beq
{\mM}^{\fo}_{1}(n,w) = {\BC}^{2} \times T^{*}{\BC\BP}^{w-1} \ \cup \ 
{\BC}^{1} \times {\BC\BP}^{n+w-1}  \ \cup {\BC}^{2} \times T^{*}{\BC\BP}^{n-1} \, , 
\label{eq:1instfold}
\eeq
the first and second components intersect along ${\BC}^{1} \times {\BC\BP}^{w-1}$
the second and the third components intersect along ${\BC}^{1} \times {\BC\BP}^{n-1}$, where ${\BC\BP}^{n-1} \cup {\BC\BP}^{w-1} \subset {\BC\BP}^{n+w-1}$
are non-intersecting ${\BC\BP}^{n-1} \cap {\BC\BP}^{w-1} = {\emptyset}$ projective subspaces. 

\subsection{Fixed point sets: butterflies and zippers}

Let us now discuss the fixed point sets of toric symmetries of the crossed and folded instantons. The torus ${\bT}_{n,w} \times U(1)^{3}_{\ec}$
acts on ${\mM}^{+}_{k}(n,w)$ and ${\mM}^{\fo}_{k}(n,w)$:
\beq
\left( B_{a}, I_{A}, J_{A} \right) \mapsto \left( e^{{\ii}t{\ec}_{a}} B_{a}, I_{A} e^{-{\ii}t {\ac}_{A}} , e^{{\ii}t {\ec}_{A}} e^{{\ii}t{\ac}_{A}} J_{A} \right)
\eeq
Here ${\ac}_{A} = {\rm diag} \left( {\ac}_{A, 1}, \ldots , {\ac}_{A, n_{A}} \right)$ where the complex numbers ${\ac}_{A, \alpha}$, ${\alpha} \in [n_{A}]$ are defined up to the overall shift
\beq
{\ac}_{A, \alpha} \sim {\ac}_{A, \alpha} + y \, , 
 \label{eq:yshift}
\eeq
with $y \in {\BC}$. Let ${\ept} = ({\ec}_{1}, {\ec}_{2}, {\ec}_{3}, {\ec}_{4})$, 
\beq
\sum_{a \in {\4}} {\ec}_{a} = 0
\eeq
We assume ${\ec}_{a} \neq 0$ for each $a \in \4$.

\subsubsection{Toric crossed instantons}
 
The fixed point set ${\mM}^{+}_{k}(n,w)^{{\bT}_{n,w} \times U(1)^{3}_{\ec}}$ is easy to describe. The infinitesimal transformation
generated by $({\ac},  {\ept})$ is compensated by the infinitesimal
$GL(k)$ transformation, generated by ${\sigma} \in {\rm End}(K)$. As in the previous section this makes $K$ a representation of ${\bT}_{n,w} \times U(1)^{3}_{\ec}$. 
The space $K$ contains two subspaces, $K_{12}$ and $K_{34}$, whose intersection $K_{12,34} = K_{12} \cap K_{34}$ 
belongs to both $P_{12}$ and $P_{34}$:
\beq
K_{12,34} \subset  P_{12} \cap P_{34} 
\eeq
\bigskip

\begin{figure}[H]
\picit{8}{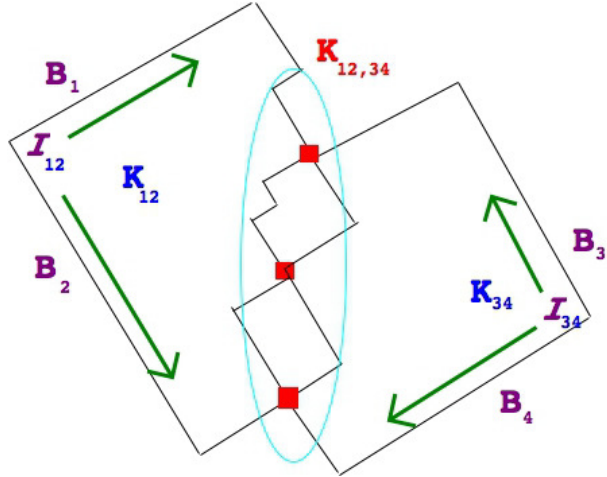}
\caption{The butterfly}
\label{fig:K12,34}
\end{figure}

\bigskip
\noindent

{}The eigenvalues of ${\sigma}$ on $K_{12}$ have the form:
\beq
{\rm Eigen} \left( {\sigma} \vert_{K_{12}} \right) =
\left\{ \, {\ac}_{12,\alpha} + {\ec}_{1} (i-1) + {\ec}_{2}(j-1) \ \Biggl\vert \ {\alpha} \in [n] \, , 
\, (i,j) \in {\lambda}^{(12,{\alpha})} \, \right\}
\label{eq:set1}
\eeq
The eigenvalues of ${\sigma}$ on $K_{34}$ have the form:
\beq
{\rm Eigen} \left( {\sigma} \vert_{K_{34}} \right) =
\left\{ \,  {\ac}_{34,\beta} + {\ec}_{3} (i-1) + {\ec}_{4}(j-1) \ \Biggl\vert \  {\beta} \in [w] \, , 
\, (i,j) \in {\lambda}^{(34,{\beta})} \, \right\}
\label{eq:set2}
\eeq
These two sets do not overlap 
 when all the parameters ${\ac}_{A, \alpha}, {\ec}_{a}$
are generic. Therefore $K_{12} \perp K_{34}$ and  the ${\bT}_{n,w} \times U(1)^{3}_{\ec}$-fixed points are isolated. These fixed points are, therefore, in one-to-one correspondence with the pairs
\beq
( {\lambda}^{12} ; {\lambda}^{34} )
\eeq
consisting of 
$n$- and $w$-tuples  
\[ {\lambda}^{12} = \left( {\lambda}^{(12,1)},  \ldots , {\lambda}^{(12,n)} \right) \, ;\qquad {\lambda}^{34} = \left(  {\lambda}^{(34,1)} , \ldots , {\lambda}^{(34, w)} \right)\ , \] of partitions, obeying
\beq
\sum_{\alpha \in [n] } \ \biggl\vert {\lambda}^{(12,{\alpha})} \biggr\vert \ + \ 
\sum_{\beta \in [w] }\ \biggl\vert {\lambda}^{(34,{\alpha})} \biggr\vert\ =\ k
\eeq
Their number is finite, therefore the set ${\mM}_{k}^{+}(n)^{{\bT}_{n,w} \times U(1)^{3}_{\ec}}$ of fixed points is compact. 

Now let us try to choose a sub-torus ${\bT}^{\prime} \subset {\bT}_{n,w} \times U(1)^{3}_{\ec}$, restricted only by the condition that
$J_{12} = J_{34} = 0$ for the ${\bT}^{\prime} $-invariant solutions of \eqref{eq:bbij}. We wish to prove that the set of ${\bT}^{\prime} $-fixed points is compact in this case as well. In the next sections we shall describe such tori in more detail.

We start by the observation that if $K_{12, 34} \neq 0$ then the two sets \eqref{eq:set1} and \eqref{eq:set2} of ${\sigma}$-eigenvalues must overlap. 
Therefore, for some $({\alpha} , {\beta} ) \in [n] \times [w]$,
and for some $(i',j') \in {\lambda}^{(12,{\alpha})}$, 
$(i'', j'') \in {\lambda}^{(34,{\beta})}$
\beq
{\ac}_{12,\alpha} + {\ec}_{1} (i'-1) + {\ec}_{2}(j'-1) = {\ac}_{34,\beta} + {\ec}_{3} (i''-1) + {\ec}_{4}(j''-1)
\label{eq:abij}
\eeq
\begin{figure}
\centerline{\picit{8}{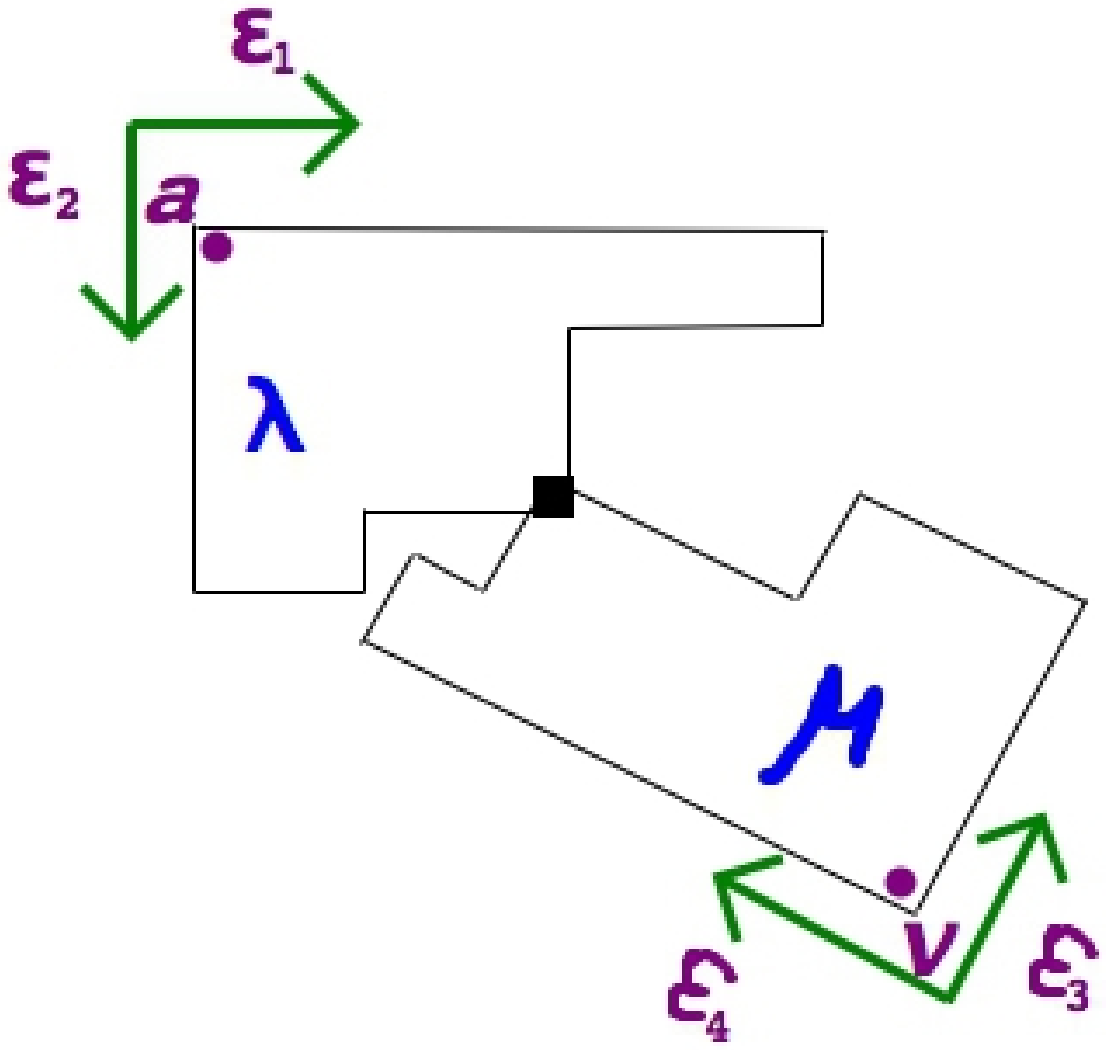}}
\label{fig:lamu2int}
\caption{${\blacksquare} = K_{12,34}$, ${\ac} = {\ac}_{12,\alpha},\ {\nu} = {\ac}_{34, \beta},\ {\lambda} = {\lambda}^{(12, {\alpha})},\ {\mu} = {\lambda}^{(34, {\beta})}$}
\end{figure}
Note that \eqref{eq:abij} is invariant under the shifts \eqref{eq:yshift}. Moreover, if  (cf. \eqref{eq:zlattice}) 
\beq
\begin{aligned}
& {\ac}_{12, \alpha'} - {\ac}_{12, \alpha''} \notin  Z_{\ept} \, , \qquad {\alpha}' \neq {\alpha}'' \\
& {\ac}_{34, \beta'} - {\ac}_{34, \beta''} \notin Z_{\ept} \, , \qquad {\beta}' \neq {\beta}'' \\
\end{aligned}
\label{eq:aabbij}
\eeq
and $0 \notin Z_{\ept}$, then
the condition \eqref{eq:abij} determines $({\alpha}, {\beta})$ and $i'j'$ and $i''j''$ uniquely, up to the shifts $(i',j', i'', j'') \mapsto (i'+k, j'+k, i''-k, j''-k)$, $k \in {\BZ}$.  The relation \eqref{eq:abij} defines the codimension $1$ subtorus ${\bT}^{\prime} \subset {\bT}_{n,w} \times U(1)^{3}_{\ec}$. Let us describe its fixed locus. 

If the condition \eqref{eq:abij} on ${\ac}, {\ept}$ is obeyed, it does not imply that $K_{12,34} \neq 0$. However, if in addition to \eqref{eq:abij} also the condition \eqref{eq:aabbij} is obeyed, then  the intersection $K_{12,34}$ is not more then one-dimensional. Let $H_{12} = P_{12; {\alpha}; i'j'} \subset P_{12}$, $H_{34} = P_{34; {\beta}; i''j''} \subset P_{34}$ be the one-dimensional eigenspaces of ${\sigma}$ corresponding to the eigenvalue \eqref{eq:abij}. If an eigenbasis of $N_{12}$ for ${\ac}_{12}$  and the eigenbasis of $N_{34}$ for ${\ac}_{34}$ are chosen, then $H_{12}$
and $H_{34}$ are endowed with the basis vectors as well (act on the eigenvector of ${\ac}_{12}$ corresponding to ${\ac}_{12, \alpha}$ by $B_{1}^{i'-1}B_{2}^{j'-1}I_{12}$ to get the basis vector of $H_{12}$). 

\medskip   

\begin{figure}[H] 
\centerline{\picit{3.5}{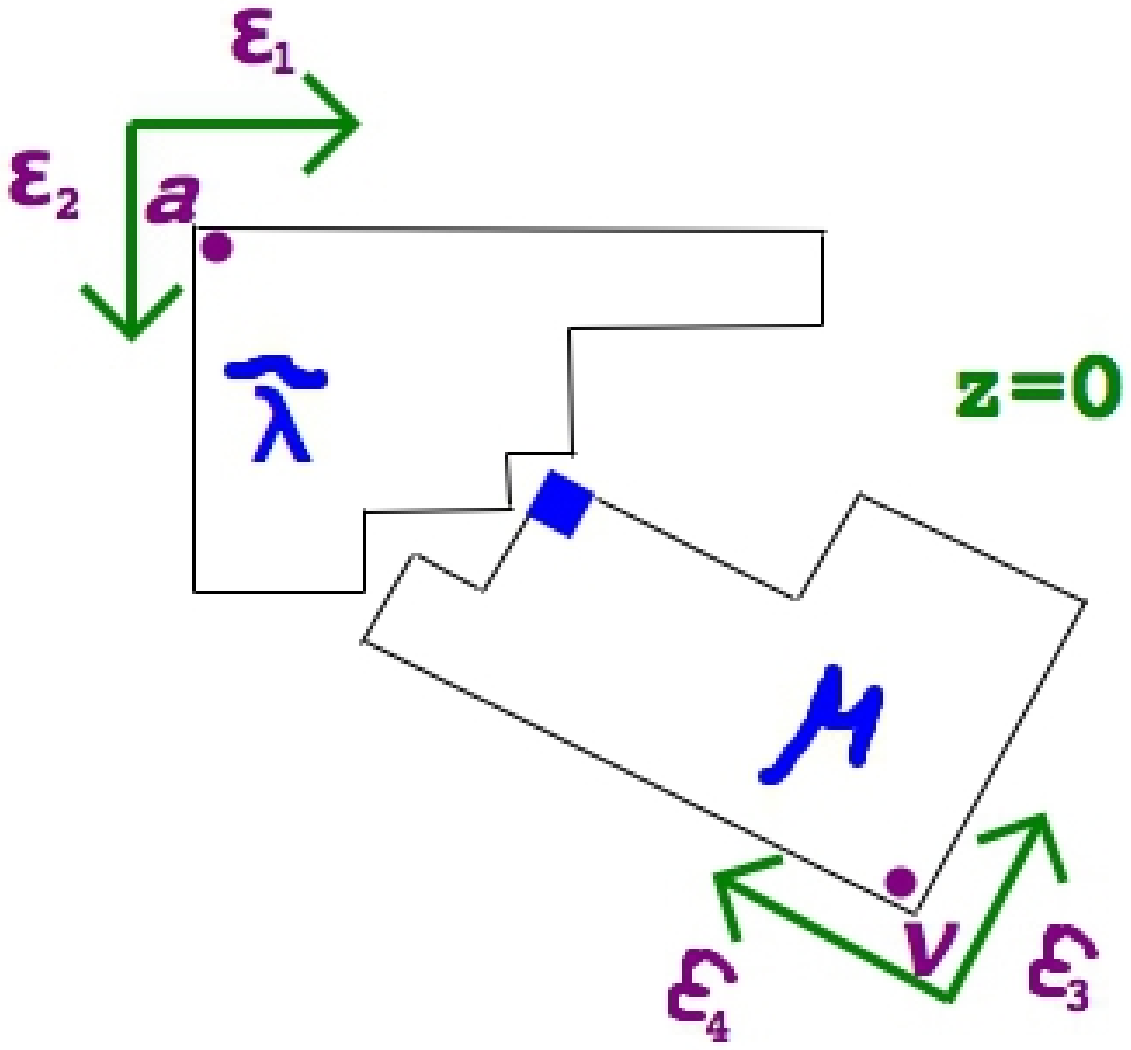}\quad \vbox{\hbox{The component of }
\hbox{the fixed point set}
\hbox{ corresponding to \eqref{eq:abij}}\hbox{ is  a copy of the complex}
\hbox{projective line: ${\BP}\left( H_{12} \oplus 
 H_{34} \right)$ .}
 \hbox{It parametrizes  rank one}\hbox{linear relations}\hbox{between
 $H_{12}$ and $H_{34}$}} \quad \picit{3.8}{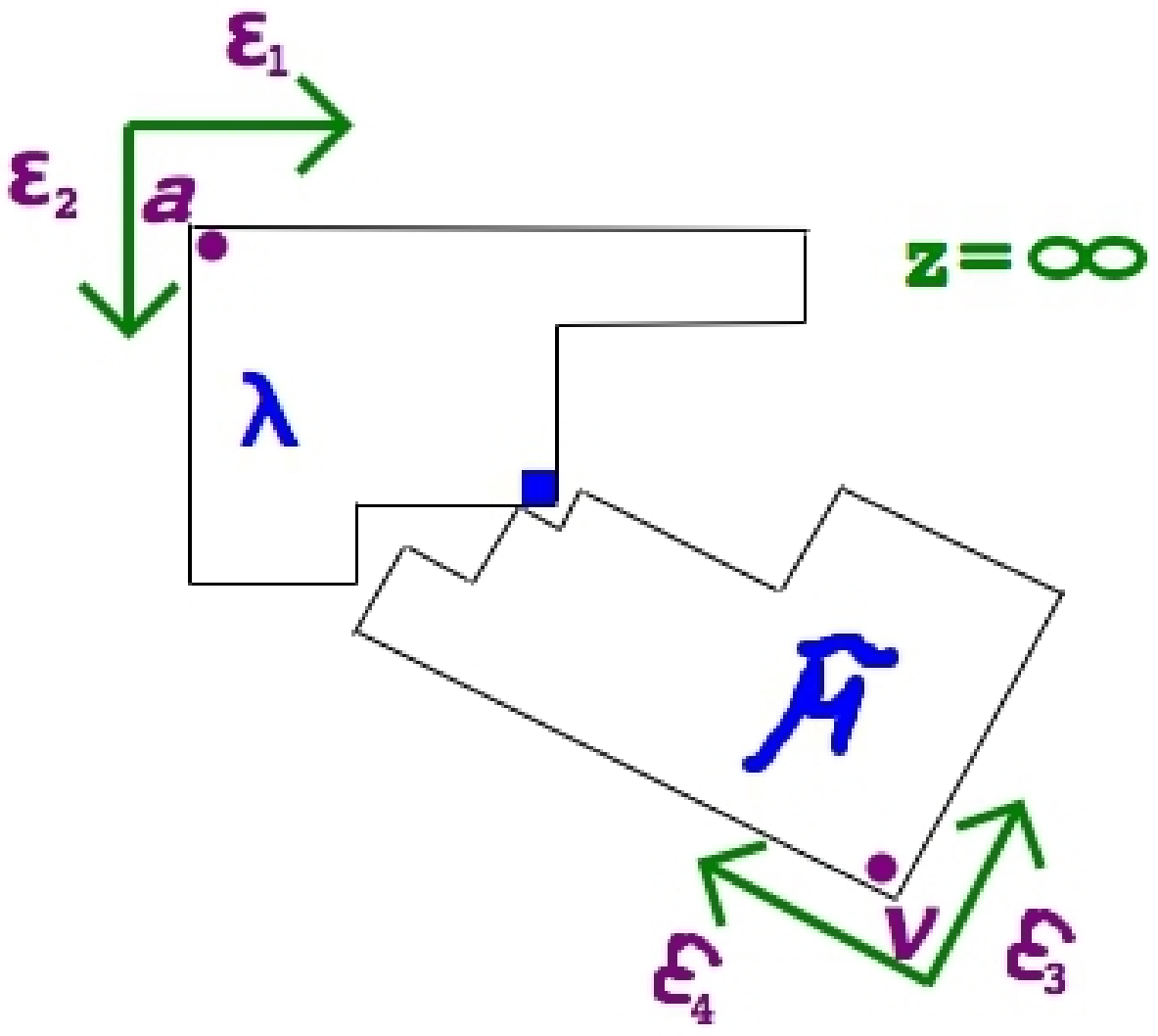}}
 \caption{The component ${\BP}\left( H_{12} \oplus 
 H_{34} \right)$}
  \label{fig:lamu}
 \medskip
 \end{figure}
  
 {} Let $z$ be the coordinate on ${\BP}\left( H_{12} \oplus 
 H_{34} \right)$ such that $z = {\infty}$ corresponds to the line $H_{12}$ while $z = 0$ corresponds to the line $H_{34}$. For $z \neq 0, \infty$ the linear spaces $H_{12}$ and $H_{34}$ coincide, $z$ being the isomorphism. When $z \to 0$ the image $(p_{1}(z=0), p_{2}(z=0))$ is the pair $({\tilde\lambda}^{12}, {\lambda}^{34})$, the image $(p_{1}(z={\infty}), p_{2}(z={\infty}))$ is the pair $({\lambda}^{12}, {\tilde\lambda}^{34})$. Here ${\tilde\lambda}^{12}$ is the $n$-tuple of partitions which differs from ${\lambda}^{12}$ in that the Young diagram of ${\tilde\lambda}^{(12, {\alpha})}$ (${\tilde\lambda}$ on the Fig. \ref{fig:lamu}) is obtained by removing the $(i',j')$
 square from the Young diagram of ${\lambda}^{(12, {\alpha})}$. Similarly, the $w$-tuple ${\tilde\lambda}^{34}$ (${\tilde\mu}$ on the Fig. \ref{fig:lamu}) is obtained by modifying ${\lambda}^{(34, {\beta})}$ by removing the box  $(i'', j'')$. 
 
In the next chapters we shall relax the condition \eqref{eq:aabbij}. In other words, we shall consider a subtorus in ${\bT}_{n, w} \times U(1)^{3}_{\ept}$.

 \subsubsection{Toric folded instantons}
  
 Now let us explore the folded instantons invariant under the action of the maximal torus
 ${\bT}_{n, w} \times U(1)^3_{\ec}$. It is easy to see that these are again the pairs $({\lambda}^{12}, {\lambda}^{13})$, with ${\lambda}^{12} = ({\lambda}^{(12,1)} , \ldots , {\lambda}^{(12, n)})$, ${\lambda}^{13} = ({\lambda}^{(13,1)} , \ldots , {\lambda}^{(13, w)})$. The spaces $K_{12}$ and $K_{13}$ do not intersect, $K = K_{12} \oplus K_{13}$.  In other words, the only ${\bT}_{n, w} \times U(1)^3_{\ec}$-invariant folded instantons are the superpositions of the ordinary instantons on ${\BC}^{2}_{12}$ and ${\BC}^{2}_{13}$, of the charges $k_{12}$ and $k_{13}$, 
 respectively, with $k = k_{12} + k_{13}$.

 Now let us consider the non-generic case, such that $K_{12, 13} \neq {\emptyset}$.  We call the corresponding fixed point ``the zipper'', see the Fig. \ref{fig:K12,13}. The codimension one subtorus for which this is possible corresponds to the relation ${\ac}_{12, \alpha} - {\ac}_{13, \beta} \notin Z_{\ept}$ between the parameters of the infinitesimal torus transformation.

 \bigskip

\begin{figure}[H]
\picit{8}{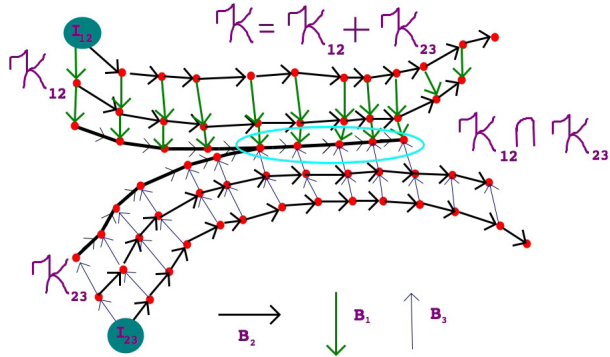}
\caption{The zipper}
\label{fig:K12,13}
\end{figure}

\bigskip
The non-empty overlap $K_{12} \cap K_{13}$ implies the sets of eigenvalues of $\sigma$ on $K_{12}$ and $K_{13}$ overlap, leading to
\beq
{\ac}_{12, {\alpha}} + {\ec}_{1}(i'-1) + {\ec}_{2}(j'-1) = {\ac}_{13, {\beta}} + {\ec}_{1} (i''-1) + {\ec}_{3}(j''-1)
\label{eq:zipeq}
\eeq
for some ${\alpha} \in [n], {\beta} \in [w], i', j', i'', j''$. Unlike the Eq. \eqref{eq:abij} the Eq. \eqref{eq:zipeq} the integers $i', i''$ are not uniquely determined. Since the left hand side of \eqref{eq:zipeq} 
is the eigenvalue of ${\sigma} \vert_{L_{1,2}}$, while the right hand side is the eigenvalue of ${\sigma} \vert_{L_{1,3}}$, we conclude:
\beq
(i',j') \in {\lambda}^{(12,{\alpha})}\, ,  (i'',j'') \in {\lambda}^{(13,{\beta})}\, , (i',j'+1) \notin {\lambda}^{(12,{\alpha})}\, ,  (i'',j''+1) \notin {\lambda}^{(13,{\beta})}
\label{eq:la12la13}
\eeq
i.e. $j' = {\lambda}^{(12, {\alpha})t}_{i'}, \ j'' = {\lambda}^{(13, {\beta})t}_{i''}$. The change $(i', i'') \mapsto (i' \pm 1, i'' \pm 1)$ maps the solution of
\eqref{eq:zipeq} to another solution of \eqref{eq:zipeq}. Let $l \geq 0$ be the maximal integer such that $j' = {\lambda}^{(12, {\alpha})}_{i'-l}, \ j'' = {\lambda}^{(13, {\beta})}_{i''-l}$, and let $l \geq m \geq 0$ be the maximal integer such that $j' = {\lambda}^{(12, {\alpha})}_{i'-m}, \ j'' = {\lambda}^{(13, {\beta})}_{i''-m}$, and
$e_{12} = B_{1}^{i'-m-1}B_{2}^{j'-1} \, I_{12, \alpha} \in K_{12, 13}$, 
$e_{13} =  B_{1}^{i''-m-1}B_{3}^{j''-1} \, I_{13, \beta} \in K_{12, 13}$. 
In other words the vectors $e_{12}$ and $e_{13}$ are linearly dependent, $e_{12} = z e_{13}$. Consequenly, the arm-lengths
$a_{i'-m,j'} = {\lambda}^{(12, {\alpha})t}_{j'}- i' + m$, $a_{i'-m,j'} = {\lambda}^{(13, {\beta})t}_{j''} - i'' + m$ must be equal:
\beq
 {\lambda}^{(12, {\alpha})t}_{j'}- i' + m = a_{i'-m,j'} = {\lambda}^{(13, {\beta})t}_{j''} - i'' + m
\label{eq:blob}
\eeq

\centerline{\picit{3.5}{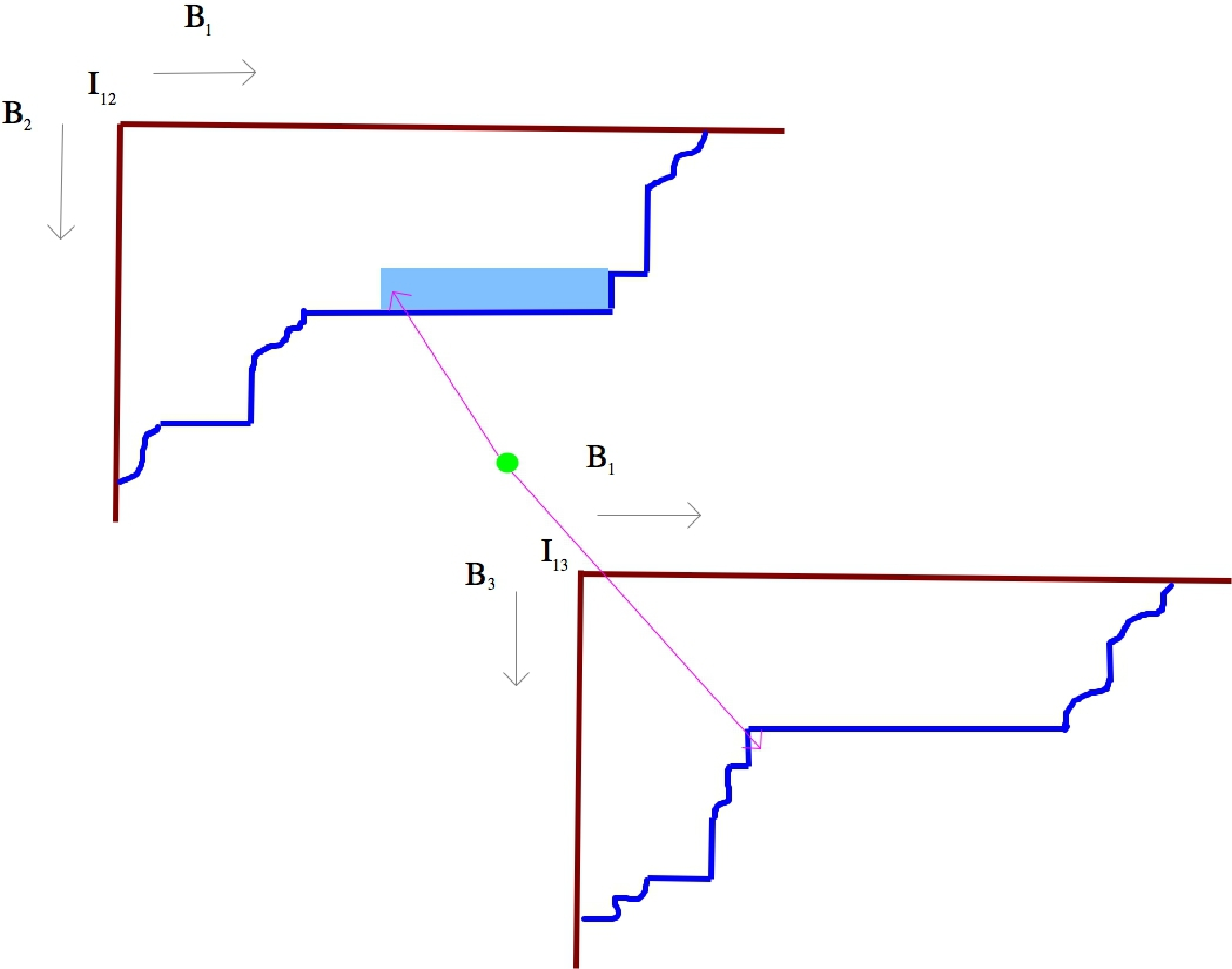}\quad \vbox{\hbox{The component of }
\hbox{the fixed point set}
\hbox{ corresponding to \eqref{eq:zipeq}}\hbox{ is  a copy of the complex}
\hbox{projective line: ${\BP}\left( {\BC}e_{12} \oplus 
 {\BC} e_{13} \right)$ .}
 \hbox{It parametrizes  rank one}\hbox{linear relations}\hbox{between
 $e_{12}$ and $e_{13}$}} \quad \picit{3.8}{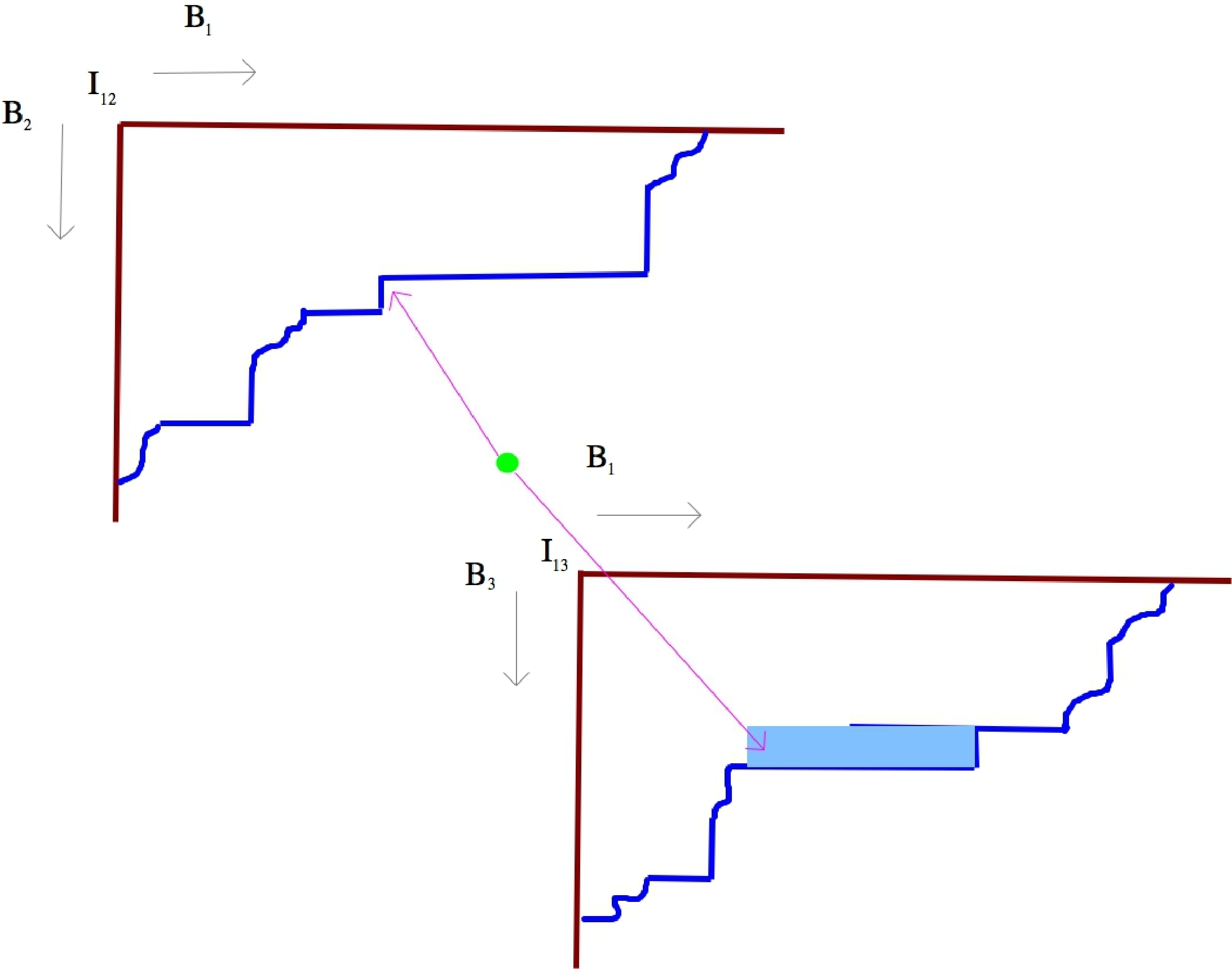}}
 \label{fig:zip34}
 \bigskip 
 {} Let $z$ be the coordinate on ${\BP}\left( {\BC}e_{12} \oplus 
 {\BC} e_{13} \right)$ such that $z = {\infty}$ corresponds to the line ${\BC}e_{12}$ while $z = 0$ corresponds to the line ${\BC}e_{13}$. Then the image $(p_{1}(z=0), p_{2}(z=0))$ is the pair $({\tilde\lambda}^{(12,{\alpha})}, {\lambda}^{(13, {\beta})})$, the image $(p_{1}(z={\infty}), p_{2}(z={\infty}))$ is the pair $({\lambda}^{(12,{\alpha})}, {\tilde\lambda}^{(13, {\beta})})$. Here we defined ${\tilde\lambda}^{(12, {\alpha})}$ to be the partition whose Young diagram is obtained by removing the block of 
 squares $(i'-m,j') \ldots ( {\lambda}^{(12, {\alpha})t}_{j'}, j')$ from the Young diagram of ${\lambda}^{(12, \alpha)}$. Similarly, the Young diagram of ${\tilde\lambda}^{(13, \beta )}$ is obtained by removing the
 block of squares $(i''-m,j'') \ldots ( {\lambda}^{(13, {\beta})t}_{j''}, j'')$ from the Young diagram of ${\lambda}^{(13, \beta )}$.

\bigskip
\centerline{
\picit{4}{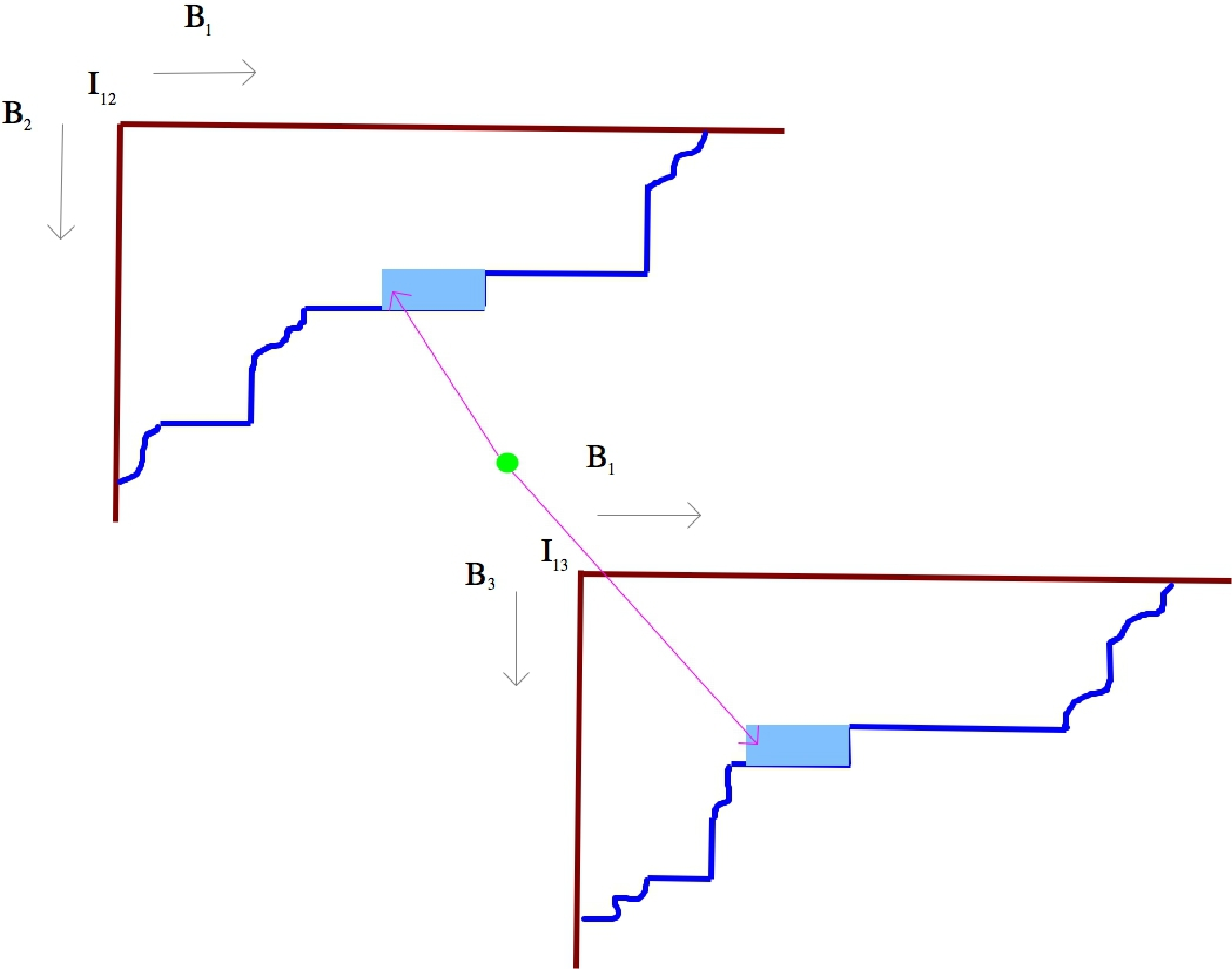}
\label{fig:Zip1}
\quad\vbox{\hbox{Note that the pair of Young diagrams}
\hbox{${\lambda}^{(12,{\alpha})}$, ${\lambda}^{(34, {\beta})}$ gives rise} \hbox{to several components
of} \hbox{the fixed point set,}
\hbox{isomorphic to ${\BC\BP}^{1}$,}
\hbox{e.g. the ones corresponding}
\hbox{to the blocks of horizontal boxes}
\hbox{of different length,}}
\quad 
\picit{4}{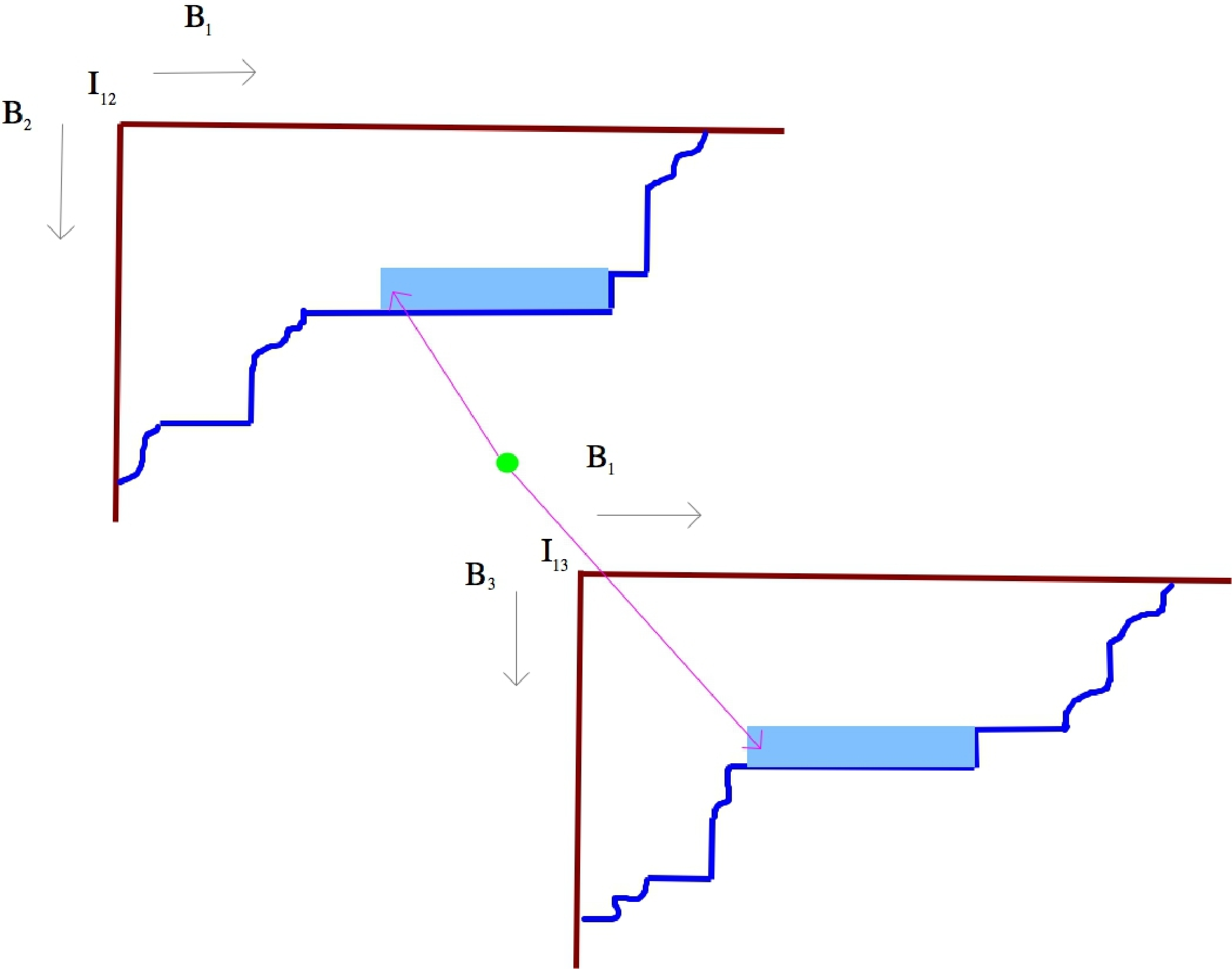}}
\label{fig:Zip2}

\bigskip\noindent
see the pictures above on the left and on the right. 
But they actually belong to moduli spaces of folded instantons
of different charges (in computing the charge $k$ we subtract the length of the block  from the sum of the sizes of Young diagrams). So despite the similarity in graphic design, these are pieces of different 
architectures.  

\bigskip
\noindent

\section{Reconstructing spiked instantons}

 In this section we describe the sewing procedure, which produces a spiked instanton out of six ordinary noncommutative  instantons.  We then use the stitching to describe the spiked instantons invariant under the toric symmetry, i.e. the $T_{\Lf}$-fixed locus. 
 
\subsection{The local K-spaces}

For $A = \{ a, b \} \in {\6}$ we define:
\beq
K_{A} = {\BC}[B_{a}, B_{b}] \, {\rm im} (I_{A})  \subset K
\label{eq:kaspa}
\eeq
By definition, this is the minimal $B_{a}, B_{b}$-invariant subspace of $K$, containing the image ${\rm im} (I_{A}) = I_{A} (N_A)$ of $N_A$.

The equations  \eqref{eq:bbij}, \eqref{eq:upab} for 
$i \geq k$ imply that
\beq
J_{B} (K_{A}) = 0, \qquad B \neq A
\label{eq:jbka}
\eeq
and
\beq
B_{b} (K_{A}) = 0, \qquad b \notin A
\label{eq:bbka}
\eeq

\subsection{Toric spiked instantons}
 
 Now let us describe the spiked instantons, invariant under the torus action. The tori in question are the subgroups
 of $\Hf$, the global symmetry group. We consider first
 the maximal torus $T_{\Hf}$ (cf. \eqref{eq:thf}) and then its subtori $T_{\Lf}$
 for various choices of the $\Lf$ data. 
 
 \subsubsection{Maximal torus}
 
 First of all, let us consider the $T_{\Hf}$-fixed points. Let
 ${\ba} = ({\ac}_{A})_{A \in {\6}}$ be the collection of diagonal matrices ${\ac}_{A} = {\rm diag}({\ac}_{A, 1}, \ldots, {\ac}_{A, n_{A}}) \in {\rm Lie}U(n_{A})$. The spiked
 instanton $[{\bB}, {\bI}, {\bJ}]$ is $T_{\Hf}$-invariant iff
 for any $\ba$ and $\be$ there exists an operator ${\sigma} \in {\rm End}(K)$, such that:
 \beq
 \begin{aligned}
&  {\ec}_{a} B_{a} = [ {\sigma}, B_{a} ] \, , \qquad a \in {\4} \\
& ({\ec}_{A} - {\ac}_{A} ) J_{A} = - J_{A} {\sigma} \, , \\
& I_{A} {\ac}_{A} = {\sigma} I_{A}  \, , \qquad A \in {\6} \\
\end{aligned}
\label{eq:thfixedspiked}
\eeq
Let $N_{A,\alpha}$, ${\alpha} \in [n_{A}]$, be the eigenspace of ${\ac}_{A}$ with the eigenvalue ${\ac}_{A,{\alpha}}$. Let $I_{A, \alpha} = I_{A}(N_{A,\alpha})$. 
We have (for $A = \{ a, b \}$, $a < b$):
\beq
K_{A} = \sum_{{\alpha} \in [n_{A}], \, i, j\geq 1} K_{A, {\alpha}}^{i,j}\ , \qquad K_{A, {\alpha}}^{i,j} = B_{a}^{i-1}B_{b}^{j-1} I_{A, {\alpha}} 
\label{eq:kaaij}
\eeq
The eigenvalue of ${\sigma}$ on $K_{A, {\alpha}}^{i,j}$ is equal to 
\beq
{\sigma}\vert_{K_{A, {\alpha}}^{i,j}} = {\ac}_{A, {\alpha}} + {\ec}_{a}(i-1) + {\ec}_{b}(j-1)
\label{eq:sigeig}
\eeq
On the other hand, Eq. \eqref{eq:thfixedspiked} implies that
the vector 
\beq
{\psi} = J_{A} (K_{A, {\alpha}}^{i,j}) \in N_{A}
\eeq
is an eigen-vector of ${\ac}_{A}$ with the eigen-value:
\beq
{\ac}_{A} {\psi} =  ({\ac}_{A, {\alpha}} + {\ec}_{a}i + {\ec}_{b}j){\psi}
\label{eq:eigja}
\eeq 
The $T_{\Hf}$-invariance means we are free to choose the parameters ${\ac}_{A,{\alpha}}$, ${\ec}_{a}$ in an arbitrary fashion. It means, that ${\ac}_{A, {\alpha}} + {\ec}_{a}i + {\ec}_{b}j \neq {\ac}_{A, {\beta}}$ for $i,j \geq 1$, ${\alpha}, {\beta} \in [n_{A}]$. Therefore $J_{A}$ vanishes on all
$K_{A, {\alpha}}^{i,j}$ subspaces, and therefore on all of $K_A$. Therefore, all $B_{a}$'s commute with each other. Also,  the eigenvalues \eqref{eq:sigeig} are different for different
$(A, {\alpha}; i,j)$. Therefore, the spaces $K_{A,\alpha}^{i,j}$
are orthogonal to each other.

Define, for $A \in {\6}, {\alpha} \in [n_{A}]$ the partition
${\lambda}^{(A, {\alpha})}$, by:
\beq
{\lambda}_{i}^{(A, {\alpha})} = {\mathrm{sup}} \, \left\{ \, j \, | \, j \geq 1, K_{A, {\alpha}}^{i,j} \neq 0 \, \right\} 
\eeq
We have:
\beq
k_{A} = \sum_{{\alpha} \in [n_{A}]} \vert {\lambda}^{(A, {\alpha})} \vert
\label{eq:kach}
\eeq
The $T_{\Hf}$-fixed points are, therefore, in one-to-one correspondence with the collections 
 \beq
 {\mathbf{\Lambda}}  = \left( {\lambda}^{(A, {\alpha})} \right)_{A \in {\6}, {\alpha} \in [n_{A}]}
 \label{eq:spikeyoung}
 \eeq
 of 
 \beq
 {\mathbf{n}} = \sum_{A \in {\6}} n_{A}
 \label{eq:ntot}
 \eeq
 Young diagrams. In the companion paper \cite{Nekrasov:2015ii} we shall be studying the statistical mechanical model, where the random variables are the collections $ {\mathbf{\Lambda}}$, while the complexified Boltzman weights are the contributions of ${\mathbf{\Lambda}}$ to the gauge partition function, defined below.

 \subsubsection{Subtori}
 Now fix the data $\Lf$ and consider the $T_{\Lf}$-invariant
 spiked instantons $[{\bB}, {\bI}, {\bJ} ]$. As usual, these come with the homomorphism $T_{\Lf} \to U(K)$ which associates the compensating $U(K)$-transformation $g_{t, \xi}$ for every $(e^{{\ii}{\xi}}, e^{{\ii}t}) \in T_{\Lf}$. 
 Since $K$ decomposes into the direct sum of weight subspaces
 \beq
 K = \bigoplus_{w, n} \, K^{w,n}
 \label{eq:kwn}
 \eeq
 where
 \beq
 g_{t, \xi} \vert_{K^{w,n}} = e^{{\ii} \langle w, \xi \rangle + {\ii} n t}
 \eeq
where $n \in {\BZ}$, while $w$ belongs to the weight lattice of $\varprod_{A\in {\6}^{-}}\, U(1)^{{\ell}_{A}-1}$. 

The relations:
\beq
\begin{aligned}
& e^{{\ii}e_{a}t} B_{a} = g_{t, \xi} B_{a} g_{t, \xi}^{-1}\, , \\
& I_{A} (N_{A, \alpha})  e^{{\ii}c_{A}( {\alpha}) t} = g_{t, \xi} I_{A} (N_{A, \alpha})  \, , \ A \in {\6}^{+} \, , {\alpha} \in [n_{A}]\\
& I_{A} (N_{A, {\iota}, \alpha})  e^{{\ii}c_{A, {\iota}}( {\alpha}) t} = g_{t, \xi} I_{A} (N_{A, {\iota}, \alpha})  \, , \ A \in {\6}^{-} \, , {\alpha} \in [n_{A}]_{\iota}\\
\end{aligned}
\label{eq:gbgb}
\eeq
imply:
\begin{multline}
I_{A} (N_{A, \alpha}) \in K^{0, c_{A}({\alpha})}\, , \ {\alpha} \in [n_{A}], \ A \in {\6}^{+}  \, , \\
I_{A} (N_{A, {\iota}_{A}, \alpha}) \in K^{0, c_{A, {\iota}_{A}}({\alpha})}\, , \ {\alpha} \in [n_{A}]_{{\iota}_{A}}, \ A \in {\6}^{-}  \, , \\
I_{A}(N_{A, {\iota} , \alpha}) \in K^{{\varpi}_{A,\iota}, c_{A,\iota}({\alpha})} \, , \ {\alpha} \in [n_{A}]_{\iota}, \ {\iota} \in {\lambda}_{A}, \ A \in {\6}^{-} \\
B_{a} (K^{w, n}) \subset K^{w, n+e_{a}}\, , \ a \in {\4} 
\end{multline}
where ${\varpi}_{A,{\iota}}$ is the fundamental weight, 
$\langle {\varpi}_{A, {\iota}}, {\xi} \rangle = {\xi}_{A,{\iota}}$. 

Finally, the $T_{\Lf}$-invariance translates to 
\begin{multline}
J_{A} (K^{w,n}) \in N_{A, {\alpha}} \, , \ A \in {\6}^{+} , \,{\alpha} \in [n_{A}] \ \Leftrightarrow n = c_{A}({\alpha}) -e_{A} \, \ , w = 0 \, ,  \\
J_{A} (K^{w,n}) \in N_{A, {\iota}_{A}, {\alpha}} \, , \ A \in {\6}^{-} , \,{\alpha} \in [n_{A}]_{{\iota}_{A}} \ \Leftrightarrow n = c_{A, {\iota}_{A}}({\alpha}) -e_{A} \, \ , w = 0 \, ,  \\
J_{A} (K^{w,n}) \in N_{A, \iota, {\alpha}} \, , \ A \in {\6}^{-}, \, {\alpha} \in [n_{A}]_{\iota} \ \Leftrightarrow n = c_{A, {\iota}}({\alpha}) -e_{A} \, \ , w = {\varpi}_{A,{\iota}} \, .  \end{multline}
which imply, with our choice of $T_{\Lf}$, that
$J_{A} = 0$. This is shown using the same arguments as we used around the Eq. \eqref{eq:eigja}.

\subsubsection{$K$-spaces for toric instantons}

Let $A \in {\6}^{+}$. The local space $K_A$ is $g_{t, \xi}$-invariant, and decomposes as
\beq
K_{A} = \bigoplus_{n}\, K_{A}^{n}
\label{eq:kan}
\eeq
with integer $n$, via
\beq
g_{t, \xi} \vert_{K_{A}^{n}} = e^{{\ii}nt}
\eeq
where $n \geq c_{A}^{-} = {\rm inf}_{\alpha \in [n_{A}]} \, c_{A}({\alpha})$,  when $e_{a}, e_{b} >0$ and $n\leq c_{A}^{+} = {\rm sup}_{\alpha \in [n_{A}]} \, c_{A}({\alpha})$ when $e_{a}, e_{b} < 0$. For $e_{a}, e_{b} > 0$ both operators $B_{a}$, $a \in A$ raise the grading. For $e_{a}, e_{b} < 0$ both operators $B_{a}$, $a \in A$ lower the grading. 
Let $k_{A, n} = {\rm dim}K_{A}^{n}$. Since $K_{A}$ is finite dimensional, $k_{A,n}$ vanish for $|n| > C_{A}$ for some some constant $C_{A} \leq k$.  

Let $A \in {\6}^{-}$. The local space $K_A$ is $g_{t, \xi}$-invariant, and decomposes as
\beq
K_{A} = \bigoplus_{n} K_{A}^{n} \, \oplus \ \bigoplus_{{\iota} \in {\lambda}_{A} ,n} K_{A,\iota}^{n}
\label{eq:kain}
\eeq
 with 
 \beq
 g_{t, \xi} \vert_{K_{A,\iota}^{n}} = e^{{\ii}n t + {\ii}{\xi}_{A,\iota}}
 \label{eq:gtxikain}
 \eeq
 for $i \in {\lambda}_{A}$, and
 \beq
 g_{t, \xi} \vert_{K_{A}^{n}} = e^{{\ii}nt}
 \label{eq:kaian}
 \eeq
 Since the eigenvalues of $g_{t, \xi}$ on $K_{A,\iota}^{n}$ for ${\iota} \in {\lambda}_A$ differ from each other and from those on $K_{B}^{n'}$ for all $B \in {\6}$, $n' \in \BZ$, the spaces $K_{A,\iota}^{n}$
 are orthogonal to $K_{B}^{n'}$ and to each other:
 \beq
 \begin{aligned}
&  K_{A, \iota'}^{n'} \perp K_{A, \iota''}^{n''}, \qquad \iota' \neq \iota''  \\
& K_{A, \iota}^{n'} \perp K_{A}^{n''} \\
\end{aligned}
\label{eq:decoperp}
\eeq
The action of ${\bB}, {\bI}$-operators respects the orthogonal decomposition \eqref{eq:decoperp}. 
 
 We now prove that the spaces $K_{A}^{n}$ and $K_{A,i}^{n}$ have an additional $U(1)$-action. Indeed, let $f_{a}, f_{b}$ be the two positive mutually prime integers, such that
 \beq
 e_{a}f_{a} + e_{b}f_{b} = 0\, , 
 \label{eq:efab}
 \eeq
 so that $e_{a} = p_{ab} f_{b}, \quad e_{b} = - p_{ab}f_{a}$
 (assuming $e_{a}>0>e_{b}$). Then the operator
 \beq
 {\es} = B_{a}^{f_{a}}B_{b}^{f_{b}}
 \label{eq:cebabb}
 \eeq
 commutes with $g_{t, \xi}$, thanks to \eqref{eq:gbgb}. Since all the eigenvalues of $B_{a}$ and $B_{b}$ vanish (again, thanks to \eqref{eq:gbgb}), the operators $B_{a}$, $B_{b}$, and $\es$ are nilpotent. By Jacobson-Morozov
 theorem, $\es$ can be included into the $sl_{2}$-triple, i.e. for each $K_{A}^{n}$, $K_{A,i}^{n}$ there are operators
 $\hs$, ${\es}^{\vee}$, such that
 \beq
 [{\es}, {\es}^{\vee}] = {\hs}, \ [{\hs}, {\es}] = 2{\es},
 \   [{\hs}, {\es}^{\vee}] = - 2{\es}^{\vee}
 \label{eq:sl2t}
 \eeq
 so that
 \beq
 K_{A}^{n} = \bigoplus_{h} K_{A}^{n,h}, \qquad
 K_{A,\iota}^{n} = \bigoplus_{h} K_{A,\iota}^{n,h}
 \label{eq:scndgr}
 \eeq
 with $h$ standing for the eigenvalue of $\hs$. 
 Now, it is not difficult to prove that the $(n,h)$-grading is equivalent to the $(i,j)$-grading, with $i,j \geq 1$:
 \beq
 \begin{aligned}
 & i = i' + (h-h'_{A}({\alpha})) f_{a} \, , \\
 & j = j' + (h-h'_{A}({\alpha})) f_{b} \, , \\
 & n = c_{A} ({\alpha}) + p_{A} ( f_{b}(i'-1) - f_{a}(j'-1)) \, , \\
 & i',j' \geq 1\, , \, i' \leq f_{a} \ {\mathrm{and/or}} \ j' \leq f_{b} \, \\
 & h'_{A}({\alpha}) = {\rm inf}  \ {\rm Spec}({\hs}\vert_{K_{A}^{n}}) \\
 \end{aligned}
 \eeq
 and $\alpha$ is uniquely determined by $n\, {\rm mod}\, p_{A} = c_{A} ({\alpha}) \, {\rm mod}\, p_{A}$. Thus,
 \beq
 K_{A} = \bigoplus_{i,j \geq 1} K_{A}^{i,j} \, , \qquad
 K_{A,\iota} = \bigoplus_{i,j \geq 1} K_{A,\iota}^{i,j}
 \label{eq:newgrad}
 \eeq
 with 
 \beq
 B_{a} (K_{A}^{i,j} ) \subset K_{A}^{i+1, j} \, , \qquad
 B_{b} (K_{A}^{i,j} ) \subset K_{A}^{i, j+1}
 \label{eq:newgradb}
 \eeq
 
Now we are ready for the final push:

\section{The compactness theorem}

We   now prove the compactness theorem which establishes the analyticity of the partition function defined in the next chapter. To this end we estimate the norm
of $({\bB}, {\bI}, {\bJ})$ whose $U(k)$-orbit is invariant with respect to the action of any minimal torus $T_{\Lf}$. 

Since $J_{A}$'s vanish, the real moment map equation 
reads as follows:
\beq
\sum_{a\in {\4}} [B_{a}, B_{a}^{\dagger}] + \sum_{A \in {\6}} I_{A}I_{A}^{\dagger} = {\zeta} {\bf 1}_{K}
\label{eq:momr}
\eeq
The trace of this equation gives 
the norm of $I_{A}$'s:
 \beq
 \sum_{A \in {\6}} \Vert I_{A} \Vert^{2} = {\zeta}k 
 \eeq 
But we need to estimate the norms $\Vert B_{a} \Vert^2$ which drop out of trace. However, it is not too difficult to chase them down. 
We have: 
\begin{multline}
\sum_{a\in {\4}} \Vert B_{a} \Vert^{2} + \sum_{A \in {\6}} \Vert I_{A} \Vert^2 = \sum_{a\in {\4}} 
{\Tr}_{K} B_{a}B_{a}^{\dagger} +
\sum_{A \in {\6}} {\Tr}_{K} I_{A}I_{A}^{\dagger} \leq \\
\sum_{A \in {\6}} {\Tr}_{K_{A}} \left( \sum_{a \in {\4}} B_{a}B_{a}^{\dagger} + \sum_{C \in {\6}} I_{C}I_{C}^{\dagger} \right)  = \sum_{A \in {\6}} \left( {\zeta} k_{A} + {\Tr}_{K_{A}} \, \sum_{a \in {\4}} B_{a}^{\dagger} B_{a} \, \right) = \\
{\zeta} \sum_{A \in {\6}}  k_{A} +  \sum_{A \in {\6}, a \in A} {\Tr}_{K_{A}}  B_{a}^{\dagger} B_{a}   \\
\label{eq:normbaka}
\end{multline}
where we used the moment map equation \eqref{eq:rmom}, projected onto $K_A$, and the Eq. \eqref{eq:bbka}. Define:
\beq
\begin{aligned}
& {\delta}_{A,n} = \frac{1}{\zeta} {\Tr}_{K_{A}^{n}} \left( \sum_{a \in {\4}} B_{a}B_{a}^{\dagger} + \sum_{A' \in {\6}} I_{A'}I_{A'}^{\dagger} \right) \\
& {\delta}_{A} = \frac{1}{\zeta} {\Tr}_{K_{A}} \, \left( \sum_{a \in {\4}}  B_{a}B_{a}^{\dagger} + \sum_{A'\in {\6}} I_{A'} I_{A'}^{\dagger} \right) = \sum_{n} {\delta}_{A,n} \\
\label{eq:delan}
\end{aligned}
\eeq
Now for $A \in {\6}^{+}$ use the decomposition \eqref{eq:kan}, and \eqref{eq:bbka} to show, that for $e_{a} \neq e_{b}$:
\beq
{\delta}_{A,n} =   
  k_{A, n} + 
 \frac{1}{{\zeta}} \sum_{a \in A} {\Tr}_{K_{A}^{n+e_{a}}} \, \left( 
 B_{a}B_{a}^{\dagger} \right) \leq  k_{A,n} + \sum_{a \in A} {\delta}_{A, n+e_{a}} 
  \label{eq:kanmom}
 \eeq
 where we very conservatively estimated:
 \beq
 {\Tr}_{K_{A}^{n}} \, \left( 
 B_{a}B_{a}^{\dagger} \right) \leq {\delta}_{A,n} \, , 
 \eeq
 for any $a \in \4$. 
 This very conservative inequality can be used to show the boundeness of ${\delta}_{A,n}$. 
 
 From now on let us assume $e_{a}> e_{b} >0$. The case of negative $e_{a}, e_{b}$ is treated analogously. 
 First of all, let introduce the sequence of generalized Fibonacci numbers $F_{n}^{p,q}$, $n \in {\BZ}$, for positive integeres $p > q > 0$, by:
 \beq
 \begin{aligned}
 & F_{n}^{p,q} = 0\, , \qquad\qquad 1- p \leq n \leq 0 \\
 & F_{1}^{p,q} = 1\, , \\
 & F_{n}^{p,q} = F_{n-p}^{p,q} + F_{n - q}^{p,q}\, , \ n > 0\\
 \end{aligned}
 \eeq
 It is easy to write the formula for $F_{n}^{p,q}$ in terms of the roots ${\lambda}_{\omega}$, ${\omega} = 0, \ldots , p-1$, of the characteristic equation
 \beq
 {\lambda}_{\omega}^{p} = {\lambda}_{\omega}^{p-q} +1 \, ,
 \label{eq:lpq}
 \eeq
 \beq
 F_{n}^{p,q} = \sum_{\omega} f_{\omega} {\lambda}_{\omega}^{n}
 \label{eq:fibnum}
 \eeq
 where the coefficients $f_{\omega}$ are to be found from the linear equations $F_{n}^{p,q} = {\delta}_{n,1}$, $1- p \leq n \leq 1$. 

Now, we can estimate ${\delta}_{A,n}$ by induction in $n$:
 \beq
 {\delta}_{A,n} \leq \sum_{n' \geq n} k_{A,n'} F_{n'+1-n}^{e_{a},e_{b}}\,  
 \label{eq:pqest}
 \eeq
This leads to the following, also very conservative, bound on ${\delta}_{A}$:
  \beq
  {\delta}_{A} < k_{A} F_{k_{A}}^{e_{a},e_{b}}
  \label{eq:kfkab}
  \eeq 
 When $e_{a} = e_{b} = e_{A}/2 \geq 1$ we can make a better estimate:
 \beq
 {\delta}_{A,n} = k_{A, n} + 
 \frac{1}{{\zeta}} \sum_{a \in A} {\Tr}_{K_{A}^{n+e_{a}}} \, \left( 
 B_{a}B_{a}^{\dagger} \right) \leq  k_{A,n} +  {\delta}_{A, n+e_{A}/2} 
  \label{eq:kanmome}
 \eeq
which, by iteration, implies:
\beq
{\delta}_{A,n} \leq k_{A,n} + k_{A, n+e_{A}/2} + k_{A, n+e_{A}} + \ldots  
\eeq
which in turns implies the upper bound on 
\beq
{\delta}_{A}  \leq \frac{1}{e_{A}} k_{A} (k_{A}+1)
\eeq
It remains to estimate ${\delta}_{A}$ for $A \in {\6}^{-}$. 
This is easy to do using the $(i,j)$-grading \eqref{eq:newgrad}. Define:
\beq
{\delta}_{A, n} = \frac{1}{\zeta}  \sum_{i+j = n+2} {\Tr}_{K_{A}^{i,j}} \left( \sum_{a \in {\4}} B_{a}B_{a}^{\dagger} + \sum_{A' \in {\6}} I_{A'}I_{A'}^{\dagger} \right) \, 
\label{eq:delaminn}
\eeq
Then, using \eqref{eq:rmom}, projected onto $K_A^{i,j}$, and \eqref{eq:newgradb}, we derive the estimate:
\beq
{\delta}_{A, n} \leq {\delta}_{A, n+1} + \sum_{i+j = n+2} {\rm dim}K_{A}^{i,j} 
\eeq
from which we get the estimate:
\beq
{\delta}_{A} \leq \frac{1}{2} k_{A}(k_{A}+1)
\label{eq:delamest}
\eeq
\section{Integration over the spiked instantons}

The moduli spaces ${\mM}_{k}^{i}({\vec n})$ are not your favorite smooth varieties. They can be stratified by smooth varieties of various dimensions. Over these smooth components the obstruction bundles keep track of the non-genericity of the equations we used to define the spaces ${\mM}_{k}^{i}({\vec n})$. 

In applications we need to compute the integrals over the spaces ${\mM}_{k}^{\infty}({\vec n})$, as well as to define and compute the equivariant indices of various twisted Dirac operators (for five dimensional theories compactified on a circle). 

Mathematically one can take the so-called virtual approach \cite{Rahul:1997}, where the fundamental cycle $[{\mM}_{k}^{\infty}({\vec n})]$ is replaced by the virtual fundamental cycle
$[{\mM}_{k}^{\infty}({\vec n})]^{\rm vir}$, which is defined as the Euler class of the bundle of equations over the smooth variety of the original variables $({\bB}, {\bI}, {\bJ})$. 
There are two difficulties with this definition: $i)$ the space of $({\bB}, {\bI}, {\bJ})$, being a vector space, is non-compact; $ii)$ the bundle of equations is infinite dimensional, unless we are in the situation with only the crossed or ordinary instantons. 

The problem $i)$ is solved by passing to the equivariant cohomology. The problem $ii)$ is cured by working with ${\mM}_{k}^{i}({\vec n})$ for large but finite $i$, and then  regularize the limit $i \to \infty$ by using the $\Gamma$-functions.

Physically, the problem is solved by the considerations of the matrix integral (matrix quantum mechanics) of the system of $k$ $D(-1)$-branes ($k$ $D0$-branes whose worldlines wrap ${\BS}^{1}$) in the vicinity of six stacks of $D3$ branes ($D4$ branes) wrapping coordinate two-planes ${\BC}^{2}_{A}$ (times a circle ${\BS}^{1}$) in the $IIB$ ($IIA$) background ${\BR}^{2} \times {\BC}^{4}$ (${\BR}^{1} \times {\BS}^{1} \times {\BC}^{4}$). 

One can also define the elliptic genus by the study of the two-dimensional gauge theory corresponding to the stack of $k$ $D1$-strings wrapping a ${\bT}^{2}$ shared by six stacks of $D5$ branes in $IIB$ string theory, wrapping ${\bT}^{2} \times {\BC}^{2}_{A}$.

\subsection{Cohomological field theory}

Let us briefly recall the physical approach. For every variable, i.e. for every matrix element of the
matrices $B_{a}, I_{A}, J_{A}$, we introduce the fermionic variables ${\Psi}^{\bB}_{a}, {\Psi}^{\bI}_{A}, {\Psi}^{\bJ}_{A}$ with the same ${\Hf} \times U(K)$ transformation properties. 

For every equation $s_{A}, {\mu}, {\sigma}_{{\bar a}A}, {\Upsilon}_{A}, {\Upsilon}_{A, B ; i}$ we introduce a pair 
of fermion-boson variables valued in the dual space: 
$({\chi}_{A}, h_{A}), ({\chi}, {\mu}), ({\chi}_{{\bar a}A}, h_{{\bar a}A}), ({\xi}_{A}, r_{A}), ({\varpi}_{A', A''; i} , y_{A', A''; i} )$. 

Finally, we need a triplet of variables $({\sigma}, {\bar \sigma}, {\eta})$ (two bosons and a fermion), valued in the Lie algebra of $U(K)$. 

Our model has the fermionic symmetry acting by:
\beq
\begin{aligned}
& {\delta}B_{a} = {\Psi}^{\bB}_{a} \, , \quad {\delta}{\Psi}^{\bB}_{a} = - [ {\sigma}, B_{a} ] + {\ec}_{a} B_{a}\\
& {\delta}I_{A} = {\Psi}^{\bI}_{A} \, , \quad {\delta}{\Psi}^{\bI}_{A}  = - {\sigma} I_{A} + I_{A} {\ac}_{A} \\
& {\delta}J_{A} = {\Psi}^{\bJ}_{A} \, , \quad {\delta}{\Psi}^{\bJ}_{A} = - {\ac}_{A} J_{A} + J_{A} {\sigma} + {\ec}_{A} J_{A}\\
\end{aligned}
\label{eq:decosyi}
\eeq
(cf. \eqref{eq:thfixedspiked}) and
\beq
\begin{aligned}
& {\delta}{\chi}_{A} = h_{A}\, , \quad {\delta}h_{A} = - [ {\sigma}, {\chi}_{A} ] + {\ec}_{A} {\chi}_{A}\, , \\
& \qquad\qquad\qquad\qquad {\delta}{\chi} = h\, , \quad {\delta}h  = - [{\sigma} , {\chi} ]\, ,  \\
& {\delta}{\chi}_{{\bar a}A} = h_{{\bar a}A} \, , \quad {\delta}h_{{\bar a}A} = - {\ac}_{A} {\chi}_{{\bar a}A} + {\chi}_{{\bar a}A} {\sigma} - {\ec}_{\bar a} {\chi}_{{\bar a}A}\, , \\
& {\delta} {\xi}_{A} = r_{A}\, \quad {\delta}r_{A} = {\ac}_{\bar A} {\xi}_{A}  - {\xi}_{A} {\ac}_{A} + {\ec}_{A} {\xi}_{A} \, , \\ 
& \qquad\qquad {\delta}{\varpi}_{A', A''; i} = y_{A', A''; i} \, ,  \\
& \qquad\qquad\qquad {\delta}y_{A', A''; i} = - {\ac}_{A'} {\varpi}_{A', A''; i}  + {\varpi}_{A', A''; i} {\ac}_{A''} 
+ (i-1) {\ec}_{a}{\varpi}_{A', A''; i} \, ,  \\
& \qquad\qquad\qquad\qquad {\rm whenever}\qquad A' \cap A'' = \{ a \}\\
\end{aligned}
\label{eq:decosyii}
\eeq
and
\beq
{\delta}{\bar\sigma} = {\eta}, \quad {\delta}{\eta} = [ {\sigma}, {\bar\sigma}], \qquad {\delta}{\sigma} = 0
\eeq
Now we can define the partition function
\begin{multline}
{\CalZ}_{k}^{i}({\vec n}; {\ept}, {\ba}) = 
\int \, e^{-{\bS}_{i}}\, \frac{D{\sigma}}{{\rm Vol}(U(k))}\, D{\bB}D{\bB}^{\dagger} D{\Psi}^{\bB}D{\Psi}^{{\bB}\dagger} \ldots D{\chi}Dh \ldots D{\bar\sigma}D{\eta} \\
 \prod_{j=1}^{i-1} D{\chi}_{A, B; j} Dh_{A, B; j}D{\bar\chi}_{A, B; j} D{\bar h}_{A, B; j} \ 
\label{eq:zkpf}
\end{multline}
where
\begin{multline}
{\bS}_{i} = {\delta} {\Xi}_{i}\, , \qquad
{\Xi}_{i} = {\Xi}^{s} + {\Xi}^{f} + {\Xi}^{K} +  \sum_{A} {\Xi}^{N}_{A} + \sum_{A', A''; \# A' \cap A'' = 1} \sum_{j=1}^{i-1} {\Xi}^{N}_{A'A'', j} \, , \\
{\Xi}^{s} = {\Tr}_{K} \, {\eta}  [ {\sigma}, {\bar\sigma} ] \, , \quad {\Xi}^{f} = {\Tr}_{K}  \, \sum_{a} \left( {\Psi}^{\bB}_{a} \left( - [ {\bar\sigma}, B_{a}^{\dagger} ] + {\bar\ve}_{a} B_{a}^{\dagger} \right) + c.c. \right) + \qquad\qquad \\
\qquad\qquad + {\Tr}_{K}  \, \sum_{A} \left(  \left( - {\bar\sigma} I_{A} + I_{A} {\bar\ac}_{A} \right) {\Psi}^{\bI\dagger}_{A} + c.c. \right) + 
{\Tr}_{K}  \, \sum_{A} \left(  {\Psi}^{\bJ\dagger}_{A} \left( J_{A} {\bar\sigma}  - {\bar\ac}_{A} J_{A} + {\bar\ve}_{A} J_{A} \right)  + c.c. \right) \, , \\
{\Xi}^{K} = {\Tr}_{K} \, {\chi} \left( - {\ii} \left( {\mu} - {\zeta} {\bf 1}_{K} \right) + h \right) +  \sum_{A} \left( {\chi}_{\bar A} \left( - {\ii} s_{A} + h_{A} \right) + \sum_{{\bar a} \in {\bar A}} {\chi}_{{\bar a}A}^{\dagger} \left( - {\ii} {\sigma}_{{\bar a}A} + h_{{\bar a}A} \right) \right) \, . \\
{\Xi}_{A}^{N} = {\Tr}_{N_{A}}\, \left(  {\xi}_{A}^{\dagger} \left( - {\ii}{\Upsilon}_{A} + r_{A} \right)  + c.c. \right)  \, , \qquad
{\Xi}_{A', A''; j}^{N} = {\Tr}_{N_{A}} \, \left(  {\varpi}_{A', A''; j}^{\dagger} \left( - {\ii} {\Upsilon}_{A', A''; j} + y_{A', A'' ; j} \right)  + c.c. \right) \\
\end{multline}
Here ${\bar\ve}_{a}$, ${\bar\ac}_{A}$ are auxiliary elements of the Lie algebra of $\Hf$, which can be chosen arbitrary,
as long as the integral \eqref{eq:zkpf} converges. 

\subsection{Localization and analyticity}

The usual manipulations with the integral \eqref{eq:zkpf}, for generic $({\ept}, {\ba})$, express it as a sum over the fixed points, which we enumerated in the Eq. \eqref{eq:spikeyoung}. Each fixed point contributes a  homogeneous (degree zero) rational function of ${\ac}_{A, {\alpha}}$'s, $1 \leq {\alpha} \leq n_{A}$ and ${\ec}_{a}$, times the product
\begin{multline}
\prod_{A \in {\6}, \, 4 \in {\bar A}} \ \prod_{{\alpha}=1}^{n_{A}}\prod_{{\beta}=1}^{n_{\bar A}} \ \left( {\ac}_{A, \alpha} - {\ac}_{{\bar A}, \beta} + {\ec}_{\bar A} \right) \quad \times \\  
 \prod_{A, B \in {\6}, \,  A \cap B = \{ c \}} \ \prod_{{\alpha}=1}^{n_{A}} \prod_{{\beta}=1}^{n_{B}} \prod_{j=1}^{i-1}\
\left( {\ac}_{A, \alpha}  - {\ac}_{B, \beta} + {\ec}_{c}(j-1) \right)
\end{multline}

The compactness theorem of the previous chapter implies, among other things,  that the partition functions ${\CalZ}_{k}^{i}({\vec n}; {\ept}, {\ba})$, 
for $i > k$, have no singularities in
\beq
{\xt}_{A} = \frac{1}{n_{A}} \sum_{{\alpha}=1}^{n_{A}} {\ac}_{A, \alpha}
\label{eq:coma}
\eeq
with fixed ${\ec}_{a}$'s and ${\tilde\ac}_{A, \alpha} = {\ac}_{A, \alpha} - {\xt}_{A}$. In other words,  they are polynomials of $\xt_{A}$.

\section{Quiver crossed instantons}

\subsection{Crossed quivers}

For oriented graph $\gamma$ let  $\Vg$ denote the set of its
vertices, and  $\Eg$ the set of edges, with $s,t : {\Eg} \to {\Vg}$ the source and the target maps.
Sometimes we write ${\gamma} = \left( {\Vg}, {\Eg}, s, t \right)$.  The crossed quiver is the data ${\gcr} = ({\gamma}^{+}, {\gamma}^{-}, p)$, where
${\gamma}^{\pm}$ are two oriented graphs, and 
 $p \in {\BZ}_{\geq 0}$ is a non-negative integer. Let ${\Xi}_{p} = {\BZ}/p{\BZ}$ be the additive group with $p$ elements, for $p >0$ and ${\BZ}$ for $p=0$.  Define
 ${\sV} = {\Vgp} \times {\Vgm} \times {\Xi}_{p}$. The group ${\Xi}_{p}$ acts on $\sV$ by translations of the third factor. The generator of $\Xi_{p}$ acts by ${\omega} = (v^{+},v^{-},n) \mapsto {\omega} +1 \equiv (v^{+},v^{-},n+1)$, with $v^{\pm} \in {\Vg}^{\pm}$. We also define $^{\pm}{\sE} = {\Egpm} \times {\Vgmp} \times {\Xi}_{p}$ and the natural maps $s, t : \ ^{\pm}{\sE} \to {\sV}$, e.g.
 $s(e, u, n) = ( s(e), u, n)$ for $e \in {\Egp}, u \in {\Vgm}$, $t(e,v,n) = (v, t(e), n)$ for $e \in {\Egm}, v \in {\Vgp}$ etc. The group ${\Xi}_{p}$ also acts on $^{\pm}{\sE}$, so we shall write:
 ${\eta} = (e,u, n) \mapsto {\eta} \pm 1 \equiv (e, u, n \pm 1)$. The source and target maps are $\Xi_{p}$-equivariant, i.e. $s({\eta}\pm 1) = s({\eta}) \pm 1$. 

\subsubsection{Paths and integrals}

We shall use the notion of a path. Define the path ${\sf p}_{{\omega}', {\omega}''}^{\pm}$ of length $\ell$ to be a sequence of pairs:
 \beq
{\sf p}_{{\omega}', {\omega}''}^{\pm} =  ({\eta}_{1}, {\sf or}_{1}), ({\eta}_{2},  {\sf or}_{2}), \ldots, ({\eta}_{\ell}, {\sf or}_{\ell})
\label{eq:path}
\eeq with ${\eta}_{j} \in\, ^{\pm}{\sE}, {\sf or}_{j} \in \{ - 1, +1 \}$,  (we call ${\sf or}_{j}$ the orientation of the edge ${\eta}_{j}$ relative to ${\sf p}_{{\omega}', {\omega}''}^{\pm}$) such that for any $j = 1, \ldots , {\ell}-1$ either 
\beq
\begin{aligned}
 \quad s({\eta}_{j+1}) = t({\eta}_{j}{\pm}1), & \qquad {\sf or}_{j} = {\sf or}_{j+1} = 1, \\
{\rm or}\ t({\eta}_{j+1}) = t({\eta}_{j}) , & \qquad {\sf or}_{j} = - {\sf or}_{j+1} = 1,  \\
{\rm or}\ s({\eta}_{j+1}) = s({\eta}_{j}) , & \qquad -{\sf or}_{j} = {\sf or}_{j+1} = 1,  \\
{\rm or}\ t({\eta}_{j+1}) = s({\eta}_{j}{\mp}1) , & \qquad -{\sf or}_{j} =  - {\sf or}_{j+1} = 1,
\end{aligned}
\label{eq:paths}
\eeq
and also 
either $s({\eta}_{1}) = {\omega}'$ (${\sf or}_{1} = 1$) or $t({\eta}_{1}\pm 1) = {\omega}'$  (${\sf or}_{1} = -1$) and also either $s({\eta}_{\ell}) = {\omega}''$ (${\sf or}_{\ell} = -1$) or $t({\eta}_{\ell}{\pm}1) = {\omega}''$ (${\sf or}_{\ell} = 1$). 

For a function $b: \, ^{\pm}{\sE} \to {\BR}$ (a $1$-chain) and a path
${\sf p}_{{\omega}', {\omega}''}^{\pm}$ define the integral 
\beq
\int_{{\sf p}_{{\omega}', {\omega}''}^{\pm}} b = \sum_{j=1}^{\ell} \, {\sf or}_{j} b ({\eta}_{j})
\label{eq:intpath}
\eeq
The function $b: ^{\pm}{\sE} \to {\BR}$ is a coboundary, $b = {\delta}c$, of a $0$-chain $c: {\sV} \to {\BR}$, iff $b({\eta}) = c(t({\eta}{\pm}1)) - c(s({\eta}))$. The integral of a coboundary obeys Stokes formula:
\beq
\int_{{\sf p}_{{\omega}', {\omega}''}^{\pm}} {\delta}c = \sum_{j=1}^{\ell} \, {\sf or}_{j} \left( c (t({\eta}_{j}\pm 1)) - c(s({\eta}_{j})) \right) = c({\omega}'') - c({\omega}')
\eeq

\subsubsection{Representations of crossed quivers}
 
Fix four dimension vectors ${\bkt}, {\bn}^{+}, {\bn}^{-}, {\mm} : {\sV} \to {\BZ}_{\geq 0}$. Let $(K_{\omega}, N_{\omega}^{\pm},  M_{\omega})_{{\omega} \in {\sV}}$ be a collection of complex vector spaces whose dimensions are given by the components of the corresponding dimension vectors, e.g. dim$K_{{\omega}} = k_{{\omega}} \equiv {\bkt}({\omega})$. We view the spaces
$N_{{\omega}}^{\pm}, M_{{\omega}}$ as fixed, e.g. with some fixed basis, while the spaces $K_{\omega}$ are varying, i.e. defined up to the automorphisms. 
We also fix a decomposition $M_{\omega} = M_{\omega}^{\prime} \oplus M^{\prime\prime}_{\omega}$ as an additional refinement of our structure. 

\subsubsection{Weight assignment for crossed quivers}

For the crossed quiver $\gamma$ and its representation let
us fix the integral data: $({\bn}_{\omega}^{\pm}, {\mm}_{\omega}', {\mm}_{\omega}'')_{{\omega} \in {\Vg}}$, ${\underline{\bf t}} = \left( t({\omega}) \right)_{{\omega} \in {\sV}}$ and ${\be} = \left( e({\eta}) \right)_{{\eta} \in {\sE}}$, 
with integers  $t({\omega}), e({\eta}) \in {\BZ}$, obeying
$t({\omega}+1) = t({\omega})$, $e({\eta}+1) = e({\eta})$, and the integral  vectors ${\bn}_{\omega}^{\pm} = \left( n_{\omega, \alpha}^{\pm} \right)_{{\alpha}=1}^{n_{\omega}^{\pm}} \in {\BZ}^{n_{\omega}^{\pm}}$ etc. The data $({\bn}^{\pm}, {\be}, {\mm}', {\mm}'')$ is defined up to the action of the lattice ${\BZ}^{\Vg}$:
a function $f: {\Vg} \to {\BZ}$ shifts the data \eqref{eq:vdata} by:
\beq
n_{\omega, \alpha}^{\pm} \mapsto n_{\omega, \alpha}^{\pm} - f ({\omega})\, , \ e \mapsto e + {\delta}f
\label{eq:shsym}
\eeq
\subsubsection{Crossed quiver instantons}

Consider the vector superspace ${\CalA}^{\gamma}_{\bkt}({\bn}^{\pm}, {\mm})$
of linear maps $({\bB}, {\bI}, {\bJ}, {\bf\Theta})$
\beq
\begin{aligned}
{\sf bosons:} \qquad
  {\bB} = (B_{\eta}^{\pm}, {\tilde B}_{\eta}^{\pm})_{{\eta} \,  \in \, ^{\pm}{\sE}}&\, , \quad {\bI} = (I_{{\omega}}^{+}, {I}_{{\omega}}^{-})_{{\omega} \in {\sV}}\ , \quad {\bJ} = (J_{\omega}^{+}, J^{-}_{\omega})_{{\omega} \in {\sV}}\, , \\
     B_{\eta}^{\pm} : K_{s({\eta})} \to K_{t({\eta}) \pm 1}\, , \quad & {\tilde B}_{\eta}^{\pm} : K_{t({\eta})} \to K_{s({\eta}) \pm 1}\, , \quad {\eta} \, \in \, ^{\pm}{\sE}\\
 I_{\omega}^{\pm}: N_{\omega}^{\pm} \to K_{\omega \pm 1}\, ,\  &  \  J_{\omega}^{\pm}: K_{\omega} \to N_{{\omega} {\pm}1}^{\pm}\ ,  \quad {\omega} \in {\sV} \\
{\sf fermions:} \qquad {\bf\Theta} = \  ( {\Theta}_{\omega}^{\prime}, {\Theta}_{\omega}^{\prime\prime} )_{{\omega} \in {\sV}}\,  , & \qquad  \\   {\Theta}_{\omega}^{\prime} \in  {\Pi}{\rm Hom}(K_{\omega}, M_{\omega}^{\prime})\, , & \qquad {\Theta}_{\omega}^{\prime\prime} \in {\Pi}{\rm Hom}( M_{{\omega}-1}^{\prime\prime}, K_{\omega})  \\
\end{aligned}
\label{eq:quivasp}
\eeq
Let ${\CalG}^{\gamma}_{\bkt}, {\CalG}^{\BC\gamma}_{\bkt}$ be the groups:
\beq
{\CalG}^{\gamma}_{\bkt} = \varprod_{\scriptscriptstyle{{\omega}\in {\sV}}}
\, U(k_{{\omega}}) \, , \qquad  {\CalG}^{\BC\gamma}_{\bkt} = \varprod_{\scriptscriptstyle{{\omega}\in {\sV}}}
\, {\mathrm GL}(k_{{\omega}}, {\BC}) \, , 
\label{eq:gqgr}
\eeq
which act on ${\CalA}^{\gamma}_{\bkt}({\bn}^{\pm}, {\mm})$ via:
\begin{multline}
\left( \, g_{\omega } \, \right)_{\scriptscriptstyle{{\omega}\in {\sV}}} \ \cdot \ ({\bB}, {\bI}, {\bJ}, {\bf\Theta}) \ = \\ 
\left( \, g_{t({\eta}) \pm 1} B_{\eta}^{\pm} g_{s({\eta})}^{-1}, \,  g_{s({\eta})\pm 1} 
{\tilde B}_{\eta}^{\pm} g_{t({\eta})}^{-1} \, ; \
g_{{\omega}\pm 1} I^{\pm}_{{\omega}} , \ J^{\pm}_{{\omega}} 
g_{\omega}^{-1} \, ; \  {\Theta}_{\omega}^{\prime} g_{\omega}^{-1} , g_{\omega} {\Theta}_{\omega}^{\prime\prime}\right)
\end{multline}
We impose the following analogues of the Eqs. 
\eqref{eq:spik}:
\beq
\begin{aligned}
{\mu}_{\omega} = {\zeta}_{\omega} {\bf 1}_{K_{\omega}}\ , \qquad &  {\omega} \in {\sV}, \qquad {\zeta}_{\omega} > 0\\
s_{e^{+}, e^{-};\, n} = 0\ , \qquad &  e^{\pm} \in {\Egpm}, \, n \in {\Xi}_{p} \\
{\tilde s}_{e^{+}, e^{-};\, n} = 0\ , \qquad &  e^{\pm} \in {\Egpm},  \, n \in {\Xi}_{p} \\
{\sf s}_{{\omega}} = 0\ , \qquad & {\omega} \in {\sV} \\
{\Sigma}_{\eta} = 0\ ,  \qquad & {\eta} \in {\sE}\, ,  \\
{\tilde\Sigma}_{\eta} = 0\ ,  \qquad & {\eta} \in {\sE}\, ,  \\
{\Upsilon}_{{\omega}} \  = \ 0 \, ,  \qquad &  {\omega} \in {\sV},  \\
\label{eq:spikq}
\end{aligned}
\eeq
where
\beq
\begin{aligned}
& {\Sigma}_{\cdot} = {\sigma}_{\cdot}  + {\hat\sigma}_{\cdot}^{ \, \dagger} \, , \\
& {\tilde\Sigma}_{\cdot} = {\tilde\sigma}_{\cdot}  -  {{\hat{\tilde\sigma}}_{\cdot}}^{\, \dagger} \,  , \\
& s_{\dots}  = {\mu}_{\dots}^{\sf 13} - \left( {\mu}_{\dots}^{\sf 24} \right)^{\dagger}\, , \\
& {\tilde s}_{\dots}  = {\mu}_{\dots}^{\sf 14} + \left( {\mu}_{\dots}^{\sf 23} \right)^{\dagger}\, , \\
& {\sf s}_{\dots}  = {\mu}_{\dots}^{\sf 12} + \left( {\mu}_{\dots}^{\sf 34} \right)^{\dagger}\, , \\
& {\Upsilon}_{{\cdot}} = {\upsilon}_{{\cdot}-1}^{+} -  \left( {\upsilon}_{{\cdot}+1}^{-} \right)^{\dagger} 
\end{aligned}
\eeq
with the linear maps
\beq
\begin{aligned}
& {\sigma}_{\eta} : N^{\pm}_{s({\eta})} \to K_{t({\eta})} \, , \qquad {\rm for}
\quad {\eta} \in\, ^{\mp}{\sE}
\\
& {\tilde\sigma}_{\eta} : N^{\pm}_{t({\eta})} \to K_{s({\eta})} \, , \qquad {\rm for}\quad {\eta} \in\, ^{\mp}{\sE}\\
& {\upsilon}_{{\omega}{\mp}1}^{\pm} : N_{{\omega}}^{\mp} \to N^{\pm}_{{\omega}} \, , \\
\end{aligned}
\label{eq:sigen}
\eeq
and
\begin{figure}
\picit{5}{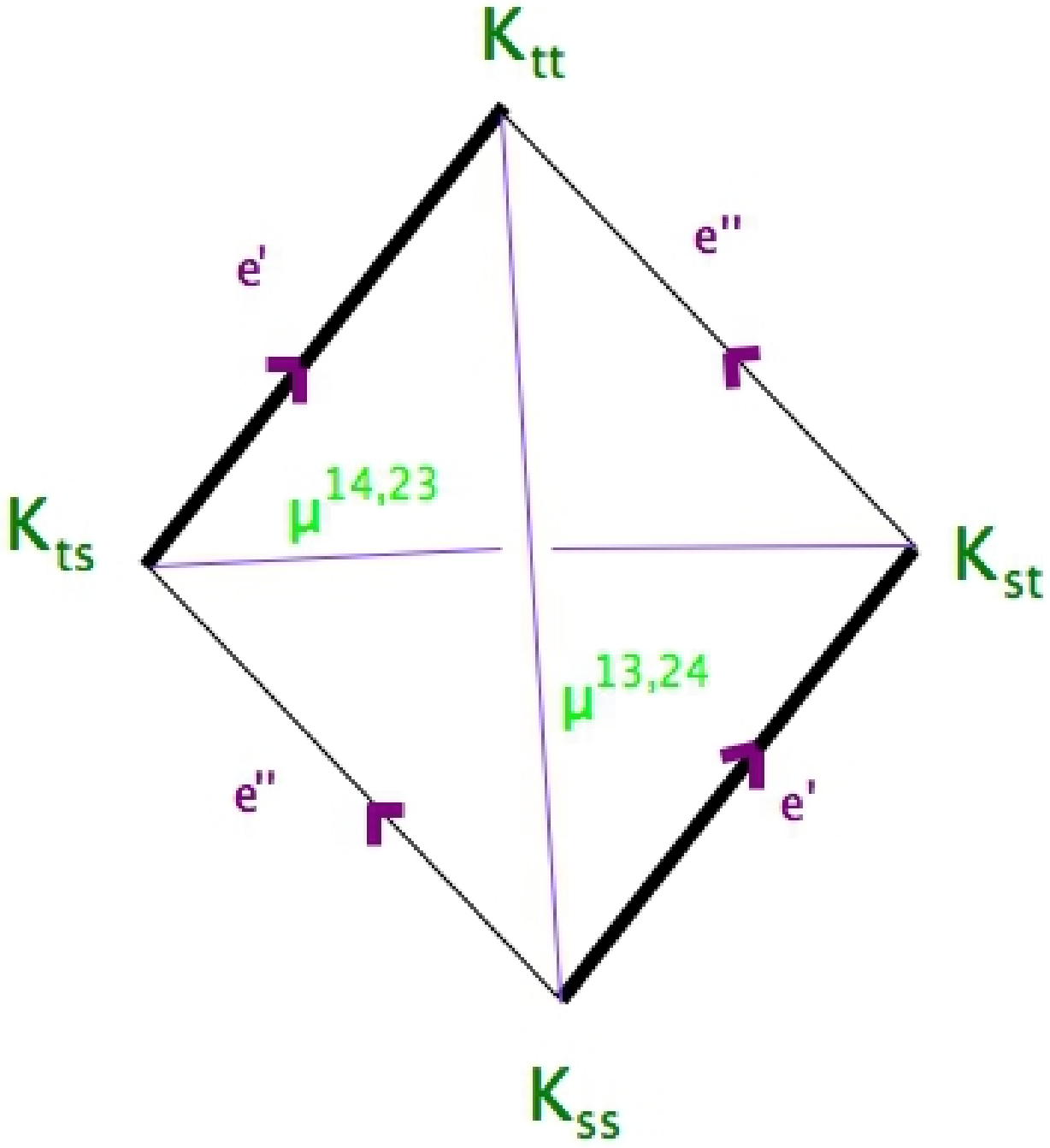}
\end{figure}
\beq
\begin{aligned}
 & {\mu}_{\omega} : K_{\omega} \to K_{\omega} \, , \\  
 & {\mu}_{{\omega}}^{\sf 12} : K_{\omega} \to K_{{\omega} + 2}\ , \\
 & {\mu}_{{\omega}}^{\sf 34} : K_{\omega} \to K_{{\omega}  - 2} \ ,  \\ 
  & {\mu}_{e^{+},e^{-} ; n}^{\sf 13} : K_{s(e^{+}), s(e^{-}), n} \to K_{t(e^{+}), t(e^{-}),n}\ , \\
   & {\mu}_{e^{+},e^{-} ; n}^{\sf 24} : K_{t(e^{+}), t(e^{-}),n} \to K_{s(e^{+}), s(e^{-}), n}\  ,\\
 &  {\mu}_{e^{+},e^{-} ; n}^{\sf 14} : K_{s(e^{+}), t(e^{-}), n} \to K_{t(e^{+}), s(e^{-}),n}\ , \\ 
 & {\mu}_{e^{+},e^{-} ; n}^{\sf 23} : K_{t(e^{+}), s(e^{-}),n} \to K_{s(e^{+}), t(e^{-}), n}\  .\\
\end{aligned}
\label{eq:momqmaps}
\eeq
The maps \eqref{eq:sigen}, for ${\eta} \in \, ^{\pm}{\sE}$, are given by, :
\beq
\begin{aligned} 
& {\sigma}_{\eta} = B_{{\eta}\pm 1}^{\mp} I^{\pm}_{s({\eta})} \ , \qquad
{\hat\sigma}_{\eta} =  J_{s({\eta}) \pm 1}^{\mp} {\tilde B}^{\pm}_{\eta} \, , \\
& {\tilde\sigma}_{\eta} = {\tilde B}^{\mp}_{{\eta} \pm 1} I^{\pm}_{t({\eta})}\ , \qquad
 {\hat{{\tilde\sigma}_{\eta}}} =  J_{t({\eta})\pm 1}^{\mp} B^{\pm}_{\eta} \, , \\
\end{aligned}
\label{eq:sigmen}
\eeq
and
\beq
{\upsilon}_{{\omega}\mp 1}^{\pm}  = J^{\pm}_{{\omega} {\mp}1} I^{\mp}_{{\omega}} \\
\eeq
The crossed quiver analogues \eqref{eq:momqmaps} of real and complex moment maps  are given by: for $\omega \in {\Vg}$,
\begin{multline}
{\mu}_{\omega} = \ I_{\omega-1}^{+}\left( I^{+}_{\omega-1} \right)^{\dagger} \ + \ I_{\omega+1}^{-}\left( I^{-}_{\omega+1} \right)^{\dagger} \  - \  \left( J^{+}_{\omega} \right)^{\dagger} J_{\omega}^{+}  \ - \  \left( J^{-}_{\omega} \right)^{\dagger} J_{\omega}^{-} \qquad - \\
\\
- \ \sum_{{\eta}\in s^{-1}({\omega}) \cap ^{+}{\sE}} \left( B_{\eta}^{+} \right)^{\dagger} B_{\eta}^{+} \ - \ \sum_{{\eta}\in t^{-1}({\omega})\cap ^{+}{\sE}} \left( {\tilde B}_{\eta}^{+} \right)^{\dagger} {\tilde B}^{+}_{\eta}\qquad - \\
- \ \sum_{{\eta}\in s^{-1}({\omega}) \cap ^{-}{\sE}} \left( B_{\eta}^{-} \right)^{\dagger} B_{\eta}^{-} \ - \ \sum_{{\eta}\in t^{-1}({\omega})\cap ^{-}{\sE}} \left( {\tilde B}_{\eta}^{-} \right)^{\dagger} {\tilde B}^{-}_{\eta}\qquad + \\
+\sum_{{\eta} \in s^{-1}({\omega}) \cap ^{+}{\sE}} {\tilde B}_{{\eta}-1}^{+} \left( {\tilde B}_{{\eta}-1}^{+} \right)^{\dagger}  + \sum_{{\eta} \in s^{-1}({\omega}) \cap ^{-}{\sE}} {\tilde B}_{{\eta}+1}^{-} \left( {\tilde B}_{{\eta}+1}^{-} \right)^{\dagger} \ +  \\
+\sum_{{\eta} \in s^{-1}({\omega}) \cap ^{+}{\sE}} B_{{\eta}-1}^{+} \left( B_{{\eta}-1}^{+} \right)^{\dagger}  + \sum_{{\eta} \in s^{-1}({\omega}) \cap ^{-}{\sE}} B_{{\eta}+1}^{-} \left( B_{{\eta}+1}^{-} \right)^{\dagger} \ ,  \\
 \label{eq:rmomq}
\end{multline} 
and
\beq
\begin{aligned}
{\mu}^{\sf 12}_{{\omega}} \ = & \qquad\qquad\qquad\qquad\qquad \picit{5}{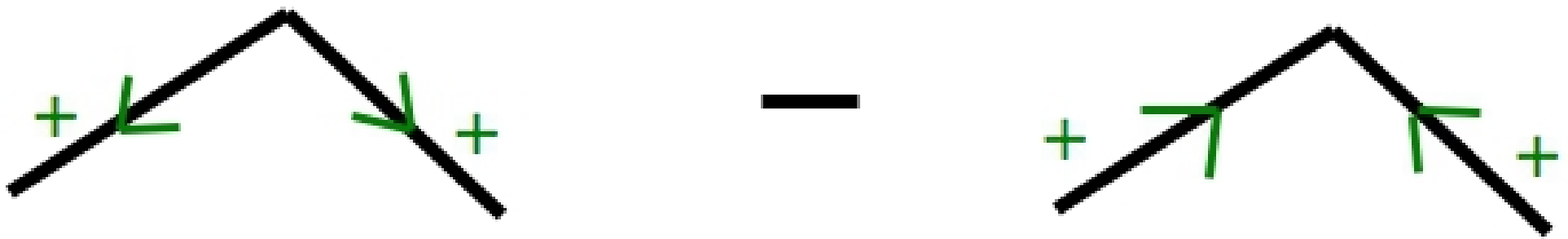} \\
& \qquad  \\
& \qquad I^{+}_{{\omega}+1} J^{+}_{{\omega}} \ + \ \sum_{{\eta} \, \in \, ^{+}{\sE} \cap t^{-1}({\omega})} B_{{\eta}+1}^{+} {\tilde B}_{\eta}^{+} - 
\sum_{{\eta} \, \in \, ^{+}{\sE} \cap s^{-1}({\omega})} 
{\tilde B}_{{\eta}+1}^{+} B^{+}_{{\eta}} \,  
\label{eq:mu12q}
\end{aligned}
\eeq
\beq
\begin{aligned}
{\mu}^{\sf 34}_{{\omega}} \ = & \qquad\qquad\qquad\qquad\qquad \ \picit{5}{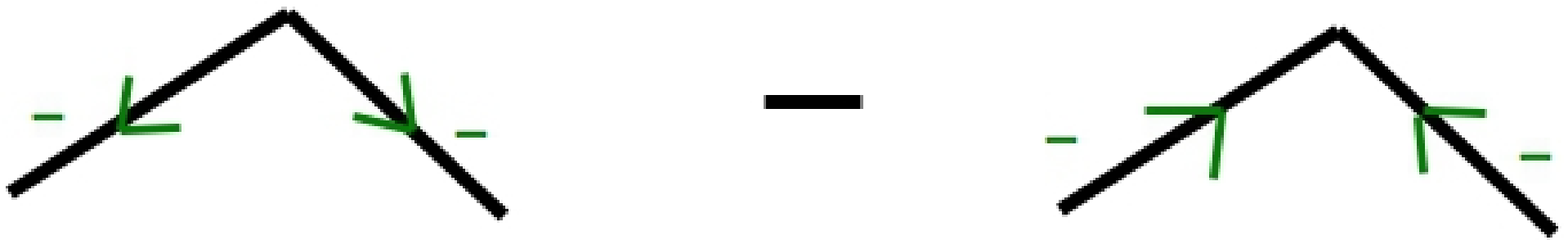} \\
&  \qquad  \\
&  \qquad I^{-}_{{\omega} -1} J^{-}_{{\omega}} \ + \ \sum_{{\eta} \, \in \, ^{-}{\sE} \cap t^{-1}({\omega})} B_{{\eta}-1}^{-} {\tilde B}_{\eta}^{-} - 
\sum_{{\eta} \, \in \, ^{-}{\sE} \cap s^{-1}({\omega})} 
{\tilde B}_{{\eta}-1}^{-} B^{-}_{{\eta}} \,  
\label{eq:mu34q}
\end{aligned}
\eeq
for $e^{\pm} \in {\Egpm}, \,  n \in {\Xi}_{p}$,
\beq
\begin{aligned}
{\mu}^{\sf 13}_{e^{+}, e^{-} ; n} \ = & \qquad\qquad\qquad\qquad\qquad \ \picit{5}{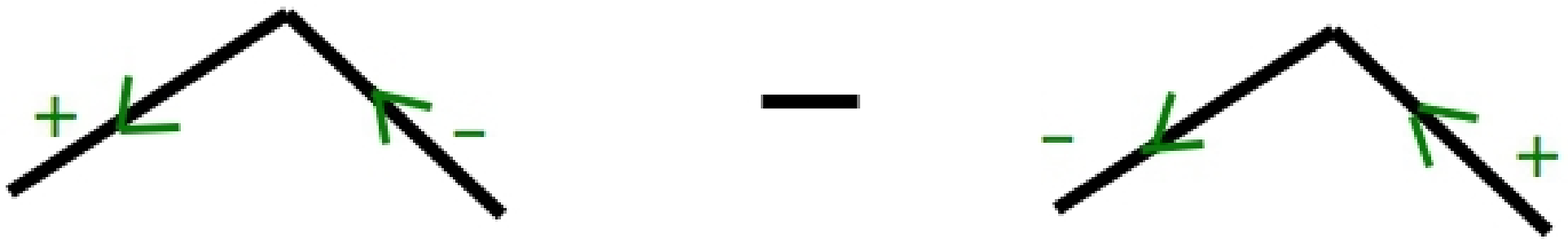} \\
&  \\
&  \qquad\qquad\qquad\qquad B_{e^{+}, t(e^{-}), n-1}^{+} B_{e^{-}, s(e^{+}),n}^{-} - 
B_{e^{-}, t(e^{+}), n+1}^{-} B_{e^{+}, s(e^{-}), n}^{+} \,  
\label{eq:mu13q}
\end{aligned}
\eeq
\beq
\begin{aligned}
{\mu}^{\sf 24}_{e^{+}, e^{-} ; n} \ = & \qquad\qquad\qquad\qquad\qquad \ \picit{5}{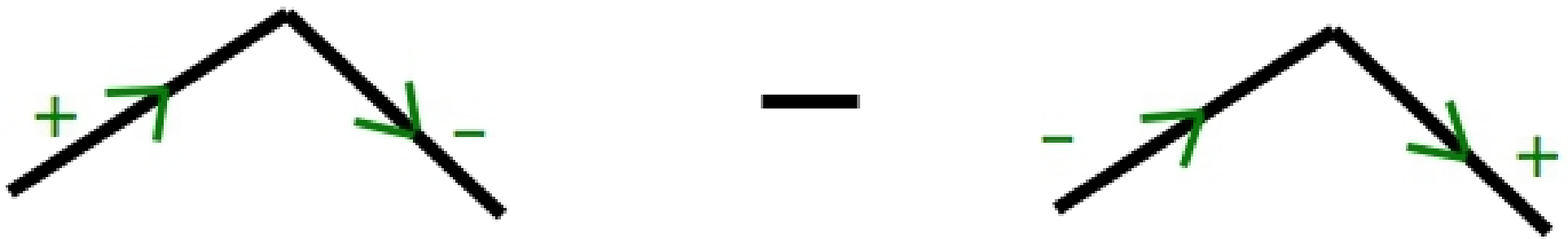} \\
&  \\
&  \qquad\qquad\qquad\qquad {\tilde B}_{e^{+}, s(e^{-}), n-1}^{+} {\tilde B}_{e^{-}, t(e^{+}),n}^{-} - 
{\tilde B}_{e^{-}, s(e^{+}), n+1}^{-} {\tilde B}_{e^{+}, t(e^{-}), n}^{+} \,  
\label{eq:mu24q}
\end{aligned}
\eeq
\beq
\begin{aligned}
{\mu}^{\sf 23}_{e^{+}, e^{-} ; n} \ = & \qquad\qquad\qquad\qquad\qquad \ \picit{5}{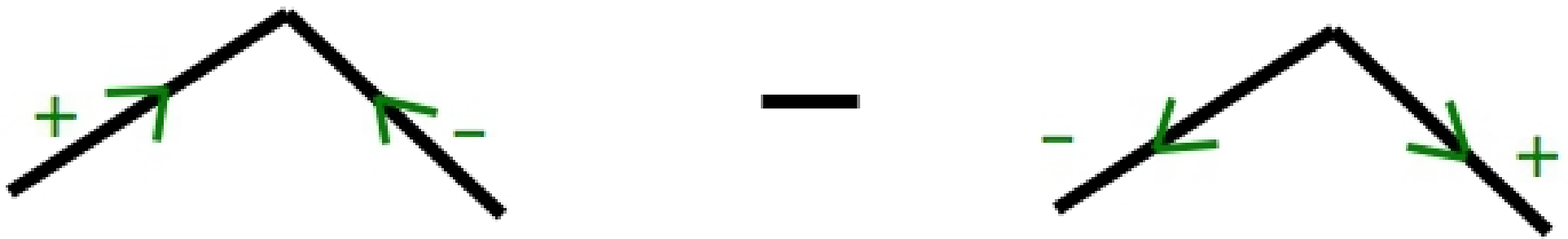} \\
&  \\
&  \qquad\qquad\qquad\qquad {\tilde B}_{e^{+}, t(e^{-}), n-1}^{+} B_{e^{-}, t(e^{+}),n}^{-} - 
B_{e^{-}, s(e^{+}), n+1}^{-} {\tilde B}_{e^{+}, s(e^{-}), n}^{+} \,  
\label{eq:mu23q}
\end{aligned}
\eeq
\beq
\begin{aligned}
{\mu}^{\sf 14}_{e^{+}, e^{-} ; n} \ = & \qquad\qquad\qquad\qquad\qquad \ \picit{5}{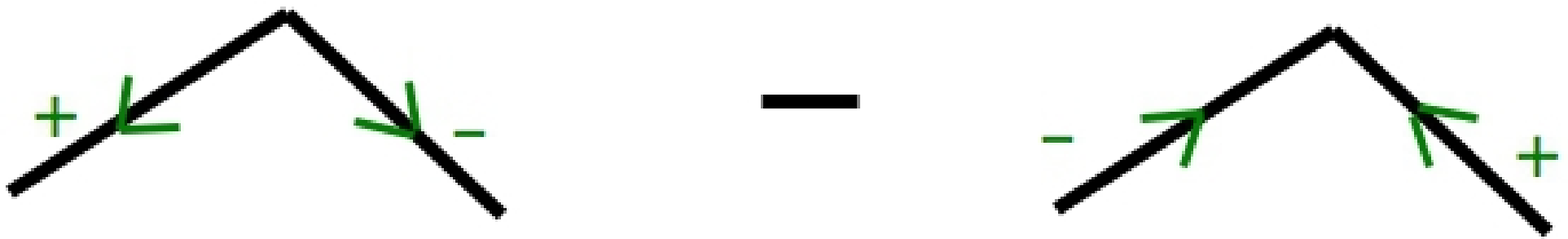} \\
&  \\
&  \qquad\qquad\qquad\qquad B_{e^{+}, s(e^{-}), n-1}^{+} {\tilde B}_{e^{-}, s(e^{+}),n}^{-} - 
{\tilde B}_{e^{-}, t(e^{+}), n+1}^{-} B_{e^{+}, t(e^{-}), n}^{+} \,  
\label{eq:mu14q}
\end{aligned}
\eeq
The moduli space of quiver crossed instantons ${\mM}^{\gamma}_{\bkt} ({\bn}^{\pm}, {\mm})$
is the space of solutions to \eqref{eq:spikq} modulo the action \eqref{eq:gqgr} of ${\CalG}_{\bkt}^{\gamma}$. 

The identity 
\begin{multline}
\sum_{{\eta} \in {\Eg}} {\Tr}_{K_{t({\eta})}} \left( {\sigma}_{\eta}{\hat\sigma}_{\eta} \right) - {\Tr}_{K_{s({\eta})}} \left(  {\tilde\sigma}_{\eta}{\hat{\tilde\sigma}}_{\eta}  \right)\ + \\
\sum_{{\omega} \in {\Vg}} {\Tr}_{N_{\omega}^{+}} \left( {\upsilon}^{+}_{\omega - 1} {\upsilon}^{-}_{\omega +1} \right) + 
 {\Tr}_{K_{{\omega}}}\, \left( {\mu}^{\sf 12}_{\omega-2}{\mu}^{\sf 34}_{\omega} \right) \ + \\
+ \sum_{e^{\pm} \in {\Egpm}, \,  n \in {\Xi}_{p}}
{\Tr}_{K_{t(e^{+}), s(e^{-}), n}}\, \left( {\mu}^{\sf 14}_{e^{+}, e^{-}; n} {\mu}^{\sf 23}_{e^{+}, e^{-}; n} \right) - {\Tr}_{K_{t(e^{+}), t(e^{-}),n}}\, \left( {\mu}^{\sf 13}_{e^{+}, e^{-}; n} {\mu}^{\sf 24}_{e^{+}, e^{-}; n} \right)\ = 0 
\end{multline}
can be used to demonstrate, by the argument identical to that in \eqref{eq:bps}, that the equations \eqref{eq:spikq}
imply the holomorphic equations
\beq
\begin{aligned}
& {\sigma}_{\eta} = 0, \\
& {\tilde\sigma}_{\eta} = 0, \\
&  {\hat\sigma}_{\eta} = 0, \\
& {\hat{\tilde\sigma}}_{\eta} = 0 \, , \\
& {\mu}^{\sf 12}_{\omega} = 0, \\
& {\mu}^{\sf 34}_{\omega} = 0\, , \\
& {\mu}^{\sf 13}_{e^{+},e^{-}; n} = 0 , \\
& {\mu}^{\sf 14}_{e^{+},e^{-}; n} = 0\, , \\
& {\mu}^{\sf 23}_{e^{+},e^{-}; n} = 0 , \\
& {\mu}^{\sf 24}_{e^{+},e^{-}; n} = 0\, , \\
& {\upsilon}^{\pm}_{\omega} = 0 
\label{eq:holospikq}
\end{aligned}
\eeq
Thus, ${\mM}^{\gamma}_{\bkt}({\bn}^{\pm}, {\mm})$ is the space of stable solutions of \eqref{eq:holospikq} modulo the action \eqref{eq:gqgr} of ${\CalG}_{\bkt}^{\BC\gamma}$.
Here, the stability condition is formulated as follows:
\beq\mathboxit{\begin{aligned}
& {\rm Any\ collection}\ (K'_{\omega})_{{\omega} \in {\sV}} \ {\rm of\ subspaces}\ K'_{\omega} \subset K_{\omega} \, , \ {\rm such\ that} \\
&  \qquad\qquad \color{red}{I^{\pm}_{{\omega}} (N_{\omega}^{\pm}) \subset K'_{\omega \pm 1 }} \\
& \qquad\qquad\qquad {\color{black}\rm and}\\
& \qquad\qquad \color{blue}{B_{\eta}^{\pm}(K'_{s({\eta})}) \subset K'_{t({\eta}) \pm 1}} \, , \quad {\rm for\ all}\ {\eta} \in\ ^{\pm}{\sE}  \\
& \qquad\qquad \color{blue}{{\tilde B}_{\eta}^{\pm}(K'_{t({\eta})}) \subset K'_{s({\eta}) \pm 1}} \, , \quad {\rm for\ all}\ {\eta} \in\ ^{\pm}{\sE}  \\
& \qquad\qquad\qquad \color{black}{\rm coincides\ with\ all\ of\ } K_{\omega} \, , \ \color{green}{K'_{\omega} = K_{\omega}}\, , \\
& \qquad\qquad\qquad  {\rm in\ other\ words\, , }\\
&\qquad\qquad\qquad\qquad\qquad\qquad\qquad\color{green} K_{\omega} = K_{\omega}^{+} + K_{\omega}^{-}
 \\
\end{aligned}}  
\label{eq:stabq}
\eeq
where $K^{\pm}_{\omega}\subset K_{\omega}$ is the subspace, generated by acting with arbitrary (noncommutative) polynomials in $B_{\eta}^{\pm}$, ${\tilde B}_{\eta}^{\pm}$'s with ${\eta} \in \ ^{\pm}{\sE}$ on the image $\sum_{\omega' \in {\sV}} I^{\pm}_{{\omega}'}(N_{{\omega}'}^{\pm})$:
\beq
K_{\omega}^{\pm} = \sum_{\omega' \in {\sV}} \, \left(  {\BC} \left[ B^{\pm}_{\eta}, {\tilde B}^{\pm}_{\eta} \right]_{\eta \in \ ^{\pm}{\sE}} \,   I^{\pm}_{{\omega}'}(N_{{\omega}'}^{\pm}) \right) \ \cap \ K_{\omega}
\label{eq:kompr}
\eeq
The space ${\mM}^{\gamma}_{\bkt}({\bn}^{\pm}, {\mm})$ is acted
upon by the group 
\beq
{\Hf}^{\gamma} =  \left( {\Hf}_{\sf ff}^{\gamma} \times {\Hf}_{\sf edg}^{\gamma} \right)
 / {\Hf}_{\sf ver}^{\gamma} \times U(1)_{\sf u}\, , 
 \label{eq:globqgr}
 \eeq
 where 
\beq
\begin{aligned}
& {\Hf}_{\sf ff}^{\gamma} =  \varprod_{{\omega}\in {\Vg}} \, 
U(N^{+}_{{\omega}})\times U(N^{-}_{{\omega}})\times U(M^{\prime}_{{\omega}})\times U(M^{\prime\prime}_{{\omega}})\, , \\  
& {\Hf}_{\sf edg}^{\gamma} = 
\varprod_{{\eta} \in {\Egp} \amalg {\Egm}} \, U(1)  \, , \quad 
 {\Hf}_{\sf ver}^{\gamma}  =  
\varprod_{{\omega}\in {\Vg}} \, U(1) \\
\end{aligned} 
\label{eq:hfgq}
\eeq
and the embedding of ${\Hf}_{\sf ver}^{\gamma}$ into ${\Hf}_{\sf ff}^{\gamma} \times {\Hf}_{\sf edg}^{\gamma}$ is given by:
\beq
\left( u_{\omega} \right)_{\omega \in \Vg} \mapsto \left( u_{\omega} \cdot {\bf 1}_{N_{\omega}^{+}} \, , \,  u_{\omega} \cdot {\bf 1}_{N_{\omega}^{-}} \, , \, u_{\omega} \cdot {\bf 1}_{M_{\omega}^{\prime}} \, , \, u_{\omega} \cdot {\bf 1}_{M_{\omega}^{\prime\prime}} \right)_{\omega \in {\Vg}} \times \left( u_{s({\eta})\pm 1}^{-1} u_{t({\eta})} \right)_{{\eta} \in ^{\pm}{\sE}}
\eeq
The group \eqref{eq:globqgr} acts on ${\mM}^{\gamma}_{\bkt}({\bn}^{\pm}, {\mm})$ in the following fashion:
\begin{multline}
\left( \left( h_{\omega}^{+}, h_{\omega}^{-}, m_{\omega}^{\prime}, m_{\omega}^{\prime\prime} \right)_{\omega \in \Vg} \times \left( u_{\eta} \right)_{{\eta} \in \Eg} \times u \right) \quad \cdot \quad
\left[ {\bB}, {\bI}, {\bJ}, {\bf\Theta} \right]  = \\
= \left[ u^{\pm 1} u_{\eta} g_{t({\eta})\pm 1} B_{\eta}^{\pm}g_{s({\eta})}^{-1}\, , \, u^{\pm 1} u_{\eta}^{-1} g_{s({\eta})\pm 1} {\tilde B}_{\eta}^{\pm}g_{t({\eta})}^{-1}\, \vert_{{\eta} \in {\Eg}} \right. \\ \qquad\qquad\qquad\qquad \left. u^{\pm 1} g_{{\omega}\pm 1} I_{\omega}^{\pm} \left( h_{\omega}^{\pm} \right)^{-1} \, , \, h_{\omega\pm 1}^{\pm} J_{\omega}^{\pm} g_{\omega}^{-1} u^{\pm 1} \, , \, m_{\omega}^{\prime} {\Theta}_{\omega}^{\prime}  , \, {\Theta}^{\prime\prime}_{\omega} \left( m_{\omega}^{\prime\prime} \right)^{-1} \, \vert_{{\omega} \in {\Vg}} \right] \, ,\\
\label{eq:globqac}
\end{multline}
where we indicated the compensating ${\CalG}^{\gamma}_{\bkt}$-transformations. It is obvious from the Eq. \eqref{eq:globqac} that 
the ${\Hf}^{\gamma}_{\sf ff} \times {\Hf}^{\gamma}_{\sf edg}$-transformations which are in the image of ${\Hf}^{\gamma}_{\sf ver}$ can be undone by a ${\CalG}^{\gamma}_{\bkt}$-transformation. 

\subsubsection{Compactness theorem in the crossed quiver case}

Let us demonstrate the compactness of the set $\left( {\mM}^{\gamma}_{\bkt}({\bn}^{\pm}, {\mm}) \right)^{{\sf T}_{\gamma}}$ of ${\sf T}_{\gamma}$-fixed points in ${\mM}^{\gamma}_{\bkt}({\bn}^{\pm}, {\mm})$, where ${\sf T}_{\gamma} \subset {\Hf}^{\gamma}$ is a subtorus of the global symmetry group. The choice of ${\sf T}_{\gamma}$ is restricted by the following requirement: it must contain a $U(1)$-subgroup, to be denoted by $U(1)_{\sf v}$, such that  1) the composition  $p \circ i$, where $i : U(1)_{\sf v} \hookrightarrow {\Hf}^{\gamma}$ is the embedding, and  $p: {\Hf}^{\gamma} \to U(1)_{\sf u}$ is the projection,  is a non-trivial homomorphism, $v \mapsto v^{k}$, $k \neq 0$, 2) the embedding into ${\Hf}^{\gamma}$ is parametrized by the collection 
\beq
v \in  U(1)_{\sf v} \mapsto \left( v^{{\bn}_{\omega}^{+}}, v^{{\bn}_{\omega}^{-}}, v^{{\mm}_{\omega}'}, v^{{\mm}_{\omega}^{''}} \right)_{\omega \in \Vg} \times \ v^{\be} \  \times v^{k}
\label{eq:vdata}
\eeq
The symmetry \eqref{eq:shsym} reflects the quotient by ${\Hf}_{\sf ver}^{\gamma}$ in \eqref{eq:globqgr}.

We shall impose an additional requirement on the data $({\bn}^{\pm}_{\Vg}, {\mm}_{\Vg}^{\prime}, {\mm}_{\Vg}^{\prime\prime}, {\be}, k)$: for any ${\omega}', {\omega}'' \in {\Vg}$
and any ${\alpha}' \in [n_{\omega'}^{\pm}]$, ${\alpha}'' \in [n_{\omega''}^{\pm}]$
\beq
n_{\omega', \alpha'}^{\pm} - n_{\omega'', \alpha''}^{\pm} \pm k ( {\ell}+1) +  \int_{{\sf path}_{{\omega}' , {\omega}''}^{\pm}} e \neq 0
\label{eq:nonpolj}
\eeq
for any path ${\sf p}_{{\omega}' , {\omega}''}^{\pm}$.  Note \eqref{eq:nonpolj} is invariant under \eqref{eq:shsym}. 
The requirement \eqref{eq:nonpolj} can be slighlty weakened, namely one can allow \eqref{eq:nonpolj} to fail for a single pair $({\omega}' , {\alpha}') = ({\omega}'', {\alpha}'')$. In what follows we insist on \eqref{eq:nonpolj}, though. 

The proof goes as follows:
Define the function $\delta$ on the Grassmanian of subspaces $V \subset \oplus_{\omega} K_{\omega}$:
\beq
{\delta}_{V} = {\Tr}_{V} \, \left( \sum_{\eta\in {\Eg}, \pm} B_{\eta}^{\pm}\left( B_{\eta}^{\pm} \right)^{\dagger} + {\tilde B}_{\eta}^{\pm}\left( {\tilde B}_{\eta}^{\pm} \right)^{\dagger} + \, \sum_{\omega, \pm} I_{\omega}^{\pm} \left( I_{\omega}^{\pm}  \right)^{\dagger} +   J_{\omega}^{\pm} \left( J_{\omega}^{\pm}  \right)^{\dagger} \right)
\label{eq:deltv}
\eeq
The function $\delta$ is monotone: ${\delta}_{V'} \leq {\delta}_{V''}$ for $V' \subset V''$. 
We have:
\beq
\Vert {\bB} , {\bI} , {\bJ}  \Vert^2 = \sum_{\omega \in {\Vg}} {\delta}_{K_{\omega}} 
\eeq
Now, the ${\sf T}_{\gamma}$-invariance implies, by the usual arguments, that the spaces $K_{\omega}^{\pm}, K_{\omega}$, for each $\omega \in {\Vg}$ are ${\sf T}_{\gamma}$-representations, 
\beq
K_{\omega}^{\pm} = \bigoplus_{{\sf w}  \in {\sf T}_{\gamma}^{\vee}} \, K_{\omega, \sf w}^{\pm}
 \otimes {\CalR}_{\sf w} 
 \label{eq:tkomrep}
 \eeq
 where ${\CalR}_{\sf w}$ are the irreps of ${\sf T}_{\gamma}$. 

First, we need to prove that $J_{\omega}^{\pm} = 0$ for the ${\sf T}_{\gamma}$-invariant $({\bB}, {\bI}, {\bJ})$. The equations  \eqref{eq:holospikq} imply that $J_{\omega}^{\pm}(K^{\mp}_{\omega}) = 0$. Let us restrict \eqref{eq:tkomrep} onto $U(1)_{\sf v}$:
\beq
K_{\omega}^{\pm} = \bigoplus_{w  \in {\BZ}} \, K_{\omega, w}^{\pm}
 \otimes {\CalR}_{w} 
 \label{eq:resttkom}
 \eeq
 where ${\CalR}_{w}$ are the irreps of $U(1)_{\sf v}$: $v \mapsto v^{w}$. 
 We have: 
 \beq
 I_{\omega}^{\pm} (N_{\omega}^{\pm}) = \bigoplus_{\alpha = 1}^{n_{\omega}^{\pm}}
 \, N_{\omega, \alpha}^{\pm} \otimes {\CalR}_{n_{\omega, \alpha}^{\pm}}
 \eeq
 and
 \beq
 B_{\eta}^{\pm} (K^{\pm}_{s({\eta}), w}) \subset K^{\pm}_{t({\eta})+1, w+k + e({\eta})}\, , \quad
  {\tilde B}_{\eta}^{\pm} (K^{\pm}_{t({\eta}), w}) \subset K^{\pm}_{s({\eta})+1, w+ k - e({\eta})}\,
 \eeq
 Thus the weights $s$ which occur in the decomposition \eqref{eq:resttkom} have the form:
 \beq
 n^{\pm}_{{\omega}', {\alpha}'} \pm k {\ell} + \int_{{\sf p}_{{\omega}' , {\omega}}^{\pm}}e
 \label{eq:sweights}
 \eeq
 for some ${\omega}' \in {\Vg}$ and some length $\ell$ path ${\sf p}_{{\omega}' , {\omega}}^{\pm}$. 
 
 Now the $U(1)_{\sf v}$ equivariance implies that $J_{\omega}^{\pm} (K_{\omega, w} \otimes {\CalR}_{w})$ belongs to the eigenspace of $v \in U(1)_{\sf v}$ in $N_{\omega \pm 1}^{\pm}$
 with the eigenvalue $v^{w \mp k}$. Since the eigenvalues of $v \in U(1)_{\sf v}$ on 
 $N_{\omega \pm 1}^{\pm}$ are given by: $v^{n_{\omega \pm 1, \alpha}^{\pm}}$ the
 non-vanishing $J_{\omega}^{\pm}$ means that for some ${\omega}', {\alpha}'$ the eigenvalue 
 \eqref{eq:sweights} coincides with $n_{\omega \pm 1, \alpha} \mp k$, which contradicts  \eqref{eq:nonpolj}. Thus, ${\bJ} = 0$. 
 
Now, use the real moment map equation to deduce:
\begin{multline}
{\delta}_{K_{\omega, w}^{\pm}} = {\zeta}_{\omega} \, {\rm dim} (K_{\omega, w}^{\pm}) \ + \\ 
+\, \sum_{{\eta} \in s^{-1}({\omega}) \cap \, ^{\pm}{\sE}} {\Tr}_{K_{t({\eta}){\pm}1, w + k + e({\eta})}^{\pm}} B_{\eta}^{\pm} \left( B_{\eta}^{\pm} \right)^{\dagger} \ + \ \sum_{{\eta} \in t^{-1}({\omega}) \cap \, ^{\pm}{\sE}} {\Tr}_{K_{s({\eta}){\pm}1, w + k - e({\eta})}^{\pm}} {\tilde B}_{\eta}^{\pm} \left( {\tilde B}_{\eta}^{\pm} \right)^{\dagger} \leq \\
\leq \  {\zeta}_{\omega} \, {\rm dim} (K_{\omega, w}^{\pm}) \ + \ \sum_{{\eta} \in s^{-1}({\omega}) \cap \, ^{\pm}{\sE}} {\delta}_{K_{t({\eta}){\pm}1, w + k + e({\eta})}^{\pm}}  \ + \ \sum_{{\eta} \in t^{-1}({\omega}) \cap \, ^{\pm}{\sE}} {\delta}_{K_{s({\eta}){\pm}1, w + k - e({\eta})}^{\pm}} \label{eq:delkompm}
\end{multline}
 Now repeat the same estimate by pushing the arguments $w'$ of the $\delta_{K_{{\omega}', w'}}$'s in the right hand side of \eqref{eq:delkompm} outside the domain where the corresponding $K_{{\omega}', w'}$ spaces are non-trivial (this is possible because the total dimension of the $K$ space is finite). 
 In this way we get an upper bound on $\delta_{K_{{\omega}', w'}}$'s and the norms of $B$'s, ${\bI}$'s and ${\bJ}$'s, as promised. 
  
\subsection{Orbifolds and defects: ADE $\times\ U(1) \ \times $ ADE}

The construction above can be motivated by the following examples. 

Recall that the moduli space ${\iM}^{+}_{k}(n, w)$
of crossed instantons has an ${\mathrm{SU}}(2) \times {\mathrm{U}}(1) \times {\mathrm{SU}}(2)$ symmetry. 

\subsubsection{Space action} 
Let $\Gamma$ be a discrete subgroup of 
${\mathrm{SU}}(2)_{12} \times {\mathrm{U}}(1)_{\Delta} \times {\mathrm{SU}}(2)_{34}$, 
\beq
{\iota}: {\Gamma} \longrightarrow {\Gr} = {\mathrm{SU}}(2)_{12} \times {\mathrm{U}}(1)_{\Delta} \times {\mathrm{SU}}(2)_{34} \, , 
\label{eq:defhom}
\eeq
\subsubsection{Framing action}
Now let us endow the spaces $N_{12}$ and $N_{34}$
with the structure of $\Gamma$-module:
\beq
\begin{aligned}
& N_{12} = \bigoplus_{{\omega} \in {\Gamma}^{\vee}} N_{\omega} \otimes {\CalR}_{\omega} \, , \\
& N_{34} = \bigoplus_{{\omega} \in {\Gamma}^{\vee}} W_{\omega} \otimes {\CalR}_{\omega} \, , \\
\end{aligned}
\label{eq:ndecona}
\eeq
in other words let us fix the homomorphisms
\beq
{\rho}_{A} : {\Gamma} \longrightarrow U(n_{A})
\label{eq:gana}
\eeq
Let us denote by ${\bn}, {\bw}$ the vectors of dimensions
$({\rm dim}N_{12, \omega})_{\omega \in \Gamma^{\vee}}$, 
$({\rm dim}N_{34, \omega})_{\omega \in \Gamma^{\vee}}$,
respectively. 

The data \eqref{eq:ndecona}, \eqref{eq:defhom} defines the embedding of the group $\Gamma$ into $\Hf$, the symmetry group of  ${\iM}^{+}_{k}(n, w)$. 

\subsubsection{New moduli spaces}
The set of ${\Gamma}$-fixed points in ${\iM}^{+}_{k}(n,w)$
splits into components 
\beq
{\iM}^{+}_{k}(n,w)^{\Gamma} = \bigcup\limits_{\bkt}\ {\iM}^{+, \gamma_{\Gamma}}_{\bkt}({\bn},{\bw})
\eeq
This is a particular case of the space ${\mM}^{\gamma}_{\bkt}({\bn},{\bw},0)$. Indeed, let
${\Vg} = {\Gamma}^{\vee}$, while ${\Eg}^{\pm}$ are defined by decomposing 
\beq
{\CalR}_{\omega} \otimes {\BC}^{2}_{\substack{12\\ 34}} = \bigoplus_{e \in s^{-1}({\omega})\cap {\Eg}^{\pm}} {\CalR}_{t(e)} \bigoplus_{e \in t^{-1}({\omega})\cap {\Eg}^{\pm}} {\CalR}_{s(e)} 
\eeq 
The requirement that $\Gamma$ preserves the ${\mathrm{U}}(k)$-orbit
of $({\bB}, {\bI}, {\bJ})$ translates to the fact that
${\Gamma}$ is unitary represented in $K$, so that
\beq
{\gamma} \cdot ({\bB}, {\bI}, {\bJ}) = (g_{\gamma}{\bB}g_{\gamma}^{-1}, g_{\gamma} {\bI}, {\bJ} g_{\gamma}^{-1}) \, , \qquad {\gamma} \mapsto g_{\gamma} \in {\mathrm{U}}(K)
\label{eq:gequiv}
\eeq
Thus, we can decompose $K$ into the irreps of $\Gamma$:
\beq
K = \bigoplus_{{\omega} \in {\Gamma}^{\vee}} K_{\omega} \otimes {\CalR}_{\omega}
\label{eq:kdeco}
\eeq
The operators ${\bB}, {\bI}, {\bJ}$ then become linear maps between the spaces $K_{\omega'}, N_{A, {\omega}''}$, which can be easily classified by unraveling the equivariance conditions \eqref{eq:gequiv}. 
 
The components ${\iM}^{+, \gamma_{\Gamma}}_{\bkt}({\bn},{\bw})$
can then be deformed by modifying the real moment map equation to
\beq
{\mu} = \sum_{{\omega} \in {\Gamma}^{\vee}} {\zeta}_{\omega} \, {\bf 1}_{K_{\omega}}
\eeq
In the particular case ${\Gamma} \subset {\mathrm{SU}}(2)_{34}$ the orbifold produces the moduli spaces of supersymmetric gauge configurations
in the 
\beq
{\Gg} = \varprod_{{\omega} \in {\Gamma}^{\vee}} \, {\mathrm{U}}(n_{\omega})
\label{eq:gagr}
\eeq
gauge theory in the presence of a point-like defect, the $qq$-character
\beq
{\CalX}_{(w_{\omega})_{{\omega} \in {\Gamma}^{\vee}}} \left( x, 
{\nu}_{{\omega}, {\beta}} \right)
\label{eq:calxw}
\eeq
The gauge theory in question is the affine ADE quiver gauge theory.

\subsubsection{Odd dimensions and finite quivers}

We can also obtain the moduli space of supersymmetric gauge field configurations in the quiver gauge theories built on finite quivers. 
The natural way to do that is to start with an affine quiver and  
send some of the gauge couplings to zero, i.e. 
by making some of dimensions $k_{\omega}$ vanishing. 

Remarkably, this procedure produces the superspace, the odd variables originating in the multiplet of the cohomological field theory. Let us explain this in more detail. Let
us consider, for simplicity, the group $\Gamma \subset SU(2)_{34}$, so that
${\Eg}^{+}$ has one element. 

The linear algebra data 
$$
\left( \, B_{1, {\omega}}\,  , \, B_{2, {\omega}}\,  , \,  I_{12, \omega}\,  , \,  J_{12, \omega}\,  , \, 
I_{34, \omega}\,  , \,  J_{34, \omega}\,  , \,  B_{e}\,  , \, {\tilde B}_{e}  \, \right)
\in {\CalA}^{\gamma}_{\bkt}({\bn}, {\bw}, 0)$$ 
is parametrized by the 
\beq
2 \sum_{{\omega}\in {\Gamma}^{\vee}} k_{\omega} \left( k_{\omega} +  n_{\omega}  +  w_{\omega} \right)  + 2 \sum_{e \in E} k_{t(e)} k_{s(e)}
\eeq
complex dimensional space.   The Eqs. \eqref{eq:spikq} plus the  ${\rm GL}_{\gamma}({\bkt})$-invariance remove
\beq
 \sum_{{\omega} \in {\Gamma}^{\vee}} \left( n_{\omega}w_{\omega} + 2 k_{\omega} (k_{\omega} + w_{\omega}) \right)  +  \sum_{e\in E} 2 k_{s(e)} k_{t(e)}  +  k_{s(e)} n_{t(e)}  + k_{t(e)} n_{s(e)} 
\eeq
dimensions (this is half the number of equations \eqref{eq:holospikq}). The result is  ${\bkt}$-linear, 
\beq
{\rm virtual \ dim}{\iM}^{+, \gamma}_{\bkt}({\bn},{\bw}) = \sum_{{\omega}\in {\Gamma}^{\vee}} \left( k_{\omega} m_{\omega} - n_{\omega} w_{\omega} \right)
\eeq
where
\beq
m_{\omega} = 2 n_{\omega} - \sum_{e \in s^{-1}({\omega})} n_{t(e)} - \sum_{e \in t^{-1}({\omega})} n_{s(e)}
\label{eq:betak}
\eeq
Now, if for all $\omega \in {\Gamma}^{\vee}$ the deficits $m_{\omega}$ are non-negative, and at least for one vertex the deficit is positive then the quiver is, in fact, a finite ADE Dynkin diagram. In this case we can add the odd variables taking values in the spaces ${\rm Hom} (K_{\omega}, M_{\omega})$ with the complex vector space $M_{\omega}$ of dimension $m_{\omega}$, and define the moduli space to be the supermanifold which is the total space of the odd vector bundle ${\Pi}{\rm Hom} (K_{\omega}, M_{\omega})$ over the previously defined bosonic moduli space. In practice this means that the integration over the ``true'' moduli space is the integral over the coarse moduli space of the equivariant Euler class of the vector bundle ${\rm Hom} (K_{\omega}, M_{\omega})$. This is what the cohomological field theory applied to the affine quiver case with the subsequent setting $k_{\omega'} = 0$ for some $\omega' \in {\Gamma}^{\vee}$ would amount to. With the ``compensator'' vector bundle in place the virtual dimension of the moduli space becomes $\bkt$-independent. This is the topological counterpart of the asymptotic conformal invariance of the gauge theory.

\section{Spiked instantons on orbifolds and defects}

Now let us go back to the general case of spiked instantons. Choose a discrete subgroup $\Gamma$
of $U(1)^{3}_{\ept}$, e.g. ${\Gamma} \approx {\Xi}_{p_{1}} \times {\Xi}_{p_{2}} \times {\Xi}_{p_{3}}$, ${\Gamma}^{\vee} \approx {\Gamma}$.  Let $t_{a}: {\Gamma} \to U(1)$, $a \in {\4}$
be the homomorphisms corresponding to the embedding $U(1)^{3}_{\ept} \subset SU(4)$. We have:
\beq
\prod_{a\in {\4}} t_{a}({\gamma}) = 1
\eeq
Let ${\CalR}_{a} \in {\Gamma}^{\vee}, a \in {\4}$ be the corresponding one-dimensional representations of $\Gamma$, e.g ${\CalR}_{1} = {\CalR}_{1,0,0}$, ${\CalR}_{4} = {\CalR}_{-1,-1,-1}$.  We shall use the additive notation, ${\CalR}_{a} = {\CalR}_{{\vec\varpi}_{a}}$
so that
${\CalR}_{\vec\omega} \otimes {\CalR}_{a} = {\CalR}_{{\vec\omega}+{\vec\varpi}_{a}}$. 
Fix the framing homomorphisms:
${\rho}_{A,\Gamma} \to U(n_{A})$:
\beq
N_{A} = \bigoplus_{{\vec\omega} \in {\Gamma}^{\vee}} N_{A,\vec\omega}\otimes
{\CalR}_{\vec\omega}
\eeq
The set of ${\Gamma}$-fixed points in ${\mM}_{k}({\vec n})$
splits into components 
\beq
{\mM}_{k}({\vec n})^{\Gamma} = \bigcup\limits_{\bkt}\ {\mM}^{\gamma_{\Gamma}}_{\bkt}
({\vec\bn})
\eeq
It describes the moduli spaces of spiked instantons in the presence of additional 
surface and point-like conical defects. The compactness theorem holds in this case.  The proof
is a simple extension of the proof of section $\bf 8$ with the spaces $K_{A}$ replaced by 
$K_{A, {\vec\omega}}$, where ${\vec\omega} = ({\omega}_{1}, {\omega}_{2}, {\omega}_{3})$, ${\omega}_{i} \in {\Xi}_{p_{i}}$:
\beq
K_{A, {\vec\omega}} = \sum_{\vec\omega' \in {\Gamma}^{\vee}} \sum_{f \in {\BC}[x,y]} f(B_{a}, B_{b}) I_{A}(N_{A, {\vec\omega}'})
\label{eq:kaome}
\eeq
where the sum is over polynomials obeying:
\beq
f(t_{a}({\gamma})x, t_{b}({\gamma})y) = {\chi}_{{\CalR}_{{\vec\omega} - {\vec\omega}'}}({\gamma})
\label{eq:chiga}
\eeq
for all $\gamma \in {\Gamma}$. As before $B_{c, {\vec\omega}'}(K_{A, {\vec\omega}}) = 0$ whenever $c \notin A$. The vector $\bkt$ encodes the dimensions of the spaces 
\beq
K_{\vec\omega} = \sum_{A \in \6} K_{A, \vec\omega}\, 
\eeq
the operators $B_{a}$ have the block form:
\beq
B_{a}(K_{\vec\omega}) \subset K_{{\vec\omega} +{\vec\varpi}_{a}}
\label{eq:bakom}
\eeq
The norms $\Vert B_{a} \Vert^2$ are estimated with the help of the quantities
\beq
{\delta}_{A, {\vec \omega}, n}  = {\Tr}_{K_{A, \vec\omega}} \, \left( \sum_{a} B_{a}B_{a}^{\dagger} + \sum_{C} I_{C}I_{C}^{\dagger} \right)
\label{eq:delaomn}
\eeq

\section{Conclusions and future directions}

In this paper we introduced several moduli spaces: ${\mM}^{+}, {\mM}^{\fo}, {\mM}^{*}$ of matrices solving quadratic equations modulo symmetries. These moduli spaces generalize the Gieseker-Nakajima partial compactification ${\iM}$ of the ADHM moduli space of $U(n)$ instantons
on ${\BR}^{4}$. We gave some motivations for these constructions and proved the compactness theorem which we shall use in the next papers
to establish useful identities on the correlation functions of supersymmetric gauge theories in four dimensions. 

In this concluding section we would like to make a few remarks. 

First of all, one can motivate the crossed instanton construction by starting the with the ordinary ADHM construction and adding the co-fields \cite{Cordes:1994sd, Cordes:1995} which mirror the embedding of the ${\CalN}=2$ super-Yang-Mills vector multiplet into the ${\CalN}=4$ super-Yang-Mills vector multiplet \cite{Vafa:1994tf}.

Secondly, we would like to find the crossed instanton analogue of the stable envelopes of \cite{MO:2012}.

Third, it would be nice to generalize the spiked instanton construction to allow more general orbifold groups ${\Gamma} \subset SU(4)$, and
more general (Lagrangian?) subvarieties in ${\BC}^{4}/{\Gamma}$.

 Now, to the serious drawbacks of our constructions.  
The purpose of the ADHM construction, after all, is the construction of the solutions to the instanton equations \[ F_{A}^{+} = 0 \]
 We didn't find the analogue of the ADHM construction for the spiked instantons. Conjecturally, the matrices $[{\bB}, {\bI},{\bJ}]$ solving the Eqs. \eqref{eq:spik} are in one-to-one correspondence with the finite action solutions to Eqs. \eqref{eq:zzbpsii}. 

Finally, we have proposed a definition of quiver crossed instantons, which are 
defined for quivers more general then the products of ADE Dynkin diagrams. It would be interesting to find the precise restrictions on these quivers compatible with the compactness theorem. 

In the forthcoming papers the compactness theorem will be used to derive the main statements of the theory of $qq$-characters \cite{Nekrasov:2015i}. While this paper was in preparation, the algebraic counterpart of our compactness theorem was studied in 
\cite{Kimura:2015rgi}. Various consequences of the compactness theorem will be studied in \cite{Nekrasov:2015ii}. Some of them have already been observed in \cite{Poghossian:2010pn, Fucito:2011pn, Nekrasov:2013xda, Aganagic:2013tta, Aganagic:2014kja, Poghosyan:2016mkh}. 

\bibliographystyle{acm}
\bibliography{Nekrasov-BPSCFT}

\end{document}